\documentclass[11pt]{amsart}
\usepackage{etex}
\newtheorem{Theorem}{Theorem}[section]
\newtheorem{Proposition}[Theorem]{Proposition}
\newtheorem{Lemma}[Theorem]{Lemma}

\newtheorem{Remark}[Theorem]{Remark}

\newtheorem{Assumption 2}[Theorem]{Assumption 2}

\numberwithin{equation}{section}

\usepackage{amsmath}
\usepackage{amssymb}
\usepackage{lmodern}
\usepackage{stmaryrd}

\let\savedegree\degree
\let\degree\relax
\usepackage{mathabx}
\let\degree\savedegree

\usepackage{gensymb}

\def\k#1{\kern#1em}
\def\Ib#1{{I\kern-.25em#1}}
\def\Ibb#1{{I\kern-.23em#1}}

\def\CC{{\mathbb C}}

\def\KK{{\mathbf K}}

\def\NN{{\mathbb N}}

\def\RR{{\mathbb{R}}}

\def\xx{\mathbf x}

\def\vci{\vrule  width.02em height1.47ex depth-.0ex}
\def\11{{\rm\k{.2}\vci\k{-.37}1}}

\def\fin{{\begin{flushright}
\it{Q.E.D.}
\end{flushright}}}

\begin{document}

\address{Universit\'e de Bordeaux, Institut de Math\'ematiques, UMR CNRS 5251, F-33405 Talence Cedex}

\email{alain.bachelot@u-bordeaux.fr}

\title{Wave asymptotics at a cosmological time-singularity: classical
  and quantum scalar fields}

\author{Alain BACHELOT}

\begin{abstract}
We investigate the propagation of the scalar waves in the FLRW
universes beginning with a Big Bang and ending with a Big Crunch, a Big
Rip, a Big Brake or a Sudden Singularity. We obtain the sharp
description of the asymptotics for the solutions of the linear Klein-Gordon equation,
and similar results for the  semilinear equation with a subcritical
exponent. We prove that the number of cosmological particle creation is finite under general assumptions on the initial Big Bang
and the final Big Crunch or Big Brake.
\end{abstract}

\maketitle

\pagestyle{myheadings}
\markboth{\centerline{\sc Alain Bachelot}}{\centerline{\sc {Wave asymptotics at a cosmological time singularity}}}
\section{Introduction}
A lot of cosmological models have a time singularity either in the past, such
as the Big Bang, or in the future, such as the Big Crunch, the Big
Rip introduced by R.R. Caldwell \cite{caldwell}, the Big Brake
introduced in \cite{gorini} or a Sudden Singularity (see {\it e.g.}
 \cite{visser}, \cite{fernandez}, \cite{nojiri},  for a classification of these time singularities). The physical literature on
this subject is very abundant. In particular we mention the works by
J. D. Barrow {\it et alii} \cite{barrow2011}, \cite{barrow2010}, \cite{barrow2015},
\cite{barrow2009}.  The non linear hyperbolic equations, mainly
the Einstein system, have been
deeply investigated near these singularities, we can cite {\it e.g.} the work
by L. Andersson and A. D. Rendall
\cite{andersson} on the Belinskii-Khalatnikov-Lifshitz asymptotics at
the singularities, the study of T. Damour, M. Henneaux, A. D. Rendall,
M. Weaver \cite{[1]}, the recent papers by I. Rodianski and J. Speck \cite{[11]},
\cite{[12]}, \cite{[13]}, \cite{[15]}, and the
Fuchsian methods that have been applied to the Gowdy spacetimes by
S. Kichenassamy and A. D. Rendall \cite{[4]}, \cite{[5]},
F. Beyer and P. G. LeFloch \cite{beyer2010}, \cite{beyer2017};
moreover the asymptotic behaviour of generic vacuum solutions in the
$T^3$-Gowdy setting is deeply investigated by H. Ringstr\"om in \cite{[7]}. There are also numerous works on the
 field equations in the smooth (without singulatity) FLRW space-times,
 in particular in the De Sitter universe (see the fundamental work of
 H. Friedrich \cite{[2]}, and also {\it e.g.} 
\cite{dodeca}, \cite{ebert2018},
\cite{galstian2009},  \cite{galstian2015}, \cite{galstian2017}, \cite{[3]},
\cite{nakamura}, \cite{[6]}, \cite{[8]}, \cite{[10]},
\cite{[14]}, \cite{[16]}). In contrast, surprisingly enough, few mathematical
papers deal with the simpler issue of
the behaviour of the linear fields near the cosmological
singularities. We can cite the works by P. T. Allen and A. D. Rendall \cite{allen} on the
cosmological 
linearized perturbations; there are also recent works:  in an
important monography   \cite{ringstrom2017}, H. Ringstr\"om
investigates the linear systems on cosmological backgrounds, in
particular on the
Kasner solutions; \cite{ringstrom2018} deals with the
Klein-Gordon equation on Bianchi backgrounds and silent singularities;
A. Alho, G. Fournodavlos, A. T. Franzen study
the wave equation near a Big Bang in \cite{alho}. In this paper, we
study the linear Klein-Gordon equation $\Box u+m^2u+\xi Ru=0$ on a
type of
spatially complete warped geometries, including the FLRW
universes for which the scale factor has  time singularities, and we obtain
accurate asymptotics of the fields at these points. These issues are
also investigated by H. Ringstr\"om who
considers in \cite{ringstrom2017} spatially compact settings whith a
different methodology.
The important case of $\xi=0$, $m=0$ is studied for some Big Bang in \cite{alho} and
\cite{ringstrom2018} (see Remark \ref{ringostarnc} below).
We also discuss
the semilinear equations in the subcritical case and we get similar
results. Concerning the linear quantum fields, we investigate the cosmological particle production between a
Big Bang and a Big Crunch or a Big Brake. In short, in this paper,  we develop the
rigorous mathematical framework for most of the results
obtained by  J. D. Barrow  {\it et alii} in the previously cited
papers. Now we describe our
setting and the results.\\

Given a $d$-dimensional complete Riemannian $C^2$  manifold
$\left(\mathbf{K},\gamma_{ij}d\xx^id\xx^j\right)$, $d\geq 3$, and a bounded
open interval $I=(t_-,t_+)\subset\RR$, we
consider the FLRW universe $(\mathcal{M},g)$ where
$\mathcal{M}:=I_t\times\mathbf{K_x}$ is endowed with
the  Lorentzian metric
\begin{equation}
 \label{g}
 g_{\mu\nu}dx^{\mu}dx^{\nu}=dt^2-a^2(t) \gamma_{ij}d{\mathbf
   x}^id{\mathbf x}^j.
\end{equation}
The scale factor $a(t)$ is a positive  function in $C^2(I)$ and we
assume that near $t_{\pm}$ we have for $k\leq 2$
\begin{equation}
 \label{condida}\
\frac{d^ka}{dt^k}(t)=\frac{d^k}{dt^k}\left[c_0^{\pm}\mid t-t_{\pm}\mid^{\eta^{\pm}_0}+c_1^{\pm}\mid t-t_{\pm}\mid^{\eta^{\pm}_1}\right]+o\left(\mid
  t-t_{\pm}\mid^{\eta^{\pm}_1-k}\right),\;\; t\rightarrow t_{\pm},
\end{equation}
where the coefficients $c_0^{\pm},\;\eta_j^{\pm}\in\RR$ satisfy :
\begin{equation}
 \label{condidac}
 c^{\pm}_0>0,\;\eta^{\pm}_0<\eta^{\pm}_1.
\end{equation}
The {\it  Big Bang} (resp. the {\it Big Crunch}) corresponds to the case
$\eta^{-}_0>0$ (resp. $\eta^{+}_0>0$). It will be useful to
distinguish the $C^0-Big\,Bang/Crunch$ for which
$\eta_0^{\pm}\in(0,1)$, from the $C^1-Big\,Bang/Crunch$  for which
$\eta_0^{\pm}\geq 1$. A {\it Sudden Singularity} at
$t_{\pm}$ is associated to $\eta^{\pm}_0=0$ and
$\eta^{\pm}_1\notin\NN$.  A {\it Big Brake} is a Sudden Singularity at
$t_+$ with
$c_1^+>0$ and $\eta^{+}_1>1$. A very severe singularity is the {\it Big Rip}, defined by
$\eta^{\pm}_0<0$, that is {\it Slow} if $\eta_0^{\pm}\in (-1,0)$ and
{\it Strong} if $\eta_0^{\pm}\leq-1$. In this paper we consider an universe with such
time singularities at $t_{\pm}$, and we
investigate the asymptotic behaviour near $t_{\pm}$  of the solution $u$ of the
Klein-Gordon type equation
\begin{equation}
 \label{kg}
 {\Box}_{g}u+m^2u+\xi R_gu=0,
\end{equation}
where ${\Box}_{g}$ is the D'Alembertian operator
\begin{equation*}
 \label{box}
 {\Box}_{g}u:=\frac{1}{\sqrt{|\det(g)|}} \partial_{\mu}\left(g^{\mu \nu}
\sqrt{|\det(g)|}\partial_{\nu}u\right)=\left[\partial_t^2+d\frac{a'(t)}{a(t)}\partial_t-\frac{1}{a^2(t)}\Delta_{\mathbf K}\right]u,
\end{equation*}
$\Delta_{\mathbf K}$ is the Laplace-Beltrami operator on
$\mathbf{K}$,
\begin{equation*}
 \label{laplace}
 \Delta_{\mathbf K}u:=\frac{1}{\sqrt{\det(\gamma)}} \partial_{i}\left(\gamma^{ij}
\sqrt{\det(\gamma)}\partial_{j}u\right),
\end{equation*}
$m\geq 0$ is the mass of the field, $\xi\in\RR$ is a given  constant,
and $R_g$ is the Ricci scalar of $g$.
We may consider (\ref{kg}) as a kind of master equation arising in
different contexts. $u$ can be a component of the
metric tensor when the first-order perturbed Einstein equations are
recast as a second order PDE, or $u$ is a scalar field involving in several
scenarios studied in cosmology, such as the quintessence (see {\it
  e.g.} 
\cite{barrow2011}, \cite{barrow2015}, \cite{barrow2009},
\cite{chimento}, \cite{elizalde} and the references therein). 

In this Klein-Gordon equation, $\xi$ that is associated with the
coupling of the field with the geometry, plays a fundamental role. The
behaviour of the field near the time singularity $t_{\pm}$ crucially
depends on its value. To discuss the different situations it is
convenient to use the conformal time $\tau$ that, given $t_0\in
(t_-,t_+)$, is defined by
\begin{equation}
 \label{tau}
\tau:=\int_{t_0}^t\frac{1}{a(s)}ds,
\end{equation}
and we introduce the Liouville transform $\varphi$ of $u$:
\begin{equation}
 \label{phi}
 \varphi(\tau,\xx):=[a(t)]^{\frac{d-1}{2}}u(t,\xx).
\end{equation}
Then $u$ is
solution of (\ref{kg}) iff $\varphi$ satisfies
\begin{equation}
 \label{kgfifi}
 \left\{\partial_{\tau}^2-\Delta_{\mathbf
     K}+\xi R_{\gamma}(\xx)+V(\tau)\right\}\varphi=0,\;\;\tau\in(\tau_-,\tau_+),
\end{equation}
where
\begin{equation}
 \label{VQ}
 V(\tau)=m^2\alpha^2(\tau)+\left(\xi-\frac{d-1}{4d}\right)Q(\tau),
\;\;
 Q(\tau):=d(d-3)\frac{\alpha'^2(\tau)}{\alpha^2(\tau)}+2d\frac{\alpha''(\tau)}{\alpha(\tau)}.
\end{equation}
\begin{equation}
 \label{alpha}
 \alpha(\tau):=a(t),\;\;\tau_{\pm}:=\int_{t_0}^{t_{\pm}}\frac{1}{a(s)}ds\in[-\infty,\infty].
\end{equation}

The behavior of the field depends on the asymptotic behavior of the
potential near the singularity. We summarize the different cases in
Figure \ref{tableV}.

\begin{figure}[h!]
$$
\begin{array}{|c|c|c|c|c|}
\hline
\rm Singularity&\eta_0^+&\xi&\tau_+&V(\tau)\\
\hline
C^1\; Big\;Crunch&(1,\infty)&\RR&+\infty&O(\tau^{-2})\\
\hline
C^1\; Big\;Crunch&1&\RR&+\infty&C(\xi-(d-1)/4d)+O(e^{-c_0^+\tau})\\
\hline
C^0\; Big\;Crunch&(0,1)&\frac{d-1}{4d}&<\infty&\sim
                                               C(\tau_+-\tau)^{2\eta_0^+/(1-\eta_0^+)}\\
\hline
C^0\; Big\;Crunch&(0,1)&\neq\frac{d-1}{4d}&<\infty&\sim
                                                   C(\tau_+-\tau)^{-2}\\
\hline
Sudden\;Singularity&0&\frac{d-1}{4d}&<\infty&\sim m^2(c_0^+)^2\\
\hline
Sudden\;Singularity&0&\neq\frac{d-1}{4d}&<\infty&\sim
                                                 m^2(c_0^+)^2+Cc_1^+(\tau_+-\tau)^{\eta_1^+-2}\\
\hline
Big\;Rip&(-\infty,0)&\frac{d-1}{4d}&<\infty&\sim
                                            C(\tau_+-\tau)^{2\eta_0^+/(1-\eta_0^+)}\\
\hline
Big\;Rip&(-\infty,0)&\neq\frac{d-1}{4d}&<\infty&\sim
                                                C(\tau_+-\tau)^{-2}\\
\hline
\end{array}
$$
\caption{Asymptotics of the potential at the singularity. In the last
  column, $C\in\RR\setminus\{0\}$ and $V(\tau)\sim
W(\tau)$ means $V(\tau)=W(\tau)+o\left(W(\tau)\right)$ as
$\tau\rightarrow \tau_+$.}
\label{tableV}
\end{figure}


In the case of a $C^1-$Big Crunch, $\eta_0^+\geq 1$, we have $\tau_+=\infty$ and we deal with a
scattering problem for a Klein-Gordon equation with a time dependent
potential $V(\tau)$. This situation is simple because $V(\tau)=O(\tau^{-2})$ if $\eta_0^+>1$ hence we have to compare the solutions of (\ref{kgfifi})
with those of
\begin{equation}
 \label{asssfr}
 \left\{\partial_{\tau}^2-\Delta_{\mathbf
     K}+\xi R_{\gamma}(\xx)\right\}\varphi=0,\;\;\tau\in[0,\infty).
\end{equation}
If $\eta_0^+=1$, $V(\tau)
=(c_0^+)^2d(d-1)\left(\xi-\frac{d-1}{4d}\right)+O\left(e^{-c_0^+\tau}\right)$
and the asymptotic dynamics is given by
\begin{equation}
 \label{asssmod}
 \left\{\partial_{\tau}^2-\Delta_{\mathbf
     K}+\xi R_{\gamma}(\xx)+(c_0^+)^2d(d-1)\left(\xi-\frac{d-1}{4d}\right)\right\}\varphi=0,\;\;\tau\in[0,\infty).
\end{equation}
Therefore the investigation of the behaviour of $u(t)$ at the
singularity $t=t_+$ is reduced to a classic problem of asymptotic
completeness of wave operators for $\tau\rightarrow+\infty$. Given $u$
solution of (\ref{kg}), we look
for $\varphi_0^+$ and $\varphi_1^+$ such that the solution $\varphi$
of (\ref{asssfr}) or (\ref{asssmod}) with initial data
$\varphi(0)=\varphi_0^+$, $\partial_{\tau}\varphi(0)=\varphi_1^+$,
satisfies in a suitable sense
\begin{equation}
 \label{assu}
 [a(t)]^{\frac{d-1}{2}}u(t,\xx)\sim\varphi(\tau,\xx),\;\;a(t)\partial_t\left[[a(t)]^{\frac{d-1}{2}}u(t,\xx)\right]\sim \partial_{\tau}\varphi(\tau,\xx),\;\;t\rightarrow t_+.
\end{equation}
Conversely, given $\varphi_0^+$, $\varphi_1^+$, we expect to find $u$
solution of (\ref{kg}) satisfying (\ref{assu}) and we want to
prove that the map
\begin{equation}
 \label{W!}
 W_+:\;(u(t_0),\partial_tu(t_0))\mapsto (\varphi_0^+,\varphi_1^+)
\end{equation}
is an isomorphism on a suitable functional space.
Numerous
papers have been devoted to this issue for wave equations with a time
dependent mass (see {\it e.g.} \cite{ebert2017} and the references
therein). We show that all the expected results hold with a convenient
cut-off for the low frequencies in order to eliminate the exponentially
increasing solutions of (\ref{asssfr}) or (\ref{asssmod}). These modes
possibly occur if $\xi R_{\gamma}(\xx)$  is negative at some
points. In this case, the spectrum of the operator $-\Delta_{\KK}+\xi
R_{\gamma}$ can intersect
$(-\infty,0)$ and we need an {\it infrared cut-off} to eliminate these
solutions.\\

The case $\eta_0^+<1$ is deeply different. We have $\tau_+<\infty$, therefore we are interested in the existence of the following limits
\begin{equation}
 \label{limu}
 \varphi_0^+:=\lim_{t\rightarrow t_+}[a(t)]^{\frac{d-1}{2}}u(t,\xx)=\lim_{\tau\rightarrow\tau_+}\varphi(\tau,\xx),
\end{equation}
\begin{equation}
 \label{limdtu}
 \varphi_1^+:=\lim_{t\rightarrow t_+}a(t)\partial_t\left([a(t)]^{\frac{d-1}{2}}u(t,\xx)\right)=\lim_{\tau\rightarrow\tau_+}\partial_{\tau}\varphi(\tau,\xx).
\end{equation}

Moreover $\alpha^2(\tau)=O\left((\tau_+-\tau)^{\gamma}\right)$ with
$\gamma=\frac{2\eta_0^+}{1-\eta_0^+}>-2$, and $Q(\tau)\sim c(\tau_+-\tau)^{-2}$. Therefore
(\ref{kgfifi}) is a Klein-Gordon equation with a time-dependent
potential $V(\tau)$ that is singular at $\tau_+$. The existence of the
previous limits is obtained in extending to $[0,\tau_+]$ the solutions
defined on $[0,\tau_+)$. Conversely, given $\varphi_0^+$,
$\varphi_1^+$, we investigate the Cauchy problem for the singular
Klein-Gordon equation (\ref{kgfifi}) with the initial data specified
at the singularity $\tau_+$. Therefore we want to prove the existence
of the operator $W_+$ defined by (\ref{W!}) and its invertibility. Such an equation has
been investigated by D. Del Santo, T. Kinoshita, M. Reissig in \cite{delsanto}, where the authors study
the behaviour of the solution near the time singularity of a
Klein-Gordon equation. It turns out that their equation corresponds, in our context, to the case for
which $\KK$ is  the Euclidean space. We shall take advantage of the
elegant techniques used in  \cite{delsanto}  to investigate our general
situation. We now briefly present our results for the future time
singularity $t_+$ (the results at $t_-$ are obtained straightforwardly
by reversing
time).\\

 First we consider the  case
\begin{equation}
 \label{xiconf}
 \xi=\frac{d-1}{4d}
\end{equation}
called {\it
   conformal coupling} (if the mass is zero we have $V=0$ and the equation (\ref{kg})
 is conformally invariant).  Then the potential $V$ in equation (\ref{kgfifi})
 is just
$$
V(\tau)=m^2\alpha^2(\tau),
$$
and we
have $V(\tau)\sim c(\tau_+-\tau)^{\gamma}$. If $\eta_0^+>-1$ we have
$\gamma>-1$ hence $V\in L^1(0,\tau_+)$. This integrability allows to
easily solve the Cauchy problem even for the initial data specified at
the singularity, therefore $W_+$ exists and is invertible. The case of
the Strong Big Rip $\eta_0^+\leq-1$ is much more delicate. We prove
that $\varphi_0^+$ is well defined but in general $\varphi_1^+$ does
not exist as we can see in explicit examples. We summarize our results
for the conformal coupling in Figure \ref{figueconf}.\\

\begin{figure}[h!]
$$
\begin{array}{|c|c|c|c|c|}
\hline
\rm Singularity&\eta_0^+&\rm Infrared\;Cut-Off&\varphi_0^+&\varphi_1^+\\
\hline
C^1\; Big\;Crunch&[1,\infty)&YES&YES&YES\\
\hline
C^0\; Big\;Crunch&(0,1)&\;&\;&\;\\
Sudden\;Singularity&0&NO&YES&YES\\
Slow\;Big\;Rip&(-1,0)&\;&\;&\;\\
\hline
Strong\;Big\;Rip&(-\infty,-1]&NO&YES&NO\\
\hline
\end{array}
$$
\caption{Results of existence of $\varphi_j^+$ for the conformal
  coupling $\xi=\frac{d-1}{4d}$. When $\varphi_0^+$ and $\varphi_1^+$
exist, $W_+$ is an isomorphism.}
\label{figueconf}
\end{figure}

The {\it non conformal coupling} corresponds to $\xi\neq
\frac{d-1}{4d}$. In this case the singularity of the potential $V$ at
$\tau_+$ is strong if $\eta_0^+\neq 0$ since
\begin{equation}
 \label{}
 V(\tau)\sim\left(\xi-\frac{d-1}{4d}\right)\left(d+1-\frac{2}{\eta_0^+}\right)d\frac{(\eta_0^+)^2}{1-\eta_0^+}(\tau_+-\tau)^{-2},\;\;\tau\rightarrow\tau_+.
\end{equation}
Let's consider the ordinary differential equation
$$
\varphi''+A^2\varphi+\frac{\frac{1}{4}-\mu^2}{(\tau_+-\tau)^2}\varphi=0,\;\;\tau<\tau_+,
$$
that is a toy model for (\ref{kgfifi}) (for example if $\KK=S^d$). The
solutions are given by
$$
\varphi(\tau)=(\tau_+-\tau)^{\frac{1}{2}}\left[C_1J_{\mu}\left(A(\tau_+-\tau)\right)+C_2Y_{\mu}\left(A(\tau_+-\tau)\right)\right],
$$
and we can see that in general the limits (\ref{limu}) and
(\ref{limdtu}) can not exist. Nevertheless, if $\mu^2< 0$
$\varphi(\tau) $ tends to zero as $\tau\rightarrow\tau_+$ (and
$\partial_{\tau}\varphi(\tau)$ can be diverging). Therefore it is
natural to investigate the case
\begin{equation}
 \label{condichi}
 \left(\xi-\frac{d-1}{4d}\right)\left(d+1-\frac{2}{\eta_0^+}\right)d\frac{(\eta_0^+)^2}{1-\eta_0^+}>\frac{1}{4}.
\end{equation}
We prove that in this case, the solution of (\ref{kgfifi}) with
$\eta_0^+<1$, $\eta_0^+\neq 0$, tends also
to zero. This result is not surprising since (\ref{condichi}) assures
that $V$  acts like an infinite
potential barrier. The case of the $C^0$-Big Crunch without hypothese
(\ref{condichi}) is open but there are recent results for $\xi=0$,
$d=3$, $m=0$: we have to mention \cite{alho} if $\KK=\RR^3$, $\gamma_{ij}=\delta^i_j$,
$a(t)=t^{\eta_0}$, $1/3<\eta_0<1$, and the general study by H. Ringstr\"om \cite{ringstrom2018}, for the Bianchi
spacetimes with silent singularities (see Remark \ref{ringostarnc} below).\\

The situation $\eta_0^+=0$ is very special since
$V(\tau)$ is less singular, and the asymptotic behaviour of the field near
a Sudden Singularity heavily depends on the lower term associated with $\eta_1^+$.

\begin{figure}[h!]
$$
\begin{array}{|c|c|c|c|c|c|c|}
\hline
\rm Singularity&\eta_0^+&\eta_1^+&\xi&\rm  Infrared\;Cut-Off&\varphi_0^+&\varphi_1^+\\
\hline
C^1\;Big\;Crunch&[1,\infty)&(\eta_0^+,\infty)&\RR&YES&YES&YES\\
\hline
C^0\;Big\;Crunch&(0,1)&(\eta_0^+,\infty)&(\rm\ref{condichi})&NO&YES\;(=0)&NO\\
\hline
Big\;Brake&0&(1,\infty)&\RR&NO&YES&YES\\
\hline
Sudden\;Singularity&0&(0,1)&\RR&NO&YES&NO\\
\hline
Big\;Rip&(-\infty,0)&(\eta_0^+,\infty)&(\rm\ref{condichi})&NO&YES\; (=0)&NO\\
\hline
\end{array}
$$
\caption{Results of existence of $\varphi_j^+$ for non conformal coupling $\xi\neq\frac{d-1}{4d}$. When $\varphi_0^+$ and $\varphi_1^+$
exist, $W_+$ is an isomorphism.}
\end{figure}

We also tackle the problem of nonlinear fields. Important results
are known for smooth universes, in particular for the De Sitter
spacetime. Besides the studies on the stability of the Einstein-nonlinear
field systems \cite{[6]}, \cite{[8]}, \cite{[10]},
\cite{[14]}, \cite{[16]}, we can cite for the analysis of non-linear
wave equations, the works by Ebert and Reissig
\cite{ebert2018}, Galstian and Yagdjian
\cite{galstian2015}, \cite{galstian2017}, Nakamura \cite{nakamura}. At
our knowledge a lot of hard issues about interacting fields on singular
Lorentzian manifolds remain to be investigated. In this paper we present just a modest
contribution to this topic. We consider the
semilinear Klein-Gordon equation
\begin{equation}
 \label{nlkg}
 {\Box}_{g}u+m^2u+\xi R_gu=-Cu\mid u\mid^{p-1}.
\end{equation}
In this work we investigate only the subcritical case
\begin{equation}
 \label{pe}
 1\leq p\leq\frac{d}{d-2}.
\end{equation}
In this simple situation we may use the usual Sobolev embeddings to
solve the Cauchy problem, and choosing $C>0$ the finite energy
solutions are global. We study the asymptotic behaviour at a Big
Bang/Crunch or at a Sudden Singularity for the conformal coupling
$\xi=\frac{d-1}{4d}$ and we obtain the same results as for the linear
equation.\\

Finally we investigate some quantum aspects in a universe
beginning with a Big Bang and ending with a Big Crunch or a Sudden
Singularity.
The study of cosmological particle creation in an expanding
universe has a long history since the
PhD of L. Parker in 1965 \cite{parker}. We have to mention the work of
S. A. Fulling \cite{fulling-79} for the spatially flat Robertson-Walker metrics
and the generalized Kasner (Bianchi type I) metrics.
For more recent works, see {\it e.g.} \cite{calderon}, \cite{degner}, \cite{fulling},
\cite{grib}.
Fulling had investigated
the creation of particles between two regular times
$t_1,t_2\in(t_-,t_+)$ on ${\bf K}=\RR^3$. Here we consider the case of the time
singularities $t_1=t_-$, $t_2=t_+$
when a Big Bang, a Big Crunch or a Big Brake occurs and ${\bf
  K}$ is a $d$-dimensional compact manifold. In
the last part, we give sufficient conditions on $d$ and $\eta_j^{\pm}$ for the number of
cosmological particle creation  to be finite. This number $\mathcal
N$  is
estimated by the Zeta function of the Laplacian on $\bf{K}$,
$$
\mathcal{N}\lesssim \sum_n\lambda_n^{-l(d)-1},\;\;l(d):=\max(2,[d/2]),
$$
where $[x]$ is the integer part of $x$ and $\lambda_n$ are the
positive eigenvalues of $-\Delta_{\bf K}$. The proof, inspired from
Fulling \cite{fulling-79}, is based on the Liouville-Green (or WKB)
approximation that needs a sufficient  regularity of the
coefficients. That explains why the results are established for a scale factor
$a\in C^l(t_-,t_+)$ satisfying (\ref{condida}) for $0\leq k\leq l$ with
$l=l(d) $ if $\xi=\frac{d-1}{4d}$ and $l=l(d)+2$ if $\xi\neq\frac{d-1}{4d}$.
\section{The Linear and Nonlinear Global Cauchy Problems}
In this part we investigate the Cauchy problem associated to the
inhomogeneous Klein-Gordon equation
\begin{equation}
 \label{KG}
 \left\{\partial_t^2+d\frac{a'(t)}{a(t)}\partial_t-\frac{1}{a^2(t)}\Delta_{\mathbf
     K}+m^2+\xi R_g(t,\xx)\right\}u=F,\;\;t\in I,\;\;{\mathbf x}\in\mathbf{K},
\end{equation}
\begin{equation}
 \label{CI}
 u(t_0,\xx)=u_0(\xx),\;\;\partial_tu(t_0,\xx)=u_1(\xx),
\end{equation}
where $ t_0$ is fixed in $I$. If $a\in C^{\infty}(I)$, the existence
and the uniqueness of the solution is well known since the seminal
works of Leray \cite{leray}, for smooth or distributions initial
data. We denote $X$ the space  $C^{\infty}(\mathbf{K})$ or
$C^{\infty}_0(\mathbf{K})$ or $\mathcal{D}'(\mathbf{K})$. Then for any
$u_0,u_1\in X$, $F\in C^{\infty}(I;X)$, the classic results of Leray
assure the existence and uniqueness of $u\in C^{\infty}(I;X)$. Here we
are interested in the finite energy solutions and the quantum fields,
and we assume just $a\in C^2(I)$. Hence we have to introduce a
suitable Hilbertian functional framework (for the $H^1$-solutions see
\cite{choquet}).
If $\xi\neq 0$ we assume that the scalar curvature $R_{\gamma}$ of
$(\mathbf{K},\gamma)$ is bounded
\begin{equation}
 \label{rb}
 R_{\gamma}\in L^{\infty}(\mathbf{K}).
\end{equation}

We denote $L^2(\mathbf K)$ the $L^2$-Lebesgue space on $\mathbf{K}$
endowed with the volume measure associated to the metric
$\gamma$, that we simply denote $d{\mathbf x}$. A well known result by
Strichartz (Theorem 2.4 in \cite{strichartz}) assures that
$\Delta_{\mathbf K}$ is essentially selfadjoint on
$C^{\infty}_0(\mathbf K)$ since $\mathbf K $ is a complete manifold, and we shall use the Sobolev spaces $H^s(\mathbf K)$,
$s\in\RR$, defined as the closure of $C^{\infty}_0(\mathbf K)$ for the norm
\begin{equation}
 \label{normhs}
 \Vert f\Vert_{H^s(\mathbf K)}:=\left\Vert \left(-\Delta_{\mathbf
       K}+1\right)^{\frac{s}{2}}f\right\Vert_{L^2(\mathbf K)}.
\end{equation}

More precisely, $H^s(\mathbf K)$ is the Bessel-potential space
introduced by Strichartz \cite{strichartz}. Alternatively, if $\mathbf K$ has a bounded geometry, we can
characterize  $H^s(\mathbf K)$ via localization and pull-back onto
$\RR^d$ (see \cite{grosse} and the references therein). Moreover, if
$s$ is a positive integer,
Triebel  proved that $H^s(\mathbf K)$ is just the
classical Sobolev space  $W^s(\mathbf K)$ introduced by Aubin, which contains all $L^2$
functions on $\mathbf K$ having bounded covariant derivatives up to
order $s$. The following Proposition states that the Cauchy problem is
well posed in these spaces. These expected results are not surprising
but due to the weak regularity of the coefficients they are not
straight consequences of well known results on the problems of
evolution with time dependent Hamiltonians and smooth coefficients (see {\it e.g.}
\cite{tanabe}). To overcome this difficulty we combine
the variational approach of Lions \cite{lions} and the Picard iteration method.

\begin{Proposition}
Let $a$ be a strictly positive function in $C^2(I)$ and $m\geq 0$.  If $\xi\neq
0$, we assume that (\ref{rb}) holds. Given $\theta\in[0,1]$, $u_0\in H^{1-\theta}(\mathbf{K})$, $u_1\in
 H^{-\theta}(\mathbf{K})$, $F\in
 L^2_{loc}\left(I; H^{-\theta}(\mathbf{K})\right)$, there
 exists a unique $u\in C^0\left(I;
   H^{1-\theta}(\mathbf{K})\right)\cap C^1\left(I;
   H^{-\theta}(\mathbf{K})\right)$ solution of (\ref{KG}) satisfying (\ref{CI}).
Moreover $(u_0,u_1,F)\mapsto u$ is a continuous mapping
of these spaces: for any $[t'_-,t'_+]\subset I$, there exists
$C>0$ such that for all $t\in[t'_-,t'_+]$ we have :
\begin{equation}
 \label{inegenerteta}
 \Vert
 u(t)\Vert^2_{H^{1-\theta}}+\Vert \partial_tu(t)\Vert^2_{H^{-\theta}}\leq C\left(
 \Vert
 u_0\Vert^2_{H^{1-\theta}}+\Vert u_1\Vert^2_{H^{-\theta}}+
 \left\vert\int_{t_0}^t\Vert {F}(s)\Vert^2_{H^{-\theta}}ds\right\vert\right).
\end{equation}

If $u$ and $\tilde{u}$ are two solutions with $\theta=\frac{1}{2}$ and $F=0$, then
\begin{equation}
 \label{charge}
 \sigma(u, \tilde{u}):=[a(t)]^d\left(\left<u(t),\partial_t \tilde{u}(t)\right>-\left<\tilde{u} (t),\partial_tu(t)\right>\right)
\end{equation}
does not depend on $t$, where $\left<,\right>$ denotes the duality bracket
between $H^{\frac{1}{2}}(\mathbf K)$ and $H^{-\frac{1}{2}}(\mathbf K)$.
 \label{propun}
\end{Proposition}

{\it Proof.}
It is convenient to eliminate the first order time derivative in
(\ref{KG}) by putting
\begin{equation*}
 \label{}
 v(t,\xx):=\left[ a(t)\right]^{\frac{d}{2}}u(t,\xx).
\end{equation*}
Then $u$ is solution of (\ref{KG}) iff $v$ satisfies
\begin{equation}
 \label{KG'}
 \left\{\partial_t^2-\frac{1}{a^2(t)}\Delta_{\mathbf
     K}+m^2+\xi
   R_g+\frac{d}{2}\left[\frac{a'^2(t)}{a^2(t)}\left(1-\frac{d}{2}\right)-\frac{a''(t)}{a(t)}\right]\right\}v=[
 a(t)]^{\frac{d}{2}}F.
\end{equation}
 Given
$[t'_-,t'_+]\subset I$ with 
$t_0\in [t'_-,t'_+]$, we solve the Cauchy problem for $v$ on this
interval in the same spaces $H^{1-\theta}$,
$H^{-\theta}$, $L^2_{loc}\left(I;H^{-\theta}\right)$. First we compute
the Ricci curvature for the metric $g$,
\begin{equation}
 \label{Rg}
 R_g(t,\xx)=\frac{1}{a^2(t)}R_{\gamma}(\xx)+2d\frac{a''(t)}{a(t)}+d(d-1)\frac{a'^2(t)}{a^2(t)},
\end{equation}
hence (\ref{KG'}) becomes
\begin{equation}
 \label{KG''}
  \left\{\partial_t^2-\frac{1}{a^2(t)}\Delta_{\mathbf
     K}+m^2+\frac{\xi}{a^2(t)}R_{\gamma}(\xx)+B(t)\right\}v=[
 a(t)]^{\frac{d}{2}}F
\end{equation}
where
\begin{equation}
 \label{}
 B(t):=\xi\left[2d\frac{a''(t)}{a(t)}+d(d-1)\frac{a'^2(t)}{a^2(t)}\right]+\frac{d}{2}\left[\frac{a'^2(t)}{a^2(t)}\left(1-\frac{d}{2}\right)-\frac{a''(t)}{a(t)}\right]
\in C^0(I).
\end{equation}
Because of the weak regularity of $B$ we cannot apply general results
to assure the existence and the uniqueness of the solution, and we
consider an intermediate problem.  We remark that  the family of sesquilinear forms on $H^1(\mathbf{K})$
$$
A(t;v,w):=\int_{\mathbf{K}}\left\{\frac{1}{a^2(t)}\nabla_{\mathbf K}v \cdot
\overline{\nabla_{\mathbf K}w} + \left(m^2+\frac{\xi}{a^2(t)}
  R_{\gamma}\right)v\overline{w}\right\}d\xx
$$
satisfies assumptions (8.1), (8.2) p. 265 and (9.37) p. 290 of
\cite{lions}. Therefore the Cauchy problem
\begin{equation}
 \label{KG'''}
 \left\{\partial_t^2-\frac{1}{a^2(t)}\Delta_{\mathbf
     K}+m^2+\frac{\xi}{a^2(t)}R_{\gamma}(\xx)\right\}v=G,
\end{equation}
is well-posed on $[t'_-,t'_+]$ for $v(t_0)\in H^{1-\theta}$,
$\partial_tv(t_0)\in H^{-\theta}$,
$G\in L^2([t'_-,t'_+];H^{-\theta})$: this a direct consequence of the
following results of chapter 3 in \cite{lions}: Theorem 8.2 if $\theta=0$, Theorem 9.4 if $\theta=1$,
Theorem 9.5 if $0<\theta<1$. In particular the finite energy solutions
of (\ref{KG'''}) satisfy the energy inequality : there exists $C>0$ such
for any $t\in[t'_-,t'_+]$ we have
\begin{equation}
 \label{inegalener}
 \Vert
 v(t)\Vert^2_{H^{1}}+\Vert \partial_tv(t)\Vert^2_{L^2}\leq C\left(
 \Vert
 v(t_0)\Vert^2_{H^1}+\Vert \partial_tv(t_0)\Vert^2_{L^2}+
 \left\vert\int_{t_0}^t\Vert G(s)\Vert^2_{L^2}ds\right\vert\right).
\end{equation}
Furthermore, taking advantage of the peculiar structure of
(\ref{KG'''}), we can show that  a similar estimate for weaker solutions of (\ref{KG'''})
holds. If $v\in C^0(I;H^{1-\theta})\cap C^1(I;H^{-\theta})$ is a
solution of (\ref{KG'''}) with $G\in L^2_{loc}(I;H^{-\theta})$, we
put $\tilde{v}:=(-\Delta_{\mathbf
  K}+1)^{-\frac{\theta}{2}}v$. Applying (\ref{inegalener}) with $\xi=0$ to
$\tilde{v}$ we get
$$
\Vert
 {v}(t)\Vert^2_{H^{1-\theta}}+\Vert \partial_t{v}(t)\Vert^2_{H^{-\theta}}\leq C\left(
 \Vert
 v(t_0)\Vert^2_{H^{1-\theta}}+\Vert \partial_tv(t_0)\Vert^2_{H^{-\theta}}+
 \left\vert\int_{t_0}^t\Vert \tilde{G}(s)\Vert^2_{L^2}ds\right\vert\right),
$$
$$
\tilde{G}:=(-\Delta_{\mathbf
  K}+1)^{-\frac{\theta}{2}}\left[G-\frac{\xi}{a^2(t)}R_{\gamma}v\right].
$$
Since we have
$$
\Vert\tilde{G}\Vert^2_{L^2}\leq 2\left(\Vert
  G\Vert^2_{H^{-\theta}}+\Vert
  \frac{\xi}{a^2(t)}R_{\gamma}v\Vert^2_{L^2}\right)\leq C'\left(\Vert
  G\Vert^2_{H^{-\theta}}+\Vert v\Vert^2_{L^2}\right) \leq C'\left(\Vert
  G\Vert^2_{H^{-\theta}}+\Vert v\Vert^2_{H^{1-\theta}}\right),
$$
we deduce using Gronwall's lemma that the solutions of (\ref{KG'''})
satisfy :
\begin{equation}
 \label{inegalteta}
 \Vert
 {v}(t)\Vert^2_{H^{1-\theta}}+\Vert \partial_t{v}(t)\Vert^2_{H^{-\theta}}\leq C''\left(
 \Vert
 v(t_0)\Vert^2_{H^{1-\theta}}+\Vert \partial_tv(t_0)\Vert^2_{H^{-\theta}}+
 \left\vert\int_{t_0}^t\Vert {G}(s)\Vert^2_{H^{-\theta}}ds\right\vert\right).
\end{equation}

Now we return to the initial value problem for (\ref{KG'}) by using the
Picard method. We solve iteratively
\begin{equation*}
 \label{}
 \left\{\partial_t^2-\frac{1}{a^2(t)}\Delta_{\mathbf
     K}+m^2+\frac{\xi}{a^2(t)}R_{\gamma}(\xx)\right\}v_0=\left[a(t)\right]^{\frac{d}{2}}F,\;\;v_0(t_0)=v(t_0),\;\partial_tv_0(t_0)=\partial_tv(t_0),
\end{equation*}
\begin{equation*}
 \label{}
 \left\{\partial_t^2-\frac{1}{a^2(t)}\Delta_{\mathbf
     K}+m^2+\frac{\xi}{a^2(t)}R_{\gamma}(\xx)\right\}v_{k+1}=-B(t)v_k,\;\;v_{k+1}(t_0)=\partial_tv_{k+1}(t_0)=0,\;k\geq
 0.
\end{equation*}
We put
$$
M^2:=C\left(\Vert
 v(t_0)\Vert_{H^{1-\theta}}^2+\Vert \partial_tv(t_0)\Vert_{H^{-\theta}}^2+\int_{t'_-}^{t'_+}\left[a(t)\right]^{d}\Vert
 F(t)\Vert_{H^{-\theta}}^2dt\right).
$$
Using inequality (\ref{inegalteta}), we easily prove by recurrence that
$$
\Vert
 v_k(t)\Vert_{H^{1-\theta}}^2+\Vert \partial_tv_k(t)\Vert_{H^{-\theta}}^2\leq
 M^2\frac{\left(C\Vert B\Vert_{L^{\infty}(t'_-,t'_+)} ^2(t-t_0)\right)^k}{k!}.
$$
Then we conclude that
$$
v(t,\xx):=\sum_{k=0}^{\infty}v_k(t,\xx)
$$
is a solution in $C^0([t'_-,t'_+];H^{1-\theta})\cap
  C^1([t'_-,t'_+];H^{-\theta})$ of the initial value problem
    associated to (\ref{KG'}). Moreover the continuous dependence with
    respect to the data is given by the inequality
\begin{equation*}
 \label{}
 \sup_{t\in[t'_-,t'_+]}\Vert
 v(t)\Vert_{H^{1-\theta}}+\Vert \partial_tv(t)\Vert_{H^{-\theta}}\leq
 C' M
\end{equation*}
where $C'>0$ only depends on the interval $[t'_-,t'_+]$.
To establish the uniqueness of the solutions of (\ref{KG}) and the energy type estimate
(\ref{inegenerteta}), we apply
(\ref{inegalteta}) with $G(t)=\left[a(t)\right]^{\frac{d}{2}}F-B(t)v$,
and we deduce using Gronwall's lemma that the solutions of
(\ref{KG'}) satisfy
$$
\Vert
 v(t)\Vert^2_{H^{1-\theta}}+\Vert \partial_tv(t)\Vert^2_{H^{-\theta}}\leq C\left(
 \Vert
 v(t_0)\Vert^2_{H^{1-\theta}}+\Vert \partial_tv(t_0)\Vert^2_{H^{-\theta}}+
 \left\vert\int_{t_0}^t\Vert
   F(s)\Vert^2_{H^{-\theta}}ds\right\vert\right),\;\;t\in [t'_-,t'_+],
$$
for some $C>0$ depending on $t'_{\pm}$. Replacing $v$ by
$a^{\frac{d}{2}}u$ we obtain (\ref{inegenerteta}) with another
suitable constant $C$.

To establish (\ref{charge}) it is sufficient to consider the case
$\theta=1$ thanks to the density of $H^1$ in $H^{\frac{1}{2}}$, and the
continuous dependence of $u$ with respect the initial data. In this
case we have $u,\tilde{u}\in C^0\left(I;H^1(\KK)\right)\cap
C^1\left(I;L^2(\KK)\right)\cap C^2\left(I;H^{-1}(\KK)\right)$. Hence
we may compute the time derivative of $\sigma(u(t),\tilde{u}(t))$ and
using (\ref{KG}) we get that it
is equal to zero.

\fin

 We introduce the propagator $T(t,t_0)$ associated to
the linear equation (\ref{KG}) with $F=0$, and defined by
$T\left(t,t_0\right)(u_0,u_1):=\left(u(t),\partial_tu(t)\right)$ where
$u$ is the solution of (\ref{KG}) satisfying (\ref{CI}). The
previous proposition assures that
\begin{equation}
 \label{estiprop}
 T\in L^{\infty}_{loc}\left(I\times I;
 \mathcal{L}\left( H^{1-\theta}(\KK)\times
   H^{-\theta}(\KK)\right)\right), \;\;\theta\in[0,1],
\end{equation}
and the unique solution of the inhomogenous equation with Cauchy data
$(u_0,u_1)$ at $t_0$  is given by the usual Duhamel
formula
\begin{equation}
 \label{duhamel}
 \left(u(t),\partial_tu(t)\right)=T(t,t_0)\left(u_0,u_1\right)+\int_{t_0}^tT(t,s)\left(0,F(s)\right)ds.
\end{equation}

It will be useful to have a result on the existence of solutions in higher order
Sobolev spaces, in particular to treat the non linear equation. We
strengthen the assumptions concerning the metric $\gamma$ by assuming that :
\begin{equation}
 \label{bg}
 \forall\alpha\in\NN^d,\;\;D^{\alpha}\gamma_{ij}\in L^{\infty}(\mathbf{K}),\;\; \gamma^{ij}\in L^{\infty}(\mathbf{K}),
\end{equation}
where $D$ represents coordinate derivatives in any {\it normal}
coordinate system. Some equivalent or sufficient conditions have been
obtained by J. Eichhorn (Corollary 2.2,  Proposition 2.3 and Theorem 2.4 of
\cite{eichhorn2}).

\begin{Proposition}
 We assume that $(\mathbf{K},\gamma)$ is a $C^{\infty}$ manifold that satisfies (\ref{bg}). Then, given $s\geq 0$, $u_0\in H^{s}(\mathbf{K})$, $u_1\in
 H^{s-1}(\mathbf{K})$, $F\in
 L^2_{loc}\left(I; H^{s-1}(\mathbf{K})\right)$, there
 exists a unique $u\in C^0\left(I;
   H^{s}(\mathbf{K})\right)\cap C^1\left(I;
   H^{s-1}(\mathbf{K})\right)$ solution of (\ref{KG}) satisfying (\ref{CI}).
Moreover $(u_0,u_1,F)\mapsto u$ is a continuous mapping
of these spaces.
 \label{propdeu}
\end{Proposition}

{\it Proof.}
First we consider the case $s\in\NN$. We write $s=2N+1-\theta$ with $N\in\NN$, $\theta\in\{0,1\}$. Proposition
\ref{propun} assures the existence of the solution $u\in C^0\left(I;
   H^{1-\theta}(\mathbf{K})\right)\cap C^1\left(I;
   H^{-\theta}(\mathbf{K})\right)$. Hence it is sufficient to
 establish that $u_N:=(-\Delta_{\mathbf K}+1)^Nu$ belongs to the same
 space. $u_N$ is a solution of (\ref{kg}) where $F$ is replaced by
 $\left[(-\Delta_{\mathbf K}+1)^N,
   R_{\gamma}\right]u+(-\Delta_{\mathbf K}+1)^NF$. Now a theorem
 of Eichhorn \cite{eichhorn} assures that all the covariant derivatives of the Riemannian
curvature tensor are bounded iff
 (\ref{bg}) is satisfied. Therefore (\ref{bg}) implies that for any
 $\alpha\in\NN^d$, $\nabla^{\alpha}R_{\gamma}\in L^{\infty}(\KK)$ and
 then the commutator $\left[(-\Delta_{\mathbf K}+1)^N,
   R_{\gamma}\right]$ is a bounded operator from $H^{\sigma}(\mathbf{K})$ to
 $H^{\sigma-2N+1}(\mathbf{K})$. Now the result for any integer follows from Proposition
\ref{propun} by recurrence on $N$. In the general case with
$s\in\RR^+\setminus\NN$ we write $s=N+\theta$, $\theta\in(0,1)$. The
previous result assures that for any $t\in I$, the map
$(u_0,u_1,F)\; \longmapsto\;(u(t),\partial_tu(t))$ is continuous from
$H^N\times H^{N-1}\times L^2_{loc}(I;H^{N-1})$ to $H^N\times H^{N-1}$,
  and from $H^{N+1}\times H^{N}\times L^2_{loc}(I;H^{N})$  to $H^{N+1}\times H^N$. Since these spaces are
    interpolation spaces, we deduce from the interpolation thorem (see Theorem 5.1 in
    \cite{lions})  that this map is continuous from $H^s\times
    H^{s-1}\times L^2_{loc}(I;H^{s-1})$ to $H^s\times
      H^{s-1}$. Moreover, we get from the theorem of interpolation of a family of
      operators (see Theorem 5.2 in \cite{lions}) that $u\in
      C^0(I;H^s)\cap C^1(I;H^{s-1})$. Finally the Banach-Steinhauss
      theorem implies that the map $(u_0,u_1,F)\; \longmapsto\;(u,\partial_tu)$ is continuous from
$H^s\times H^{s-1}\times L^2_{loc}(I;H^{s-1})$ to $C^0(I;H^s)\cap C^1(I;H^{s-1})$.
\fin

Finally we consider the global Cauchy problem for the semilinear
Klein-Equation  (\ref{KG}) where $F$ is the subcritical nonlinearity
\begin{equation}
 \label{FF}
F(u)=-Cu\mid u\mid^{p-1},\;\;1\leq p\leq\frac{d}{d-2},\;\;C\geq 0.
\end{equation}
To be able to use Sobolev embedding we have to be strengthened the
assumptions on the metric. We recall that $(\mathbf{K},\gamma)$ is a $C^{\infty}$ {\it bounded geometry manifold}, if the
following conditions are satisfied: (1) the injectivity radius is
strictly positive, (2) every covariant derivative of the Riemannian
curvature tensor is bounded, {\it i.e.}, thanks to the Eichhorn
theorem, (\ref{bg}) is satisfied.


\begin{Theorem}
 We assume that $(\mathbf{K},\gamma)$ is a $C^{\infty}$ {\it bounded
   geometry manifold}. Given $u_0\in H^1(\KK)$, $u_1\in L^2(\KK)$, there exists
   a unique $u\in C^0\left(I;H^1(\KK)\right)\cap C^1\left(I;L^2(\KK)\right)$
   solution of (\ref{KG}), (\ref{FF}), (\ref{CI}). Moreover, there
   exists a function $K\in C^0(I^2\times \RR^4)$ such that for any solutions $u$ and $\hat{u}$, we have for all $t,t_0\in I$,
\begin{equation}
 \label{lipou}
\begin{split}
\Vert
 u(t)&-\hat{u}(t)\Vert_{H^1}+\Vert \partial_tu(t)-\partial_t\hat{u}(t)\Vert_{L^2}\\
&\leq K\left(t,t_0,\Vert u(t_0)\Vert_{H^1}, \Vert \partial_tu(t_0)\Vert_{L^2},\Vert \hat{u}(t_0)\Vert_{H^1},\Vert \hat{u}(t_0)\Vert_{L^2}\right)\left(\Vert
 u(t_0)-\hat{u}(t_0)\Vert_{H^1}+\Vert \partial_tu(t_0)-\partial_t\hat{u}(t_0)\Vert_{L^2}\right).
\end{split}
\end{equation}
 \label{theonl}
\end{Theorem}

{\it Proof.}
Due to the assumption on $(\mathbf{K},\gamma)$, the Sobolev embeddings
hold and in particular  we have: 
$$
H^1(\KK)\subset L^{\frac{2d}{d-2}}(\KK).
$$
We deduce that
 there exists $c_p>0$ such that
 for any $u,\hat{u}\in H^1(\KK)$, 
\begin{equation}
 \label{lipche}
 \Vert F(u)-F(\hat{u})\Vert_{L^2}\leq c_p\Vert u-\hat{u}\Vert_{H^1}\left(\Vert
   u\Vert_{H^1}+\Vert \hat{u}\Vert_{H^1}\right)^{p-1}.
\end{equation}
Now the proof of the local existence is a classic
routine.
Given the Cauchy data $(u_0,u_1)$ at $t_0$, we consider the nonlinear operator $\mathcal{F}$ defined for
$(u,v)\in C^0(I;H^1\times L^2)$ by
\begin{equation}
 \label{FFF}
\mathcal{F}(u,v)(t):=T(t,t_0)(u_0,u_1)-\int_{t_0}^tT(t,s)\left(0,Cu(s)\mid u(s)\mid^{p-1}\right)ds.
\end{equation}
Thanks to  (\ref{estiprop}) and (\ref{lipche}), we have for $t\in[t_0-\epsilon,t_0+\epsilon]$
$$
\Vert \mathcal{F}(u,v)(t)\Vert_{H^1\times L^2}\leq
M_{\epsilon}\left(R+\epsilon Cc_p \vvvert (u,v)\vvvert_{\epsilon}^p\right),
$$
$$
\Vert \mathcal{F}(u,v)(t)-\mathcal{F}(\hat{u},\hat{v})(t)\Vert_{H^1\times L^2}\leq
\epsilon M_{\epsilon}C c_p \left(\vvvert (u,v)\vvvert_{\epsilon}+\vvvert (\hat{u},\hat{v})\vvvert_{\epsilon}\right)^{p-1}\vvvert (u,v)-(\hat{u},\hat{v})\vvvert_{\epsilon},
$$
where $M_{\epsilon}:=\sup_{t,t'\in[t_0-\epsilon,t_0+\epsilon]}\Vert
T(t,t')\Vert_{\mathcal{L}(H^1\times L^2)}$, $R:=\Vert (u_0,u_1)\Vert_{H^1\times
    L^2} $, and 
$$
\vvvert (u,v)\vvvert_{\epsilon}:=\sup_{t\in[t_0-\epsilon,t_0+\epsilon]}\Vert
(u(t),v(t))\Vert_{H^1\times L^2}.
$$
We deduce that for $\epsilon>0$ small enough so that
$$
\epsilon C c_p\left(2M_{\epsilon}R\right)^p\leq R,
$$
then $\mathcal{F}$ is defined
from $B_{\epsilon}\left(RM_{\epsilon}\right)$ into itself, where
$$
B_{\epsilon}(\rho):=\left\{(u,v)\in C^0\left([t_0-\epsilon,t_0+\epsilon];H^1\times L^2\right),\;\;\vvvert(u,v)-T(.,t_0)(u_0,u_1)\vvvert_{\epsilon}\leq\rho\right\}.
$$
Furthermore, if we also choose $\epsilon>0$ small enough so that
$$
\epsilon  C c_p(4R)^{p-1}M_{\epsilon}^p<1,
$$
then $\mathcal{F}$ is a strict contraction on
$B_{\epsilon}\left(RM_{\epsilon}\right)$. The unique fixed point $(u,v)$ in this
space satisfies $v=\partial_tu$ and $u$ is a solution of
(\ref{KG}), (\ref{FF}), (\ref{CI}) on $[t_0-\epsilon,t_0+\epsilon]$.\\

To obtain  global existence, we recall an abstract result of
Strauss (\cite{strauss66}, Theorem 4.1). If $u\in C^0([0,T];H^1)\cap
C^1([0,T];L^2)$ is a solution of 
$$
\partial_t^2u+A(t)u=G
$$
where $A(t)$ is a densely defined selfadjoint operator on $L^2$ with $A\in C^1\left([0,T];\mathcal{L}(H^1,H^{-1})\right)$,  and $G\in
C^0([0,T];L^2)$, then
\begin{equation}
\label{enerstra}
 \begin{split}
 \Vert\partial_tu(t)\Vert_{L^2}^2+<A(t)u(t),u(t)>_{H^{-1},H^1}&\\
&=\Vert\partial_tu(0)\Vert_{L^2}^2+<A(t)u(0),u(0)>_{H^{-1},H^1}\\
&+\int_0^t<A'(s)u(s),u(s)>_{H^{-1},H^1}ds+2\Re\int_0^t(G(s),\partial_tu(s))_{L^2}ds.
\end{split}
\end{equation}
We apply this equality to
$A(t):=a^{-2}(t)\left[-\Delta_{\KK}+1\right]$,
$G:=\left(a^{-2}(t)-m^2-\xi
  R_g\right)u-da'(t)a^{-1}(t)\partial_tu-Cu\mid u\mid^{p-1}$.
We note that for any $v\in C^1(I;H^1(\KK))$ we have
$$
\Vert v(t)\Vert_{L^{p+1}}^{p+1}-\Vert
v(t_0)\Vert_{L^{p+1}}^{p+1}=(p+1)\Re\int_{t_0}^tv(s)\mid v(s)\mid^{p-1}\overline{\partial_tv(s)}ds.
$$
By density, this equality holds also for $v\in C^0(I,H^1)\cap
C^1(I;L^2)$. We deduce that a local solution $u\in C^0((t'_-,t'_+),H^1)\cap
C^1((t'_-,t'_+);L^2)$, $[t'_-,t'_+]\subset I$,  satisfies
\begin{equation*}
 \begin{split}
 \Vert \partial_tu(t)\Vert_{L^2}^2+\frac{1}{a^2(t)}\Vert
u(t)\Vert_{H^1}^2+\frac{2C}{p+1}\Vert &u(t)\Vert_{L^{p+1}}^{p+1}\\
&\leq
\Vert \partial_tu(t_0)\Vert_{L^2}^2+\frac{1}{a^2(t_0)}\Vert
u(t_0)\Vert_{H^1}^2+\frac{2C}{p+1}\Vert
u(t_0)\Vert_{L^{p+1}}^{p+1}\\
&+\left\vert\int_{t_0}^t h(s)\left[\Vert \partial_tu(s)\Vert_{L^2}^2+\frac{1}{a^2(s)}\Vert
u(s)\Vert_{H^1}^2 \right]ds\right\vert,
\end{split}
\end{equation*}
where $h\in C^0(I)$ only depends on $a,a', R_g,d,m,\xi$. The Gronwall
lemma implies that there exists $k\in C^0(I^2\times\RR^2)$ such that
\begin{equation}
 \label{ester}
 \Vert u(t)\Vert_{H^1}+\Vert \partial_tu(t)\Vert_{L^2}\leq k\left(t,t_0, \Vert u(t_0)\Vert_{H^1},\Vert \partial_tu(t_0)\Vert_{L^2}\right).
\end{equation}
Therefore $u\in L^{\infty}\left((t'_-,t'_+);H^1\right)$, $\partial_tu\in
L^{\infty}\left((t'_-,t'_+);L^2\right)$ and the integral equation
$(u(t), \partial_tu(t))=\mathcal{F}(u,\partial_tu)(t)$ assures that  $u\in C^0\left([t'_-,t'_+];H^1\right)\cap
C^1\left([t'_-,t'_+)];L^2\right)$. We conclude by the classic argument of
unique continuation that $u$ is globally defined on $I$.\\

To prove (\ref{lipou}), we use (\ref{lipche}) and (\ref{FFF}) to get
\begin{equation*}
 \begin{split}
\Vert
u(t)-\hat{u}(t)\Vert_{H^1}+\Vert \partial_tu(t)-&\partial_t\hat{u}(t)\Vert_{L^2}\\
\leq&
c(t,t_0)\left(
\Vert
u(t_0)-\hat{u}(t_0)\Vert_{H^1}+\Vert \partial_tu(t_0)-\partial_t\hat{u}(t_0)\Vert_{L^2}\right)\\
&+\left\vert\int_{t_0}^tc(t,s)\Vert
  u(s)-\hat{u}(s)\Vert_{H^1}\left(
\Vert u(s)\Vert_{H^1}^{p-1}+\Vert \hat{u}(s)\Vert_{H^1}^{p-1}\right)ds\right\vert
\end{split}
\end{equation*}
for some $c\in C^0(I^2)$. Now (\ref{lipou}) is a consequence of
(\ref{ester}) and of
Gronwall's lemma again.
\fin

These results of existence and uniqueness of the solutions being
acquired, the fundamental problems lie in finding uniform energy
estimates on $(t_-,t_+)$, and in investigating the asymptotic
behaviours at $t_{\pm}$. To overcome the difficulties linked to the
variable speed due to the term $a^{-2}(t)\Delta_{\KK}$, it is
  convenient to replace the cosmic time $t$ by the conformal time
  $\tau$ defined by (\ref{tau}).
In this time coordinate, the metric (\ref{g}) is expressed with the
scale factor $\alpha$ given by (\ref{alpha}), as
\begin{equation}
 \label{gc}
 g_{\mu\nu}dx^{\mu}dx^{\nu}=\alpha^2(\tau)\left[d\tau^2-\gamma_{ij}d{\mathbf
   x}^id{\mathbf x}^j\right],\; \mathcal{M}=(\tau_-,\tau_+)_{\tau}\times\mathbf{K_x},
\end{equation}
and the  Ricci scalar is given by
\begin{equation*}
 \label{}
 R_g=\frac{1}{\alpha^2(\tau)}R_{\gamma}+2d\frac{\alpha''(\tau)}{\alpha^3(\tau)}+d(d-3)\frac{\alpha'^2(\tau)}{\alpha^4(\tau)}.
\end{equation*}
Hence $u$ is a solution of (\ref{KG}) iff its Liouville transform
$\varphi$ defined by (\ref{phi})
is a solution of 
\begin{equation}
 \label{KGc'}
 \left\{\partial_{\tau}^2-\Delta_{\mathbf
     K}+m^2\alpha^2(\tau)+\xi R_{\gamma}+d\left(\xi-\frac{d-1}{4d}\right)\left((d-3)\frac{\alpha'^2(\tau)}{\alpha^2(\tau)}+2\frac{\alpha''(\tau)}{\alpha(\tau)}\right)\right\}\varphi=[
 \alpha(\tau)]^{\frac{d+3}{2}}F,
\end{equation}
and the initial data (\ref{CI}) on $u$ are transformed by an
isomorphism ${\mathcal L}$ on $H^s\times H^{s-1}$, 
 \begin{equation}
 \label{ufifi}
 {\mathcal L}:\;(u_0,u_1)\mapsto (\varphi_0,\varphi_1),\;\;\varphi_0=\left[a(t_0)\right]^{\frac{d-1}{2}}u_0,\;\;\varphi_1=\frac{d-1}{2}a'(t_0)\left[a(t_0)\right]^{\frac{d-1}{2}}u_0+\left[a(t_0)\right]^{\frac{d+1}{2}}u_1.
\end{equation}
In the case of the conformal coupling $\xi=\frac{d-1}{4d},$ this
equation takes a simpler and  more tractable form. The linear equation
(\ref{kg}) is equivalent to
\begin{equation}
 \label{KGsimple}
 \left\{\partial_{\tau}^2-\Delta_{\mathbf
     K}+m^2\alpha^2(\tau)+\frac{d-1}{4d} R_{\gamma}\right\}\varphi=0,
\end{equation}
and the semilinear equation (\ref{nlkg}) with (\ref{pe}), becomes
\begin{equation}
 \label{nlkg'}
  \left\{\partial_{\tau}^2-\Delta_{\mathbf
     K}+m^2\alpha^2(\tau)+\frac{d-1}{4d} R_{\gamma}\right\}\varphi=-C[
 \alpha(\tau)]^{\nu}\varphi\mid \varphi\mid^{p-1},
\end{equation}
\begin{equation}
 \label{nuuu}
\nu:=\frac{d+3-p(d-1)}{2}\in\left[\frac{d-3}{d-2},2\right].
\end{equation}
Thus for the conformal coupling, we deal with a Klein-Gordon equation with variable
mass $m\alpha(\tau)$ that, as $\tau\rightarrow\tau_{\pm}$, tends to zero
for a Big Bang/Crunch, to the infinity for a Big Rip, or a constant (and $\alpha'$
is diverging) for a Sudden Singularity. Therefore the asymptotic
properties of $u$ heavily depend on the behaviour of the scale factor
$a(t)$ near the singularities $t_{\pm}$. In the Appendix we present
the asymptotics of $\alpha(\tau)$ near $\tau_{\pm}$.



\section{At a Big Bang/Crunch or a Sudden Singularity (Conformal Coupling)}

In this part we investigate the asymptotics at a Big Bang/Crunch or a
Sudden Singularity, of a massive scalar field with the conformal coupling
(\ref{xiconf}), solving the linear or nonlinear Klein-Gordon equation
(\ref{KG}) with $F$ given by (\ref{FF}). 
We consider only the case of a future time singularity at $t_+$, the case
at $t_-$ is easily deduced by reversing time. First we remark
that for $0\leq \eta^0_+<1$, we have $\tau_+<\infty$ and $\alpha\in
C^0([0,\tau_+])$. In this simple situation, it is sufficient to prove
that the solution of (\ref{KGsimple}) is well defined at $\tau_+$ and
the Cauchy problem is well posed for initial data specified at
$\tau_+$. In contrast, for $\eta_0^+\geq 1$, we have $\tau_+=+\infty$ and since $\alpha(\tau)\rightarrow 0$ as $\tau\rightarrow \infty$, we
compare the solutions of (\ref{KGsimple}) near $\tau_+$  with those of the massless equation
\begin{equation}
 \label{KGmo}
 \left\{\partial_{\tau}^2-\Delta_{\mathbf
     K}+\frac{d-1}{4d} R_{\gamma}\right\}\varphi=0.
\end{equation}
Therefore we deal with a scattering problem. The Cauchy problem is well posed for
this equation: given $\varphi_0\in H^{1-\theta}(\KK)$,
$\varphi_1\in H^{-\theta}(\KK)$ with $\theta\in[0,1]$,  there exists a $\varphi\in
C^0\left((\tau_-,\tau_+);H^{1-\theta}(\KK)\right)\cap
C^1\left((\tau_-,\tau_+);H^{-\theta}(\KK)\right)$ solution of
 (\ref{KGmo}) satisfying
\begin{equation}
 \label{cidi}
 \varphi(0)=\varphi_0,\;\;\partial_{\tau}\varphi(0)=\varphi_1.
\end{equation} 
We remark that in some cases,  the
solutions of (\ref{KGmo}) can be unbounded as $\tau\rightarrow
\infty$: the Kato Rellich theorem  and (\ref{rb}) assure that  $-\Delta_{\mathbf
  K}+\frac{d-1}{4d}R_{\gamma}$ endowed with the domain
$H^2(\mathbf{K})$ is selfadjoint on $L^2$, hence  if its spectrum intersects
$(-\infty,0]$ the norm of some solutions can increase as $\tau\rightarrow+\infty$. To avoid this
situation we introduce an ``infrared'' cut-off. Given $\delta\geq 0$,
we introduce
\begin{equation}
 \label{Rpos}
 H^{2,\delta}(\mathbf K):=\left\{\varphi\in H^2(\mathbf K);\;\chi_{(-\infty,\delta)}\left(-\Delta_{\mathbf
  K}+\frac{d-1}{4d}R_{\gamma}\right)\varphi=0\right\}
\end{equation}
where $\chi_{\Omega}$ is the characteristic function of
$\Omega\subset\RR$, and for $s\leq 2$ we define $H^{s,\delta}(\mathbf
K)$ as the closure of $H^{2,\delta}(\mathbf K)$ for the norm (\ref{normhs}).\\

First, we investigate the asymptotic behaviours of the free fields
solving the linear Klein-Gordon equation.

\begin{Theorem}[Big Crunch, Sudden Singularity]
We assume  that (\ref{condida}), (\ref{condidac}) and (\ref{rb}) and hold.
 Given $\theta\in[0,1]$, $u_0\in H^{1-\theta}(\mathbf{K})$, $u_1\in
 H^{-\theta}(\mathbf{K})$, we consider the  unique $u\in C^0\left(I;
   H^{1-\theta}(\mathbf{K})\right)\cap C^1\left(I;
   H^{-\theta}(\mathbf{K})\right)$ solution of (\ref{kg}),
 (\ref{xiconf}),  (\ref{CI}).\\

If $0\leq\eta_0^+<1$, there exist unique $\varphi_0^+\in
 H^{1-\theta}(\KK)$,  $\varphi_1^+\in
 H^{-\theta}(\KK)$, such that
\begin{equation}
 \label{asszer}
 \left\Vert
 \left[a(t)\right]^{\frac{d-1}{2}}u(t,.)-\varphi_0^+\right\Vert_{H^{1-\theta}}+\left\Vert a(t)\partial_t\left[\left[a(t)\right]^{\frac{d-1}{2}}u(t,.)\right]-\varphi_1^+\right\Vert_{H^{-\theta}}\longrightarrow
   0,\;\;t\rightarrow t_+,
\end{equation}
and the map $W_+:(u_0,u_1)\mapsto (\varphi_0^+,\varphi_1^+)$ is an isomorphism
on $H^{1-\theta}(\mathbf{K})\times H^{-\theta}(\mathbf{K})$. For any
$\delta\in\RR$, $H^{1-\theta,\delta}(\mathbf{K})\times
H^{-\theta,\delta}(\mathbf{K})$ is invariant under $W_+$.\\

If $1\leq \eta_0^+$, given $0<\delta$, $u_0\in H^{1-\theta,\delta}(\mathbf{K})$, $u_1\in
 H^{-\theta,\delta}(\mathbf{K})$, there exist unique $\varphi_0^+\in
 H^{1-\theta,\delta}(\KK)$,  $\varphi_1^+\in
 H^{-\theta,\delta}(\KK)$, such that
\begin{equation}
 \label{asss}
 \left\Vert
 \left[a(t)\right]^{\frac{d-1}{2}}u(t,.)-\varphi(\tau)\right\Vert_{H^{1-\theta}}+\left\Vert a(t)\partial_t\left[\left[a(t)\right]^{\frac{d-1}{2}}u(t,.)\right]-\partial_{\tau}\varphi(\tau)\right\Vert_{H^{-\theta}}\longrightarrow
   0,\;\;t\rightarrow t_+,
\end{equation}
where $\varphi$ is the solution of (\ref{KGmo}) satisfying
$\varphi(0)=\varphi_0^+$, $\partial_{\tau}\varphi(0)=\varphi_1^+$, and
$\tau$ is defined by (\ref{tau}). The map $W_+:(u_0,u_1)\mapsto (\varphi_0^+,\varphi_1^+)$ is an isomorphism
on $H^{1-\theta,\delta}(\mathbf{K})\times
H^{-\theta,\delta}(\mathbf{K})$.\\

If $0\leq \eta_0^+$, $\theta=\frac{1}{2}$ and $\sigma$ is given by
(\ref{charge}),  we have
\begin{equation}
 \label{siguema}
 \sigma(u,\tilde{u})=\left<\varphi_0^+,\tilde{\varphi}_1^+\right>-\left<\tilde{\varphi}_0^+,{\varphi}_1^+\right>
\end{equation}
where $(\tilde{\varphi}_0^+,\tilde{\varphi}_1^+)=W_+(\tilde{u}(t_0),\partial_t\tilde{u}(t_0))$.
 \label{theobb}
\end{Theorem}


{\it Proof.}
We introduce the group $U_0(\tau)$ associated to equation
(\ref{KGmo}) and defined by
$U_0(\tau)(\varphi_0,\varphi_1)=(\varphi(\tau),\partial_{\tau}\varphi(\tau))$
if $\varphi\in C^0(\RR_{\tau};H^{1-\theta}(\KK))\cap
C^1(\RR_{\tau};H^{-\theta}(\KK))$ is the solution of (\ref{KGmo}) with
initial data $\varphi(0)=\varphi_0$,
$\partial_{\tau}\varphi(0)=\varphi_1$. For any $a,b\in\RR$,
$a<b$, there exists $C_{b-a}>0$ such that 
\begin{equation}
 \label{estipropuzero}
 \sup_{\tau\in[a,b]}\Vert U_0(\tau)\Vert_{\mathcal{L}(H^{1-\theta}\times H^{-\theta})}\leq C_{b-a}<\infty.
\end{equation}
Given $u_0\in H^{1-\theta}(\KK)$, $u_1\in H^{-\theta}(\KK)$, we choose
$(\varphi_0,\varphi_1)={\mathcal L}_{t_0}(u_0,u_1)$ given by (\ref{ufifi}).
Now the solution of (\ref{KGsimple})  with initial data $(\varphi_0,\varphi_1)$ satisfies for $\tau\in(\tau_-,\tau_+)$
\begin{equation}
 \label{integq}
 (\varphi(\tau),\partial_{\tau}\varphi(\tau))=U_0(\tau)(\varphi_0,\varphi_1)-m^2\int_{0}^{\tau}U_0(\tau-\sigma)(0,\alpha^2(\sigma)\varphi(\sigma))d\sigma.
\end{equation}
Then the Gronwall lemma implies that for $\tau\in[0,\tau_+)$
\begin{equation}
 \label{estigron}
 \Vert \varphi(\tau)\Vert_{H^{1-\theta}}+
\Vert \partial_{\tau}\varphi(\tau)\Vert_{H^{-\theta}}\leq
C_{\tau_+}\left(\Vert \varphi_0\Vert_{H^{1-\theta}}+
\Vert \varphi_1\Vert_{H^{-\theta}}\right)\exp\left(C_{\tau_+}m^2\int_{0}^{\tau}\alpha^2(\sigma)d\sigma\right).
\end{equation}
For $\eta_0^+\in[0,1)$ we have $\alpha\in C^0([0,\tau_+])$ hence
\begin{equation}
 \label{estofi}
 (\varphi,\partial_{\tau}\varphi)\in
L^{\infty}\left((0,\tau_+);H^{1-\theta}\times H^{-\theta}\right),
\end{equation}
therefore (\ref{integq}) assures that
$(\varphi(\tau),\partial_{\tau}\varphi(\tau))$ has a limit as
$\tau\rightarrow\tau_+$, and $\varphi_0^+$, $\varphi_1^+$ defined by
\begin{equation}
 \label{soleq}
 (\varphi_0^+,\varphi_1^+)=U_0(\tau_+)(\varphi_0,\varphi_1)-m^2\int_{0}^{\tau_+}U_0(\tau_+-\sigma)(0,\alpha^2(\sigma)\varphi(\sigma))d\sigma
\end{equation}
satisfy (\ref{asszer}). Conversely, given
$(\varphi_0^+,\varphi_1^+)\in H^{1-\theta}\times H^{-\theta}$, we can
easily solve, by Picard iteration, the integral equation
\begin{equation}
 \label{integqp}
 (\varphi(\tau),\partial_{\tau}\varphi(\tau))=U_0(\tau-\tau_+)(\varphi_0^+,\varphi_1^+)-m^2\int_{\tau_+}^{\tau}U_0(\tau-\sigma)(0,\alpha^2(\sigma)\varphi(\sigma))d\sigma
\end{equation}
for $\tau\in(\tau_-,\tau_+], $. In other words, the Cauchy problem for
(\ref{KGsimple})  is well posed
in $C^0\left((\tau_-,\tau_+];H^{1-\theta}\times H^{-\theta}\right)$
with initial data specified at $\tau_+$. Now it is sufficient to
define $(\varphi_0,\varphi_1)$ from (\ref{soleq}) and we conclude that
$(u_0,u_1)$ given by (\ref{ufifi}) satisfy (\ref{asszer}). To obtain
the invariance of $H^{1-\theta,\delta}\times H^{-\theta,\delta}$ it is
sufficient to remark that $U_0(\tau)$ lets invariant this space.\\

If $\eta_0^+\geq 1$, we have
$\tau_+=+\infty$ and $\Vert
U_0(\tau)\Vert_{\mathcal{L}(H^{1-\theta}\times H^{-\theta})}$ can be
unbounded on $[0,\infty)$, hence we shall introduce an infrared
cut-off. First we have to define equivalent norms on $H^s(\KK)$. We
consider the operators 
$$
\Lambda_0:=-\Delta_{\KK}+1,\;\;
A_0:=-\Delta_{\KK}+\frac{d-1}{4d}R_{\gamma},\;\;
\Lambda_1:=A_0+1+\frac{d-1}{4d}\Vert
R_{\gamma}\Vert_{L^{\infty}}.
$$
By the Kato-Rellich theorem $A_0$, $\Lambda_1$ are selfadjoint operators with
the same domain as  $\Lambda_0$:
$$
H^2(\KK)={\mathcal D}(A_0)={\mathcal D}(\Lambda_0)={\mathcal D}(\Lambda_1).
$$
Since $1\leq \Lambda_j$ we may
define $\Lambda_j^{\theta}$ for $\theta\in[0,1]$ by the spectral
theorem and it is well known from complex interpolation  (see
\cite{lions}, Remark 2.3, p.10) that
\begin{equation}
 \label{equinorm}
H^{2\theta}(\KK)={\mathcal D}\left(\Lambda^{\theta}_0\right)={\mathcal D}\left(\Lambda_1^{\theta}\right)\;with\;equivalent\;norms. 
\end{equation}
Given $\delta>0$, we introduce the truncated at low frequency operator
\begin{equation}
 \label{}
 A_{\delta}:=\chi_{[\delta,\infty)}\left(A_0\right)A_0,\;\; \mathcal{D}(A_{\delta})=H^{2,\delta}(\KK),
\end{equation}
which is selfadjoint on $L^2_{\delta}:=H^{0,\delta}(\KK)$ and $\delta
I\leq A_{\delta}$. We remark
that the norms
$$
\Vert\Lambda_1^{\theta}u\Vert_{L^2},\;\; \Vert A_{\delta}^{\theta}u\Vert_{L^2}
$$
are equivalent on $H^{2\theta,\delta}(\KK)$, $\theta\in[0,1]$, hence
by duality, the same result holds for $\theta\in[-1,0]$. We conclude
with (\ref{equinorm}) that $\Vert A_{\delta}^{\theta}u\Vert_{L^2} $ is
 equivalent to the $H^{2\theta}$-norm on $H^{2\theta,\delta}$ given by $\Vert A_0^{\theta}u\Vert_{L^2} $. We remark
that the group $U_0(\tau)$ leaves invariant
$H^{1-\theta,\delta}\times H^{-\theta,\delta}$ and on this space the
solution of  (\ref{KGmo}), (\ref{cidi}) can be expressed as
$$
\varphi(\tau)=\cos\left(\tau A_{\delta}^{\frac{1}{2}}\right)\varphi_0+A_{\delta}^{-\frac{1}{2}}\sin\left(\tau A_{\delta}^{\frac{1}{2}}\right)\varphi_1.
$$
As a
consequence the restriction of $U_0(\tau)$ on
$H^{1-\theta,\delta}\times H^{-\theta,\delta}$ is uniformly bounded:
Given $\theta\in[0,1]$, $\delta>0$, there exists $C_{\delta}>0$ such
that
\begin{equation}
 \label{enerdel}
 \sup_{\tau\in\RR}\Vert U_0(\tau)\Vert_{\mathcal{L}(H^{1-\theta,\delta}\times H^{-\theta,\delta})}\leq C_{\delta}<\infty.
\end{equation}
We deduce from (\ref{integq}) that the propagator associated to
(\ref{KGsimple}) leaves invariant $H^{1-\theta,\delta}\times
H^{-\theta,\delta}$ and since $\alpha\in L^1(0,\infty)$, the Gronwall
lemma assures that the solution of
(\ref{KGsimple}) with initial data in $H^{1-\theta,\delta}\times H^{-\theta,\delta}$ satisfies for $\sigma,\tau\in[0,\infty)$
\begin{equation}
 \label{estprog}
 \left\Vert
   \left(\varphi(\tau),\partial_{\tau}\varphi(\tau)\right)\right\Vert_{H^{1-\theta,\delta}\times
   H^{-\theta,\delta}}\leq C_{\delta} \left[\left\Vert\left(\varphi(\sigma),\partial_{\tau}\varphi(\sigma)\right)\right\Vert_{H^{1-\theta,\delta}\times
   H^{-\theta,\delta}}+\exp\left(m^2C_{\delta}\int_0^{\infty}\alpha^2(\tau')d\tau'\right)\right].
\end{equation}
Therefore $U_0(-\tau)(0,\alpha^2(\tau)\varphi(\tau))$ belongs to
$L^1\left(0,\infty; H^{1-\theta,\delta}\times
   H^{-\theta,\delta} \right)$ and the following limit exists
$$
(\varphi_0^+,\varphi_1^+):=\lim_{\tau\rightarrow \infty}U_0(-\tau) \left(\varphi(\tau),\partial_{\tau}\varphi(\tau)\right)=(\varphi_0,\varphi_1)-m^2\int_0^{\infty}U_0(-\tau)\left(0,\alpha^2(\tau)\varphi(\tau)\right)d\tau
$$
and satisfies (\ref{asss}). Conversely, given
$(\varphi_0^+,\varphi_1^+)\in H^{1-\theta,\delta}\times
   H^{-\theta,\delta}$, we construct $(\varphi_0,\varphi_1)$ using the
   principle of the
   Cook method (see {\it e.g.} \cite{reed-simon3}): the time integrability of the perturbation. We introduce the propagator $U(\tau,\sigma)$
   associated to (\ref{KGsimple}) defined by
   $U(\tau,\sigma)(\varphi_0,\varphi_1)=(\varphi(\tau),\partial_{\tau}\varphi(\tau))$
   if $\varphi$ is the the solution of  (\ref{KGsimple}) with initial
   data at time $\sigma$,
   $(\varphi(\sigma),\partial_{\tau}(\sigma))=(\varphi_0,\varphi_1)$. (\ref{estprog})
   assures that there exists $C'_{\delta}<\infty$ such that
\begin{equation}
 \label{bornage}
 \sup_{0\leq \sigma,\tau}\left\Vert
   U(\tau,\sigma)\right\Vert_{\mathcal{L}(H^{1-\theta}\times
   H^{-\theta})}\leq C'_{\delta}.
\end{equation}
If $(\varphi_0^+,\varphi_1^+)\in H^{2,\delta}\times
   H^{1,\delta}$, Proposition \ref{propdeu} implies that $\sigma\mapsto U(0,\sigma)U_0(\sigma)
   (\varphi_0^+,\varphi_1^+)$ belongs to $C^1\left([0,\infty); H^{1,\delta}\times
   H^{0,\delta}\right)$ and
$$
\frac{d}{d\sigma} U(0,\sigma)U_0(\sigma)
   (\varphi_0^+,\varphi_1^+)=m^2\alpha^2(\sigma)U(0,\sigma)(0,\varphi_0(\sigma)),\;\;U_0(\sigma)  (\varphi_0^+,\varphi_1^+)=(\varphi_0(\sigma),\partial_{\tau}\varphi_0(\sigma)),
$$
hence
\begin{equation*}
 \label{}
\left\Vert U(0,\tau)U_0(\tau)
   (\varphi_0^+,\varphi_1^+)-U(0,\sigma)U_0(\sigma)
   (\varphi_0^+,\varphi_1^+)\right\Vert_{H^{1-\theta}\times
   H^{-\theta}}\leq m^2C_{\delta}C'_{\delta}\left\vert\int_{\sigma}^{\tau}\alpha^2(\tau')d\tau'\right\vert.
\end{equation*}
By density, this inequality holds for  $(\varphi_0^+,\varphi_1^+)\in H^{1-\theta,\delta}\times
   H^{-\theta,\delta}$. Therefore the following limit exists in this space,
$$
(\varphi_0,\varphi_1):=\lim_{\tau\rightarrow\infty}U(0,\tau)U_0(\tau)
   (\varphi_0^+,\varphi_1^+)
$$
and satisfies (\ref{asss}).

Finally to prove (\ref{siguema}) we remark that if
$\tilde{\varphi}(\tau)=\left[a(t)\right]^{\frac{d-1}{2}}\tilde{u}(t)$,
we have
$$
\sigma(u,\tilde{u})=\left<\varphi(\tau),\partial_{\tau}\tilde{\varphi}(\tau)\right>-\left<\tilde{\varphi}(\tau),\partial_{\tau}{\varphi}(\tau)\right>=
\lim_{\tau\rightarrow\infty}\left<\varphi(\tau),\partial_{\tau}\tilde{\varphi}(\tau)\right>-\left<\tilde{\varphi}(\tau),\partial_{\tau}{\varphi}(\tau)\right>=
\left<\varphi_0^+,\tilde{\varphi}_1^+\right>-\left<\tilde{\varphi}_0^+,{\varphi}_1^+\right>.
$$

\fin

We now investigate the non-linear problem.
We recall that by the Liouville transform, the semi-linear Klein-Gordon equation
(\ref{KG}), (\ref{FF}) is equivalent to (\ref{nlkg'}).

\begin{Theorem}
We assume that $(\mathbf{K},\gamma)$ is a $C^{\infty}$ {\it bounded
   geometry manifold}. Let $a$ be
a strictly positive function in $C^2(I)$ satisfying (\ref{condida}),
(\ref{condidac}), and $m\geq 0$.
 Given $u_0\in H^{1}(\mathbf{K})$, $u_1\in
 L^2(\mathbf{K})$, we consider the  unique $u\in C^0\left(I;
   H^{1}(\mathbf{K})\right)\cap C^1\left(I;
   L^2(\mathbf{K})\right)$ solution of (\ref{KG}),  (\ref{CI}), (\ref{FF}).\\

We suppose that $0\leq\eta_0^+<1$, and if $\eta_0^+=0$ we assume
$c_1^+>0$. Then there exist unique $\varphi_0^+\in
 H^{1}(\KK)$,  $\varphi_1^+\in
 L^2(\KK)$, such that
\begin{equation}
 \label{asszernl}
 \left\Vert
 \left[a(t)\right]^{\frac{d-1}{2}}u(t,.)-\varphi_0^+\right\Vert_{H^{1}}+\left\Vert a(t)\partial_t\left[\left[a(t)\right]^{\frac{d-1}{2}}u(t,.)\right]-\varphi_1^+\right\Vert_{L^2}\longrightarrow
   0,\;\;t\rightarrow t_+.
\end{equation}

If $1\leq \eta_0^+$, we assume that
there exists $\delta>0$ such that
\begin{equation}
 \label{Rdelta}
 \delta\leq R_{\gamma},
\end{equation}
and the nonlinearity satisfies
\begin{equation}
 \label{condipi}
\left\{
\begin{array}{lcl}
if\; \eta_0^+=1\;and\;d=3,\;\; 1\leq p<3,\\
 if\; \eta_0^+=1\;and\;d\geq4,\;\;1\leq p\leq
 \frac{d}{d-2},\\
 if\; 1<\eta_0^+<d-2,\;\;1\leq p\leq\frac{d}{d-2},\\
 if\; \eta_0^+=d-2,\;\;1\leq p<\frac{d}{d-2},\\
if\; \eta_0^+>d-2,\;\;1\leq
  p<1+\frac{2}{d-1}\left(1+\frac{1}{\eta_0^+}\right).
\end{array}
\right.
\end{equation}
Then, given $u_0\in H^{1}(\mathbf{K})$, $u_1\in
 L^2(\mathbf{K})$, there exist unique $\varphi_0^+\in
 H^{1}(\KK)$,  $\varphi_1^+\in
 L^2(\KK)$, such that the solution $u$ satisfies
\begin{equation}
 \left\Vert
 \left[a(t)\right]^{\frac{d-1}{2}}u(t,.)-\varphi(\tau)\right\Vert_{H^{1}}+\left\Vert a(t)\partial_t\left[\left[a(t)\right]^{\frac{d-1}{2}}u(t,.)\right]-\partial_{\tau}\varphi(\tau)\right\Vert_{L^2}\longrightarrow
   0,\;\;t\rightarrow t_+,
\end{equation}
where $\varphi$ is the solution of (\ref{KGmo}) with initial data
$\varphi(0)=\varphi_0^+$, $\partial_{\tau}\varphi(0)=\varphi_1^+$.\\

For any $\eta_0^+\geq 0$, the map 
\begin{equation}
 \label{OMEGA}
 \Omega_+:(u_0,u_1)\mapsto (\varphi_0^+,\varphi_1^+)
\end{equation}
 is a bicontinuous bijection
on $H^{1}(\mathbf{K})\times
L^2(\mathbf{K})$ and there exists a continuous function $\Psi$ such
that
\begin{equation}
 \label{lipschoucou}
 \Vert \Omega_+(u_0,u_1)-\Omega_+(\hat{u}_0,\hat{u}_1)\Vert_{H^1\times
   L^2}\leq\Vert (u_0,u_1)-(\hat{u}_0,\hat{u}_1)\Vert_{H^1\times L^2}\Psi\left(\Vert (u_0,u_1) \Vert_{H^1\times L^2}+\Vert(\hat{u}_0,\hat{u}_1)\Vert_{H^1\times L^2}\right),
\end{equation}
\begin{equation}
 \label{lipschoucoucou}
 \Vert \Omega^{-1}_+(\varphi^+_0,\varphi^+_1)-\Omega^{-1}_+(\hat{\varphi}^+_0,\hat{\varphi}^+_1)\Vert_{H^1\times
   L^2}\leq\Vert (\varphi^+_0,\varphi^+_1)-(\hat{\varphi}^+_0,\hat{\varphi}^+_1)\Vert_{H^1\times L^2}\Psi\left(\Vert (\varphi^+_0,\varphi^+_1) \Vert_{H^1\times L^2}+\Vert(\hat{\varphi}^+_0,\hat{\varphi}^+_1)\Vert_{H^1\times L^2}\right).
\end{equation}
 \label{}
\end{Theorem}

\begin{Remark}
 Obviously many issues on the asymptotics
 for the semi-linear Klein-Gordon equation remain open, and the assumptions
 (\ref{condipi}) could seem to be too strong. Nevertheless we remark that
 the existence of a free asymptotic field fails in an important limit
 case.  If $\KK$ is a compact manifold with
 $R_{\gamma}=6$, $d=3$, $\eta_0^+=1$, $C=1$, $p=3$, $\xi=1/6$, the semilinear Klein-Gordon equation satisfied
 by a field constant on $\KK$ is the quartic oscillator. Then given $\varphi_0\in \RR^*$, $\varphi_1=0$, the solution
 $\varphi(\tau,\xx)=\varphi(\tau)$ of (\ref{nlkg'}) with
 $\varphi(0)=\varphi_0$, $\varphi'(0)=0$,  satisfies the Duffin
 equation $\varphi''+\varphi+\varphi^3=0$ and $\varphi$ is expressed
 with the Jacobi function ${\rm cn}(z,k)$ (see \cite{nist}, p. 565):
$$
\varphi(\tau)=\varphi_0 {\rm
  cn}\left(\tau\sqrt{1+\frac{\varphi_0^2}{2}}, \frac{1}{\sqrt{2+\frac{2}{\varphi_0^2}}}\right).
$$
Hence $\varphi$ is periodic with the period
$$
T(\varphi_0)=\frac{4}{\sqrt{1+\frac{\varphi_0^2}{2}}}\int_0^{\pi/2}\frac{1}{\sqrt{1-\frac{\varphi_0^2}{2(\varphi_0^2+1)}\sin^2\theta}}d\theta.
$$
Using inequality (19.9.1) of \cite{nist}, we can estimate this period
for $\varphi_0\neq 0$,
$$
T(\varphi_0)\leq \frac{2\pi}{\sqrt{1+\frac{\varphi_0^2}{2}}}\left(1+\frac{1}{\pi}\ln\left(1+\frac{\varphi_0^2}{\varphi_0^2+2}\right)\right)<2\pi,
$$
therefore we conclude that $\varphi$ can not be asymptotic to a free field
solution of  (\ref{KGmo}) which,  in our case,  is reduced to the simple
harmonic oscillator $\varphi''+\varphi=0$.
\end{Remark}

{\it Proof.}
If $0\leq \eta_0^+<1$, the existence of the isomorphism $\Omega_+$ is
assured iff the global Cauchy problem associated to (\ref{nlkg'}) with
initial data given at time $\tau_0$ fixed in $[0,\tau_+]$ is well
posed in $C^0\left([0,\tau_+];H^1(\KK)\right)\cap
C^1\left([0,\tau_+];L^2(\KK)\right)$. Since Theorem \ref{theonl} assures that the global Cauchy problem
is well posed on $[0,\tau_+)$, it is sufficient to study it on
$[\tau'_0,\tau_+]$ for some $\tau'_0<\tau_+$. We choose $\tau'_0$ such
that $\alpha'(\tau)\leq 0$ on $[\tau'_0,\tau_+)$. Such a choice is
possible since $c_0^+>0$ and, if $\eta_0^+=0$, $c_1>0$. In the proof of the previous
theorem, we have proved that the Cauchy problem for the linear
equation (\ref{KGsimple}) is solved by a propagator
$$
U(\tau,\tau_0)\in
C^0\left([0,\tau_+]\times[0,\tau_+];\mathcal{L}\left(H^1\times L^2\right)\right).
$$
Therefore given $\varphi_0\in H^1$, $\varphi_1\in L^2$,
$\tau_0\in[\tau'_0,\tau_+]$, we have to
solve the integral equation \begin{equation}
 \label{}
 (\varphi(\tau),\partial_{\tau}\varphi(\tau))=U(\tau,\tau_0)(\varphi_0,\varphi_1)+\int_{\tau_0}^{\tau}U(\tau,\sigma)(0,-C[\alpha(\sigma)]^{\nu}\varphi(\sigma)\mid\varphi(\sigma)\mid^{p-1})d\sigma.
\end{equation}
Using the Lipschitz property (\ref{lipche}), it is easy to get a
unique local solution defined near $\tau_0$ for
$\mid\tau-\tau_0\mid\leq\epsilon$ small enough. To obtain the global
existence, we apply (\ref{enerstra}) with $A(\tau)=-\Delta_{\mathbf
     K}+1+m^2\alpha^2(\tau)$ and $G=\varphi-\frac{d-1}{4d} R_{\gamma}\varphi-C[
 \alpha(\tau)]^{\nu}\varphi\mid \varphi\mid^{p-1}$. We get for any $\tau,\tau_0$
\begin{equation}
 \label{strastra}
 \begin{split}
\Vert\partial_{\tau}\varphi(\tau)\Vert^2_{L^2}&+\Vert\varphi(\tau)\Vert^2_{H^1}+m^2\alpha^2(\tau)\Vert\varphi(\tau)\Vert^2_{L^2}+\frac{2C}{p+1}[
 \alpha(\tau)]^{\nu}\Vert\varphi(\tau)\Vert^{p+1}_{L^{p+1}}\\
&=\Vert\partial_{\tau}\varphi(\tau_0)\Vert^2_{L^2}+\Vert\varphi(\tau_0)\Vert^2_{H^1}+m^2\alpha^2(\tau_0)\Vert\varphi(\tau_0)\Vert^2_{L^2}+\frac{2C}{p+1}[
 \alpha(\tau_0)]^{\nu}\Vert\varphi(\tau_0)\Vert^{p+1}_{L^{p+1}}\\
&+2\int_{\tau_0}^{\tau}\alpha'(\sigma)\left[m^2\alpha(\sigma)\Vert\varphi(\sigma)\Vert^2_{L^2}+\frac{C\nu}{p+1}\left[\alpha(\sigma\right]^{\nu-1}\Vert\varphi(\sigma)\Vert^{p+1}_{L^{p+1}}\right]d\sigma\\
&+2\Re\int_{\tau_0}^{\tau}\left<\left(1-\frac{d-1}{4d}R_{\gamma}\right)\varphi(\sigma),\partial_{\tau}\varphi(\sigma)\right>_{L^2}d\sigma.
\end{split}
\end{equation}
Since $\alpha'\leq 0$ on $[\tau'_0,\tau_+)$ we deduce that
\begin{equation*}
 \label{}
 \begin{split}
\Vert\partial_{\tau}\varphi(\tau)\Vert^2_{L^2}&+\Vert\varphi(\tau)\Vert^2_{H^1}+m^2\alpha^2(\tau)\Vert\varphi(\tau)\Vert^2_{L^2}+\frac{2C}{p+1}[
 \alpha(\tau)]^{\nu}\Vert\varphi(\tau)\Vert^{p+1}_{L^{p+1}}\\
&\leq\Vert\partial_{\tau}\varphi(\tau_0)\Vert^2_{L^2}+\Vert\varphi(\tau_0)\Vert^2_{H^1}+m^2\alpha^2(\tau_0)\Vert\varphi(\tau_0)\Vert^2_{L^2}+\frac{2C}{p+1}[
 \alpha(\tau_0)]^{\nu}\Vert\varphi(\tau_0)\Vert^{p+1}_{L^{p+1}}\\
&+\int_{\tau_0}^{\tau}\left(1+\Vert R_{\gamma}\Vert_{L^{\infty}}\right)\left(\Vert\partial_{\tau}\varphi(\sigma)\Vert^2_{L^2}+\Vert\varphi(\sigma)\Vert^2_{H^1}\right)d\sigma.
\end{split}
\end{equation*}
We conclude with the Gronwall lemma that
$\Vert\varphi(\tau)\Vert_{H^1}+\Vert\partial_{\tau}\varphi(\tau)\Vert_{L^2}$
is bounded and so by the continuation principle, $\varphi$ is a
solution well defined on $[\tau'_0,\tau_+]$. We have also proved that
there exists a continuous function $H$ such that for any solution
$\varphi$ we have
\begin{equation}
 \label{bornout}
 \forall
 \tau,\tau_0\in[0,\tau_+],\;\Vert\varphi(\tau)\Vert_{H^1}+\Vert \partial_{\tau}\varphi(\tau)\Vert_{L^2}\leq
 H\left(\Vert\varphi(\tau_0)\Vert_{H^1}+\Vert \partial_{\tau}\varphi(\tau_0)\Vert_{L^2}\right).
\end{equation}
Given two solutions, we write
\begin{equation*}
 \label{}
\begin{split}
 (\varphi(\tau),\partial_{\tau}\varphi(\tau))-
 (\hat\varphi(\tau),\partial_{\tau}\hat\varphi(\tau))=&U(\tau,\tau_0)\left[
   (\varphi(\tau_0),\partial_{\tau}\varphi(\tau_0))-
   (\hat\varphi(\tau),\partial_{\tau}\hat\varphi(\tau_0))\right]\\
&-C\int_{\tau_0}^{\tau}\left[\alpha(\sigma)\right]^{\nu}U(\tau,\sigma)\left(0,\varphi(\sigma)\mid\varphi(\sigma)\mid^{p-1}-\hat\varphi(\sigma)\mid\hat\varphi(\sigma)\mid^{p-1}\right)d\sigma,
\end{split}
\end{equation*}
hence by using the Lipschitz estimate (\ref{lipche}) we get
\begin{equation*}
 \label{}
\begin{split}
 \Vert(\varphi(\tau),\partial_{\tau}\varphi(\tau))-
 (\hat\varphi(\tau),\partial_{\tau}\hat\varphi(\tau))\Vert_{H^1\times L^2}&\lesssim\Vert
   (\varphi(\tau_0),\partial_{\tau}\varphi(\tau_0))-
   (\hat\varphi(\tau),\partial_{\tau}\hat\varphi(\tau_0))\Vert_{H^1\times
     L^2}\\
&+\int_{\tau_0}^{\tau}\Vert\varphi(\sigma)-\hat\varphi(\sigma)\Vert_{H^1}\left(\Vert\varphi(\sigma) \Vert_{H^1}^{p-1}+\hat\varphi(\sigma)\Vert_{H^1}^{p-1}\right)d\sigma.
\end{split}
 \end{equation*}
We use (\ref{bornout}) and the Gronwall Lemma to obtain
(\ref{lipschoucou}) and \ref{lipschoucoucou}).\\

We now consider the case $\eta_0^+\geq 1$. Theorem \ref{theonl} assures that
given $u_0\in H^1$, $u_1\in L^2$
the nonlinear equation (\ref{nlkg'}) has a unique solution $\varphi\in
C^0\left([0,\infty);H^1\right)\cap C^1\left([0,\infty);L^2\right)$ with
initial data $(\varphi_0,\varphi_1)={\mathcal L}(u_0,u_1)$ given
by (\ref{ufifi}). To establish the existence of $\Omega_+$
it is sufficient to prove the
existence of the nonlinear operator $\frak{W_+}$ defined by
\begin{equation}
 \label{frakw}
 \frak{W_+}:\;(\varphi_0,\varphi_1)\mapsto \left(\psi_0^+,\psi_1^+\right):=\lim_{\tau\rightarrow\infty}U(0,\tau)\left(\varphi(\tau),\partial_{\tau}\varphi(\tau)\right)\;in\;H^1\times
L^2,
\end{equation}
then, using operator $W_+$ given by the previous Theorem, we put
\begin{equation}
 \label{omegaw}
 \Omega_+:=W_+ \circ{\mathcal L}^{-1}\circ \frak{W_+}\circ{\mathcal L}.
\end{equation}
We have
$$
U(0,\tau)\left(\varphi(\tau),\partial_{\tau}\varphi(\tau)\right)=
(\varphi_0,\varphi_1)-C\int_0^{\tau}\left[\alpha(\sigma)\right]^{\nu}U(0,\sigma)\left(0,\varphi(\sigma)\mid\varphi(\sigma)\mid^{p-1}\right)d\sigma.
$$
From (\ref{bornage}) and (\ref{Rdelta}) we have
\begin{equation}
 \label{estpropag}
 \sup_{0\leq\sigma,\tau<\infty}\Vert U(\tau,\sigma)\Vert_{\mathcal{L}(H^1\times L^2)}<\infty.
\end{equation}
Now we apply (\ref{enerstra}) with $A(\tau)=-\Delta_{\mathbf
     K}+m^2\alpha^2(\tau) +\frac{d-1}{4d} R_{\gamma}$ and $G=-C[
 \alpha(\tau)]^{\nu}\varphi\mid \varphi\mid^{p-1}$. We get for any
 $\tau,\tau_0\geq 0$,
\begin{equation}
 \label{strostro}
 \begin{split}
\Vert\partial_{\tau}\varphi(\tau)\Vert^2_{L^2}&+\Vert\nabla_{\KK}\varphi(\tau)\Vert^2_{L^2}+m^2\alpha^2(\tau)\Vert\varphi(\tau)\Vert^2_{L^2}+\frac{d-1}{4d}\Vert
R_{\gamma}\varphi(\tau)\Vert^2_{L^2}+\frac{2C}{p+1}[
 \alpha(\tau)]^{\nu}\Vert\varphi(\tau)\Vert^{p+1}_{L^{p+1}}\\
&=\Vert\partial_{\tau}\varphi(\tau_0)\Vert^2_{L^2}+\Vert\nabla_{\KK}\varphi(\tau_0)\Vert^2_{L^2}+m^2\alpha^2(\tau_0)\Vert\varphi(\tau_0)\Vert^2_{L^2}+\frac{d-1}{4d}\Vert
R_{\gamma}\varphi(\tau_0)\Vert^2_{L^2}\\
&+\frac{2C}{p+1}[
 \alpha(\tau_0)]^{\nu}\Vert\varphi(\tau_0)\Vert^{p+1}_{L^{p+1}}+2\int_{\tau_0}^{\tau}\alpha'(\sigma)\left[m^2\alpha(\sigma)\Vert\varphi(\sigma)\Vert^2_{L^2}+\frac{C\nu}{p+1}\left[\alpha(\sigma\right]^{\nu-1}\Vert\varphi(\sigma)\Vert^{p+1}_{L^{p+1}}\right]d\sigma.
\end{split}
\end{equation}
Thanks to (\ref{Rdelta}) there exists $C_{\delta}>0$ such that for any
$\tau,\tau_0\geq 0$,
\begin{equation}
 \label{esterprout}
 \begin{split}
\Vert\partial_{\tau}\varphi(\tau)\Vert^2_{L^2}&+C_{\delta}^{-1}\Vert\varphi(\tau)\Vert^2_{H^1}+\frac{2C}{p+1}[
 \alpha(\tau)]^{\nu}\Vert\varphi(\tau)\Vert^{p+1}_{L^{p+1}}\\
&\leq\Vert\partial_{\tau}\varphi(\tau_0)\Vert^2_{L^2}+C_{\delta}\Vert\varphi(\tau_0)\Vert^2_{H^1}+\frac{2C}{p+1}[
 \alpha(\tau_0)]^{\nu}\Vert\varphi(\tau_0)\Vert^{p+1}_{L^{p+1}}\\
&+2\int_{\tau_0}^{\tau}\alpha'(\sigma)\left[m^2\alpha(\sigma)\Vert\varphi(\sigma)\Vert^2_{L^2}+\frac{C\nu}{p+1}\left[\alpha(\sigma\right]^{\nu-1}\Vert\varphi(\sigma)\Vert^{p+1}_{L^{p+1}}\right]d\sigma.
\end{split}
\end{equation}
Since $\alpha>0$ on $[0,\infty)$ and $\alpha'/\alpha\in L^1_{loc}([0,\infty))$, the Gronwall lemma
assures the existence of a continuous function $C_0$ such that for any
$\tau,\tau_0\in[0,T]$ we have
\begin{equation}
 \label{estimoT}
 \Vert\varphi(\tau)\Vert_{H^1}+\Vert\partial_{\tau}\varphi(\tau)\Vert_{L^2}\leq
 C_0(T)\left(\Vert\varphi(\tau_0)\Vert_{H^1}+\Vert\partial_{\tau}\varphi(\tau_0)\Vert_{L^2}
+\Vert \varphi(\tau_0)\Vert_{L^{p+1}}^{\frac{p+1}{2}}\right).
\end{equation}

As previously, we choose $\tau'_0$ such that $\alpha'(\tau)<0$ for
$\tau\geq\tau'_0$.
We take $\tau_0=\tau'_0$ in (\ref{esterprout}) so we have for
$\tau\geq \tau'_0$
\begin{equation}
\label{estapri}
 \begin{split}
\Vert\partial_{\tau}\varphi(\tau)\Vert^2_{L^2}&+C_{\delta}^{-1}\Vert\varphi(\tau)\Vert^2_{H^1}+\frac{2C}{p+1}[
 \alpha(\tau)]^{\nu}\Vert\varphi(\tau)\Vert^{p+1}_{L^{p+1}}\\
&\leq\Vert\partial_{\tau}\varphi(\tau'_0)\Vert^2_{L^2}+C_{\delta}\Vert\varphi(\tau'_0)\Vert^2_{H^1}+\frac{2C}{p+1}[
 \alpha(\tau'_0)]^{\nu}\Vert\varphi(\tau'_0)\Vert^{p+1}_{L^{p+1}}.
\end{split}
\end{equation}
We deduce from (\ref{estimoT}), (\ref{estapri}) and  the Sobolev
estimate, that  there exists a continuous function
$H$ such that
\begin{equation}
 \label{solestH}
 \sup_{\tau\geq 0}\left(\Vert\partial_{\tau}\varphi(\tau)\Vert_{L^2}+\Vert\varphi(\tau)\Vert_{H^1}\right)\leq H\left(\Vert\partial_{\tau}\varphi(0)\Vert_{L^2}+\Vert\varphi(0)\Vert_{H^1}\right),
\end{equation}
\begin{equation}
 \label{estnl}
 \sup_{\tau\geq 0}\Vert
 \varphi(\tau)\mid\varphi(\tau)\mid^{p-1}\Vert_{L^2}\leq H\left(\Vert\partial_{\tau}\varphi(0)\Vert_{L^2}+\Vert\varphi(0)\Vert_{H^1}\right).
\end{equation}
On the other hand (\ref{nuuu}), (\ref{condipi}), (\ref{appun}) and
(\ref{appgequn}) imply that
\begin{equation}
 \label{lunalf}
 \left[\alpha(\tau)\right]^{\nu}\in
L^1(0,\infty).
\end{equation}
Therefore thanks to (\ref{estpropag}) and
(\ref{estnl}), we may define $(\psi_0^+,\psi_1^+)$ as
\begin{equation}
 \label{}
 (\psi_0^+,\psi_1^+):=(\varphi_0,\varphi_1)-C\int_0^{\infty}\left[\alpha(\sigma)\right]^{\nu}U(0,\sigma)(0,\varphi(\sigma)\mid\varphi(\sigma)\mid^{p-1})d\sigma.
\end{equation}

To invert $\Omega_+$, given $(\varphi^+_0,\varphi^+_-)\in H^1\times L^2$,
it is sufficient to solve in $C^0\left([0,\infty);H^1\times L^2\right)$ the integral equation
\begin{equation}
 \label{eqinf}
 (\varphi(\tau),\partial_{\tau}\varphi(\tau))=U(\tau,0)(\psi_0^+,\psi_1^+)+C\int_{\tau}^{\infty}\left[\alpha(\sigma)\right]^{\nu}U(\tau,\sigma)(0,\varphi(\sigma)\mid\varphi(\sigma)\mid^{p-1})d\sigma,
\end{equation}
with
$$
(\psi_0^+,\psi_1^+)={\mathcal L}\circ (W_+)^{-1}(\varphi_0^+,\varphi_1^+),
$$
and we put
\begin{equation}
 \label{wwm}
 W_+^{-1}(\varphi_0^+,\varphi_1^+):={\mathcal L}^{-1}(\varphi(0),\partial_{\tau}\varphi(0)).
\end{equation}
We remark that we may just solve (\ref{eqinf}) on some interval
$[T,\infty)$ and then we consider the unique solution of  the Cauchy problem on $[0,T]$
with initial data at $T$ given by
$(\varphi(T),\partial_{\tau}\varphi(T))$. We introduce the
Banach space $X_T$ of the bounded continuous $H^1\times L^2$-valued functions
on $[T,\infty)$. Thanks to
(\ref{estpropag}) and (\ref{lunalf}), the map
$$
\mathcal{G}: (\varphi,\psi)\in X_T\longmapsto \mathcal{G} (\varphi,\psi) (\tau):=U(\tau,0)(\psi_0^+,\psi_1^+)+C\int_{\tau}^{\infty}\left[\alpha(\sigma)\right]^{\nu}U(\tau,\sigma)(0,\varphi(\sigma)\mid\varphi(\sigma)\mid^{p-1})d\sigma
$$
is well defined. Moreover if $T$ is large enough, $\mathcal{G}$ is a strict contraction
on
$$
\left\{(\varphi,\psi)\in X_T;\;\;\sup_{\tau\geq
    T}\Vert(\varphi(\tau),\psi(\tau))\Vert_{H^1\times L^2}\leq 2 \sup_{\tau\geq
    T}\Vert U(\tau,0)(\psi_0^+,\psi_1^+)\Vert_{H^1\times L^2}\right\}.
$$
The fixed point $(\varphi,\psi)=\mathcal{G} (\varphi,\psi)$ satisfies
$\psi=\partial_{\tau}\varphi$ and $\varphi$ is a solution of
(\ref{eqinf}) on $[T,\infty)$. Finally we extend $\varphi$ into a
solution on $[0,\infty)$ by solving the Cauchy problem with initial
data given at time $T$ by
$(\varphi(T),\partial_{\tau}\varphi(T))$. Therefore we have
constructed a solution of (\ref{eqinf}) and $W_+^{-1}$ is well
defined by (\ref{wwm}). We remark that $T$ only depends on
$\Vert(\psi_0^+,\psi_1^+)\Vert_{H^1\times L^2}$. Therefore we
conclude with (\ref{estpropag}) and (\ref{estimoT}) that there exists
a continuous function $K$ such that
\begin{equation}
 \label{solestK}
 \sup_{\tau\geq
   0}\left(\Vert\partial_{\tau}\varphi(\tau)\Vert_{L^2}+\Vert\varphi(\tau)\Vert_{H^1}\right)\leq
 K\left(\Vert\psi_0^+\Vert_{H^1}+\Vert\psi_1^+\Vert_{L^2}\right).
\end{equation}

To establish the Lipchitz properties of $W_+$ and $W_+^{-1}$ we
write
$$
 (\varphi_0^+,\varphi_1^+)-(\hat\varphi_0^+,\hat\varphi_1^+)=(\varphi_0-\hat\varphi_0,\varphi_1-\hat\varphi_1)-C\int_0^{\infty}\left[\alpha(\sigma)\right]^{\nu}U(0,\sigma)(0,\varphi(\sigma)\mid\varphi(\sigma)\mid^{p-1}-\hat\varphi(\sigma)\mid\hat\varphi(\sigma)\mid^{p-1})d\sigma,
$$
then (\ref{lipschoucou}) and   (\ref{lipschoucoucou}) follow from (\ref{lipche}),
(\ref{estpropag}), (\ref{solestH}), (\ref{solestK}) and
(\ref{lunalf}).
\fin

\section{At a Big Rip (conformal coupling)}
In this part we investigate the behaviour of the solutions of the
linear Klein-Gordon equation with a conformal coupling (\ref{kg}), (\ref{xiconf}),
if a Big Rip occurs at $t_+$, that is to say we assume that the scale
factor satisfies (\ref{condida}) and (\ref{condidac}) with
$\eta_0^+<0$. Using the conformal time $\tau$ we deal with the
Klein-Gordon equation (\ref{KGsimple}) with the variable mass $m \alpha(\tau)$
that tends to infinity as
$\tau\rightarrow\tau_+$.  More precisely (\ref{applequn}) assures that
for some $C>0$, 
\begin{equation}
 \label{alfunbr}
\alpha(\tau)\sim C\left(\tau_+-\tau\right)^{\frac{\eta_0^{+}}{1-\eta_0^{+}}}.
\end{equation}
We remark that the potential $m^2\alpha^2(\tau)$ in (\ref{KGsimple})
belongs to $L^1(0,\tau_+)$ iff  $-1<\eta_0^+<0$ (Slow Big Rip).  We start by
considering this case.

\begin{Theorem}[Slow Big Rip]
We assume  that (\ref{condida}), (\ref{condidac}) and (\ref{rb})
hold. We suppose that $-1<\eta_0^+<0$.
 Given $\theta\in[0,1]$, $u_0\in H^{1-\theta}(\mathbf{K})$, $u_1\in
 H^{-\theta}(\mathbf{K})$, we consider the  unique $u\in C^0\left(I;
   H^{1-\theta}(\mathbf{K})\right)\cap C^1\left(I;
   H^{-\theta}(\mathbf{K})\right)$ solution of (\ref{kg}),
 (\ref{xiconf}), (\ref{CI}). Then
 there exist unique $\varphi_0^+\in
 H^{1-\theta}(\KK)$,  $\varphi_1^+\in
 H^{-\theta}(\KK)$, such that
\begin{equation}
 \label{assszer}
 \left\Vert
 \left[a(t)\right]^{\frac{d-1}{2}}u(t,.)-\varphi_0^+\right\Vert_{H^{1-\theta}}+\left\Vert a(t)\partial_t\left[\left[a(t)\right]^{\frac{d-1}{2}}u(t,.)\right]-\varphi_1^+\right\Vert_{H^{-\theta}}\longrightarrow
   0,\;\;t\rightarrow t_+,
\end{equation}
and the map $W_+:(u_0,u_1)\mapsto (\varphi_0^+,\varphi_1^+)$ is an isomorphism
on $H^{1-\theta}(\mathbf{K})\times H^{-\theta}(\mathbf{K})$.
\label{peititrip}
\end{Theorem}

{\it Proof.} The proof is similar to those of the case
$0\leq\eta_0^+<1$ but $\alpha$ is no more continuous on
$[0,\tau_+]$. Nevertheless thanks to (\ref{alfunbr}), we have
$\alpha^2(\tau)\in L^1(0,\tau_+)$ therefore we obtain
(\ref{estofi}) from (\ref{estigron}) again, and
$(\varphi_0^+,\varphi_1^+)$ is well defined by (\ref{soleq}). Conversely, given
$(\varphi_0^+,\varphi_1^+)\in H^{1-\theta}\times H^{-\theta}$, we can
easily solve the integral equation (\ref{integqp}) for
$\tau\in[\tau_0,\tau_+]$ for some $\tau_0<\tau_+$. It is sufficient to
take $\tau_+-\tau_0$ small enough in order the right member of
(\ref{integqp}) to be a strict contraction on
$C^0([\tau_0,\tau_+];H^{1-\theta}\times H^{-\theta})$. Then
$(\varphi(0),\partial_{\tau}\varphi(0)$ is obtained by unique continuation.

\fin

If $\eta_0^+\leq-1$ we may not expect such a result. Here are some
examples showing that $W_+$ cannot be defined as an isomorphism, more
precisely, in these cases,  $\varphi(\tau)$ has a limit as
$\tau\rightarrow\tau_+$ but in contrast,
$\partial_{\tau}\varphi(\tau)$ is diverging. We suppose that $\KK$ is a compact manifold and we
assume that there exists $\Phi_0\in L^2(\KK)$ solution of
$-\Delta_{\KK}\Phi_0+\frac{d-1}{4d}R_{\gamma}\Phi_0=0$ (for instance,
$\KK$ is a compact flat manifold and $\Phi_0=1$). Then the solution
$\varphi$ of
$\left[\partial^2_{\tau}-\Delta_{\KK}+m^2\alpha^2(\tau)+\frac{d-1}{4d}R_{\gamma}\right]\varphi=0$,
$\varphi(0)=c_0\Phi_0$, $\partial_{\tau}\varphi(0)=c_1\Phi_0$, is
given by $\varphi(\tau,\xx)=u(\tau)\Phi_0(\xx)$ where $u$ is a
solution of the differential equation
\begin{equation}
 \label{edo}
 u''+m^2\alpha^2(\tau)u=0.
\end{equation}

First we investigate the case $ a(t)=2(t_+-t)^{-1}$. We have
\begin{equation*}
 \label{}
 \alpha(\tau)=(\tau_+-\tau)^{-\frac{1}{2}}.
\end{equation*}
Then the solution of (\ref{edo}) is
expressed with Bessel functions,
$$
u(\tau)=\sqrt{\tau_+-\tau}\left[a J_1\left(2m
    \sqrt{\tau_+-\tau}\right)+b  Y_1\left(2m
    \sqrt{\tau_+-\tau}\right)\right]\Phi_0(\xx),\;\;a,b\in\CC,
$$
hence as $\tau\rightarrow\tau_+$,
\begin{equation*}
 \label{}
 \varphi(\tau,\xx)\sim-\frac{b}{m\pi}\Phi_0(\xx),\;\;\partial_{\tau}\varphi(\tau,\xx)\sim-\frac{mb}{2\pi}\ln(\tau_+-\tau)\Phi_0(\xx).
\end{equation*}
We remark that the asymptotics are characterized by $b$, which is not
sufficient to obtain $c_0$, $c_1$. Moreover the time derivative is not
bounded and the following limits exist:
\begin{equation}
 \label{estiplouc}
 \lim_{\tau\rightarrow\tau_+}\varphi(\tau)\;\;and\;\;  \lim_{\tau\rightarrow\tau_+}\left(1+\int_0^{\tau}\alpha^2(\sigma)d\sigma\right)^{-1}\partial_{\tau}\varphi(\tau)\;\;in\;\;L^2(\KK).
\end{equation}

Secondly we investigate the case $a(t)=9(t_+-t)^{-2}$ for which:
\begin{equation*}
 \label{}
 \alpha(\tau)=(\tau_+-\tau)^{-\frac{2}{3}}.
\end{equation*}
Then the solutions $u$ of (\ref{edo}) are
$$
u(\tau)=ae^{3m(\tau_+-\tau)^{\frac{1}{3}}i}\left(1-3m(\tau_+-\tau)^{\frac{1}{3}}i\right)+be^{-3m(\tau_+-\tau)^{\frac{1}{3}}i}\left(1+3m(\tau_+-\tau)^{\frac{1}{3}}i\right),\;\;a,b\in\CC.
$$
We have the following asymptotics as $\tau\rightarrow\tau_+$:
\begin{equation*}
 \label{}
 \varphi(\tau,\xx)\sim
 (a+b)\Phi_0(\xx),\;\;\partial_{\tau}\varphi(\tau,\xx)\sim -3m^2(a+b)(\tau_+-\tau)^{-\frac{1}{3}}\Phi_0(\xx).
\end{equation*}
Hence the asymptotics are given by $a+b$ that is not sufficient to
get $c_0$, $c_1$. Moreover (\ref{estiplouc}) is satisfied again.

Now we choose $a(t)=64(t_+-t)^{-3}$, hence
\begin{equation*}
 \label{}
 \alpha(\tau)=(\tau_+-\tau)^{-\frac{3}{4}},
\end{equation*}
and the solutions of  (\ref{edo}) are
\begin{equation*}
 \label{}
\begin{split}
 u(\tau)&=a\left[2m(\tau_+-\tau)^{\frac{1}{2}}J_0\left(4m(\tau_+-\tau)^{\frac{1}{4}}\right)-(\tau_+-\tau)^{\frac{1}{4}}J_1\left(4m(\tau_+-\tau)^{\frac{1}{4}}\right)\right]\\
&+b\left[2m(\tau_+-\tau)^{\frac{1}{2}}Y_0\left(4m(\tau_+-\tau)^{\frac{1}{4}}\right)-(\tau_+-\tau)^{\frac{1}{4}}Y_1\left(4m(\tau_+-\tau)^{\frac{1}{4}}\right)\right],\;\;a,b\in\CC,
\end{split}
\end{equation*}
and we have as $\tau\rightarrow\tau_+$
\begin{equation*}
 \label{}
 \varphi(\tau,\xx)\sim\frac{b}{2m\pi}\Phi_0(\xx),\;\;\partial_{\tau}\varphi(\tau,\xx)\sim-\frac{mb}{\pi}(\tau_+-\tau)^{-\frac{1}{2}}\Phi_0(\xx).
\end{equation*}
Therefore the previous conclusions hold. To avoid the possibility that
the eigenvalue zero plays a peculiar role, we consider an
eigenfunction $\Phi_{\lambda}$ associated to some eigenvalue
$\lambda^2>0$ of $-\Delta_{\KK}+\frac{d-1}{4d}R_{\gamma}$, and we take
$a(t)=2(t_+-t)^{-1}$.
Then the solution
$\varphi$ of
$\left[\partial^2_{\tau}-\Delta_{\KK}+m^2\alpha^2(\tau)+\frac{d-1}{4d}R_{\gamma}\right]\varphi=0$,
$\varphi(0)=c_0\Phi_{\lambda}$, $\partial_{\tau}\varphi(0)=c_1\Phi_{\lambda}$, is
given by $\varphi(\tau,\xx)=u(\tau)\Phi_{\lambda}(\xx)$ where $u$ is a
solution of the ODE
\begin{equation}
 \label{edlo}
 u''+\lambda^2u+\frac{m^2}{\tau_+-\tau}u=0.
\end{equation}
The solutions are expressed with Whittaker functions (with the
notations of \cite{nist}):
\begin{equation}
 \label{}
 u(\tau)=aM_{\mu,\frac{1}{2}}(-2i\lambda(\tau_+-\tau))+bW_{\mu,\frac{1}{2}}(-2i\lambda(\tau_+-\tau)),\;\;\mu:=\frac{m^2}{2\lambda}i,\;\;a,b\in\CC.
\end{equation}
We deduce that
\begin{equation*}
 \label{}
 \varphi(\tau)\sim\frac{b}{\Gamma\left(1-\frac{m^2}{2\lambda}i\right)}\Phi_{\lambda}(\xx),\;\;\partial_{\tau}\varphi(\tau,\xx)\sim-\frac{bm^2}{\Gamma\left(1-\frac{m^2}{2\lambda}i\right)}\ln(\tau_+-\tau)\Phi_{\lambda}(\xx).
\end{equation*}

All these examples show that the following Theorem concerning the
Strong Big Rip ($\eta_0^+\leq -1$), is optimal.

\begin{Theorem}[Strong Big Rip]
We assume  that (\ref{condida}), (\ref{condidac}) and (\ref{rb})
hold. We suppose  $\eta_0^+\leq-1$.
 Given $\theta\in[0,1]$, $u_0\in H^{1-\theta}(\mathbf{K})$, $u_1\in
 H^{-\theta}(\mathbf{K})$, we consider the  unique $u\in C^0\left(I;
   H^{1-\theta}(\mathbf{K})\right)\cap C^1\left(I;
   H^{-\theta}(\mathbf{K})\right)$ solution of (\ref{kg}),
 (\ref{xiconf}), (\ref{CI}). Then
 there exists  unique $\varphi_0^+\in
 H^{1-\theta}(\KK)$ such that
\begin{equation}
 \label{asssrip}
\left[ a(t)\right]^{\frac{d-1}{2}}u(t,.)\rightarrow\varphi_0^+\;in\;H^{1-\theta}-weak*,\;\; \left[ a(t)\right]^{\frac{d-1}{2}}u(t,.)\rightarrow\varphi_0^+\;in\;H^{-\theta}\;\;t\rightarrow t_+,
\end{equation}
\begin{equation}
 \label{dtripou}
 (t_+-t)^{-\eta_0^+-1}\partial_t\left(\left[a(t)\right]^{\frac{d-1}{2}}u\right)\in
 C^0\cap L^{\infty}([t_0,t_+);H^{-\theta}).
\end{equation}

\label{bigrip}
\end{Theorem}

The proof of the Theorem is based on an elegant method used in
\cite{delsanto}. The following Lemmas, an abstract energy estimate and the
resolution of the Ricatti equation, are directly inspired by this paper.

Given a densely defined selfadjoint operator $B$ on a Hilbert space
$(\mathcal H,\Vert.\Vert)$, satisfying $Id\leq B$, we denote
${\mathcal H}^{\theta}$, $\theta\leq 2$,
the closure of $Dom(B^2)$ for the norm $\Vert u\Vert_{\theta}:=\Vert B^{\theta}u\Vert$.
\begin{Lemma}
 \label{enerkoko}
Given $A\in C^1([\tau_0,\tau_+))$ and $G\in C^0\left([\tau_0,\tau_+);\mathcal{L}(\mathcal{H})\cap \mathcal{L}(\mathcal{H}^1)\right)$, any $u\in
C^0\left([\tau_0,\tau_+); {\mathcal H}^{1-\theta}\right)\cap
C^1\left([\tau_0,\tau_+); {\mathcal H}^{-\theta}\right)$,
$\theta\in[0,1]$,  solving
\begin{equation}
 \label{EQUAKA}
 \partial_{\tau}^2u+B^2u+(A'(\tau)-A^2(\tau))u=G(\tau)u,
\end{equation}
 satisfies for all $\tau\in[\tau_0,\tau_+)$
\begin{equation}
 \label{ernestine}
 \Vert u(\tau)\Vert_{1-\theta}+\frac{1}{1+\mid
 A(\tau)\mid}\Vert\partial_{\tau}u(\tau)\Vert_{-\theta}\leq
2\left[\Vert u(\tau_0)\Vert_{1-\theta}+\frac{1}{1+\mid
  A(\tau_0)\mid}\Vert\partial_{\tau}u(\tau_0)\Vert_{-\theta}\right]e^{\int_{\tau_0}^{\tau}\mid A(\sigma)\mid+\Vert
  G(\sigma)\Vert_{\mathcal{L}(\mathcal{H}) d\sigma}}. 
\end{equation}
\end{Lemma}

{\it Proof.}
We remark that thanks to the assumptions on $G$, the Cauchy problem is well posed in
$\mathcal{H}^s\times{\mathcal H}^{s-1}$ for $s\in[0,2]$ therefore it
is sufficient to consider $u\in
C^0\left([\tau_0,\tau_+);\mathcal{H}^2\right)\cap
C^1\left([\tau_0,\tau_+);\mathcal{H}^1\right)\cap
C^2\left([\tau_0,\tau_+);\mathcal{H}\right)$, and the general case
follows from the usual argument of density and continuity. Given $\theta\in[0,1]$, we introduce the modified energy
\begin{equation}
 \label{}
 \mathcal{E}(\tau):=\left(\Vert B^{1-\theta}u(\tau)\Vert^2+\Vert B^{-\theta}\left(\partial_{\tau}u(\tau)+A(\tau)u(\tau)\right)\Vert^2\right)^{\frac{1}{2}}.
\end{equation}
We have
\begin{equation}
 \label{compare}
 \Vert u(\tau)\Vert_{1-\theta}\leq  \mathcal{E}(\tau),\;\;
 \Vert \partial_{\tau}u(\tau)\Vert_{-\theta}\leq  (1+\mid A(\tau)\mid)\mathcal{E}(\tau).
\end{equation}
We compute
\begin{equation*}
 \begin{split}
 \mathcal{E}(\tau)\frac{d}{d\tau}
 \mathcal{E}(\tau)&=A(\tau)\left[-\Vert
   B^{1-\theta}u(\tau)\Vert^2+\Vert
   B^{-\theta}\partial_{\tau}u(\tau)+A(\tau)B^{-\theta}u(\tau)\Vert^2\right]\\
&+\Re
<B^{-\theta}\partial_{\tau}u(\tau)+A(\tau)B^{-\theta}u(\tau),B^{-\theta}G(\tau)u(\tau)>\\
&\leq \mid A(\tau)\mid\mathcal{E}^2(\tau)+\Vert G(\tau)\Vert_{\mathcal{L}(\mathcal{H})}\mathcal{E}^2(\tau).
\end{split}
\end{equation*}
We deduce from the Gronwall lemma:
\begin{equation}
 \label{kattof}
 \mathcal{E}(\tau)\leq \mathcal{E}(\tau_0) e^{\int_{\tau_0}^{\tau}\mid A(\sigma)\mid+\Vert
  G(\sigma)\Vert_{\mathcal{L}(\mathcal{H})} d\sigma}.
\end{equation}
Now (\ref{ernestine}) follows from (\ref{compare}) and (\ref{kattof}).

\fin


\begin{Lemma}
 \label{lemricat}
Let $V$ be a non-negative, continuous function on $[\tau_1,\tau_+)$.
We assume that $V(\tau)=O\left(\mid\tau_+-\tau\mid^{\gamma}\right)$ with
$\gamma>-2$. Then, given $M>1$, there exists $\tau_0\in[\tau_1,\tau_+)$ such that the Ricatti
 equation
\begin{equation}
 \label{ricatt}
 A'(\tau)-A^2(\tau)=V(\tau),\;\;\tau\in[\tau_0,\tau_+),
\end{equation}
has a unique solution such that $A(\tau_0)=0$, moreover $A$ satisfies
\begin{equation}
 \label{encadra}
 \int_{\tau_0}^{\tau}V(\sigma)d\sigma\leq A(\tau)\leq 2M\int_{\tau_0}^{\tau}V(\sigma)d\sigma.
\end{equation}
\begin{equation}
 \label{INTOTO}
 \int_{\tau_0}^{\tau_+}A(\tau)d\tau\leq\frac{1}{2M}.
\end{equation}

\end{Lemma}

{\it Proof.}
Since
$V(\tau)=O\left(\mid\tau_+-\tau\mid^{\gamma}\right)$ with
$\gamma>-2$,  we can choose $\tau_0\in[\tau_1,\tau_+)$ such that
\begin{equation*}
 \label{}
 \int_{\tau_0}^{\tau_+}\left(\int_{\tau_0}^{\tau}V(\sigma)d\sigma\right)d\tau\leq\frac{1}{4M^2}.
\end{equation*}
We iteratively define the sequence $A_n(\tau)$ by $A_0(\tau)=0$ and
\begin{equation*}
 \label{}
 A_{n+1}(\tau)=\int_{\tau_0}^{\tau}V(\sigma)d\sigma+\int_{\tau_0}^{\tau}A_n^2(\sigma)d\sigma.
\end{equation*}
We can easily prove by recurrence that for $n\geq1$ we have
$$
\int_{\tau_0}^{\tau}V(\sigma)d\sigma\leq A_n(\tau)\leq A_{n+1}(\tau).
$$
Moreover if we assume that
$$
A_n(\tau)\leq 2M\int_{\tau_0}^{\tau}V(\sigma)d\sigma,
$$
we evaluate
\begin{equation*}
 \label{}
\begin{split}
 A_{n+1}(\tau)\leq&
 \int_{\tau_0}^{\tau}V(\sigma)d\sigma+4M^2\int_{\tau_0}^{\tau}\left(\int_{\tau_0}^{\sigma}V(\rho)d\rho\right)^2d\sigma\\
\leq&
\int_{\tau_0}^{\tau}V(\sigma)d\sigma+4M^2\left(\int_{\tau_0}^{\tau}V(\rho)d\rho\right)\left(\int_{\tau_0}^{\tau_+}\left(\int_{\tau_0}^{\sigma}V(\rho)d\rho\right)d\sigma\right)\\
\leq&2M \int_{\tau_0}^{\tau}V(\sigma)d\sigma.
\end{split}
\end{equation*}
We conclude that $A_n(\tau)$ is an increasing sequence bounded by $2M
\int_{\tau_0}^{\tau}V(\sigma)d\sigma$. Therefore
$A:=\lim_nA_n$ satisfies (\ref{encadra}) and by the Beppo Levi theorem
it is a solution of
$$
A(\tau)=\int_{\tau_0}^{\tau}V(\sigma)d\sigma+\int_{\tau_0}^{\tau}A^2(\sigma)d\sigma.
$$
We conclude that $A$ belongs to  $C^{\infty}([\tau_0,\tau_+))$ and it is a
solution of (\ref{ricatt}) (parenthetically the first Dini theorem assures
that $A_n$ converges locally uniformly to $A$ on $[\tau_0,\tau_+)$)
. Moreover we have
$$
\int_{\tau_0}^{\tau_+}A(\tau)d\tau\leq 2M\int_{\tau_0}^{\tau_+}\left(\int_{\tau_0}^{\tau}V(\sigma)d\sigma\right)d\tau\leq\frac{1}{2M}.
$$

\fin


{\it Proof of the Theorem.}
(\ref{alfunbr}) assures that
$\alpha^2(\tau)=O\left(\mid\tau_+-\tau\mid^{\gamma}\right)$ with
$\gamma>-2$. We use the potential $A$ given by  the previous Lemma
\ref{lemricat} with
$V(\tau)=m^2\alpha^2(\tau)$ and we apply
Lemma \ref{enerkoko} with $B^2=-\Delta_{\KK}+1$,
$G=1-\frac{d-1}{4d}R_{\gamma}$ to obtain
\begin{equation*}
 \label{}
 \varphi\in C^0\cap
 L^{\infty}([\tau_0,\tau_+);H^{1-\theta}),\;\;(1+A)^{-1}\partial_{\tau}\varphi\in
 C^0\cap L^{\infty}([\tau_0,\tau_+);H^{-\theta}).
\end{equation*}
Since $\alpha^2(\tau)\sim
C(\tau_+-\tau)^{\frac{2\eta_0^+}{1-\eta_0^+}}$ and $\eta_0^+\leq-1$, we get from
(\ref{encadra}) that
\begin{equation}
 \label{dtripoufi}
 (\tau_+-\tau)^{\frac{\eta_0^++1}{\eta_0^+-1}}\partial_{\tau}\varphi\in
 C^0\cap L^{\infty}([\tau_0,\tau_+);H^{-\theta}),
\end{equation}
and we deduce that
\begin{equation*}
 \label{}
 \partial_{\tau}\varphi\in
 L^{1}([\tau_0,\tau_+);H^{-\theta}).
\end{equation*}
We conclude with the Corollary 2.1 of \cite{strauss66} that
\begin{equation}
 \label{assripfi}
 \varphi\in C^0([\tau_0,\tau_+],H^{-\theta})\cap
 C^0_w([\tau_0,\tau_+], H^{1-\theta}),
\end{equation}
where $C_w(I,X)$ denotes the space of the $X$-valued functions defined
on an interval $I$ that are weakly continuous. Coming back to $u(t)$,
(\ref{asssrip}) follows from (\ref{assripfi}) and (\ref{dtripou}) is
equivalent to  (\ref{dtripoufi}).

\fin

\section{At a Sudden Singularity (Non Conformal Coupling)}

We consider the case of Sudden Singularity, $\eta_0^+=0$,
$\eta_1^+\in(0,1)\cup(1,\infty)$, and a general coupling $\xi\in\RR$,
$\xi\neq \frac{d-1}{4d}.$ 
In
conformal time, we consider equation (\ref{kgfifi})
where thanks to (\ref{appzero}) the potential $V\in
C^0(\tau_-,\tau_+)$ given by (\ref{VQ}) satisfies
\begin{equation}
 \label{vssg}
 V(\tau)\sim
m^2(c_0^+)^2+
\left(\xi-\frac{d-1}{4d}\right)2d\eta_1^+(\eta_1^+-1)c_0^{\eta_1^+-1}c_1^+\mid\tau-\tau_+\mid^{\eta_1^+-2},\;\;\tau\rightarrow\tau_+.
\end{equation}

In the case of a Big Brake ($\eta_1^+>1$), we have  $V\in L^1([0,\tau_+))$ and the situation
is analogous to those of the conformal coupling.

\begin{Theorem}[Big Brake]
We assume  that (\ref{condida}), (\ref{condidac}) and (\ref{rb}) and
hold. We suppose that $\eta_0^+=0$, $\eta_1^+>1$.
 Given $\theta\in[0,1]$, $u_0\in H^{1-\theta}(\mathbf{K})$, $u_1\in
 H^{-\theta}(\mathbf{K})$, we consider the  unique $u\in C^0\left(I;
   H^{1-\theta}(\mathbf{K})\right)\cap C^1\left(I;
   H^{-\theta}(\mathbf{K})\right)$ solution of (\ref{kg}),  (\ref{CI}). Then
 there exist unique $\varphi_0^+\in
 H^{1-\theta}(\KK)$,  $\varphi_1^+\in
 H^{-\theta}(\KK)$, such that
\begin{equation}
 \label{asszerg}
 \left\Vert
 \left[a(t)\right]^{\frac{d-1}{2}}u(t,.)-\varphi_0^+\right\Vert_{H^{1-\theta}}+\left\Vert a(t)\partial_t\left[\left[a(t)\right]^{\frac{d-1}{2}}u(t,.)\right]-\varphi_1^+\right\Vert_{H^{-\theta}}\longrightarrow
   0,\;\;t\rightarrow t_+,
\end{equation}
and the map $W_+:(u_0,u_1)\mapsto (\varphi_0^+,\varphi_1^+)$ is an isomorphism
on $H^{1-\theta}(\mathbf{K})\times H^{-\theta}(\mathbf{K})$.
\label{teossgpet}
\end{Theorem}

{\it Proof.} The proof is similar to those of the case
$-1<\eta_0^+<1$ with $\xi=\frac{d-1}{4d}$ since  (\ref{vssg})
assures that
$V\in L^1(0,\tau_+)$ if $\eta_1^+>1$. We follow the proof of the
Theorem \ref{peititrip} by replacing $m^2\alpha^2(\tau)$ by $V(\tau)$.
\fin


The case of a  Sudden Singularity with $\eta_1^+\in(0,1)$, is analogous
to the Strong Big Rip ($\eta_0^+\leq-1$) with the conformal coupling
$\xi=\frac{d-1}{4d}$ since
$V(\tau)=O\left(\mid\tau-\tau_+\mid^{\gamma}\right)$, $\gamma>-2$.

\begin{Theorem}[Sudden Singularity]
We assume  that (\ref{condida}), (\ref{condidac}) and (\ref{rb}) and
hold. We suppose  $\eta_0^+=0$, $\eta_1^+\in (0,1)$.
 Given $\theta\in[0,1]$, $u_0\in H^{1-\theta}(\mathbf{K})$, $u_1\in
 H^{-\theta}(\mathbf{K})$, we consider the  unique $u\in C^0\left(I;
   H^{1-\theta}(\mathbf{K})\right)\cap C^1\left(I;
   H^{-\theta}(\mathbf{K})\right)$ solution of (\ref{kg}), (\ref{CI}). Then
 there exists  unique $\varphi_0^+\in
 H^{1-\theta}(\KK)$ such that
\begin{equation}
 \label{WEEEEK}
\left[ a(t)\right]^{\frac{d-1}{2}}u(t,.)\rightarrow\varphi_0^+\;in\;H^{1-\theta}-weak*,\;\; \left[ a(t)\right]^{\frac{d-1}{2}}u(t,.)\rightarrow\varphi_0^+\;in\;H^{-\theta}\;\;t\rightarrow t_+,
\end{equation}
\begin{equation}
 \label{TTTTTTTTT}
 (t_+-t)^{1-\eta_1^+}\partial_t\left(\left[a(t)\right]^{\frac{d-1}{2}}u\right)\in
 C^0\cap L^{\infty}([t_0,t_+);H^{-\theta}).
\end{equation}

\label{SSG}
\end{Theorem}

{\it Proof.}
If $\xi\neq\frac{d-1}{4d}$, (\ref{vssg}) assures that $V(\tau)$ has a
constant sign near $\tau_+$. If $V(\tau)\geq 0$ on some interval
$[\tau_1,\tau_+)$, we can mimic the proof of the Theorem \ref{bigrip} for
the Strong Big Rip with the conformal coupling, by replacing
$m^2\alpha^2(\tau)$ by $V(\tau)$. In contrast, if $V(\tau)\leq 0$ on some interval
$[\tau_1,\tau_+)$, we cannot apply Lemma \ref{lemricat}. Fortunately,
we can solve the Riccatti equation for a non-positive  right-hand side
and  Lemma \ref{lemricat} is replaced by the following:

\begin{Lemma}
Let $W$ be a non-negative, continuous function on $[\tau_1,\tau_+)$.
We assume that $W(\tau)=O\left(\mid\tau_+-\tau\mid^{\gamma}\right)$ with
$\gamma>-2$. Then, given $M>1$, there exists $\tau_0\in[\tau_1,\tau_+)$ such that the Ricatti
 equation
\begin{equation}
 \label{ricattm}
 A'(\tau)-A^2(\tau)=-W(\tau),\;\;\tau\in[\tau_0,\tau_+),
\end{equation}
has a unique solution such that $A(\tau_0)=0$, moreover $A$ satisfies
\begin{equation}
 \label{encadram}
 -\int_{\tau_0}^{\tau}W(\sigma)d\sigma\leq A(\tau)\leq -\left(1-\frac{1}{2M}\right)\int_{\tau_0}^{\tau}W(\sigma)d\sigma.
\end{equation}
\begin{equation}
 \label{glom}
 \int_{\tau_0}^{\tau_+}\mid A(\tau)\mid d\tau\leq\frac{1}{2M}.
\end{equation}

\end{Lemma}

{\it Proof.}
Since $W(\tau)=O\left(\mid\tau_+-\tau\mid^{\gamma}\right)$ with
$\gamma>-2$,  we can choose $\tau_0\in[\tau_1,\tau_+)$ such that
\begin{equation}
 \label{prom}
 \int_{\tau_0}^{\tau_+}\left(\int_{\tau_0}^{\tau}W(\sigma)d\sigma\right)d\tau\leq\frac{1}{2M}.
\end{equation}
We iteratively define the sequence $A_n(\tau)$ by $A_0(\tau)=0$ and
\begin{equation}
 \label{}
 A_{n+1}(\tau)=-\int_{\tau_0}^{\tau}W(\sigma)d\sigma+\int_{\tau_0}^{\tau}A_n^2(\sigma)d\sigma.
\end{equation}
We obviously have for any $n\geq 0$
\begin{equation*}
 \label{}
 -\int_{\tau_0}^{\tau}W(\sigma)d\sigma\leq A_{n}(\tau)
\end{equation*}
and we prove that for any $n\geq 1$,
\begin{equation*}
 \label{}
 A_n(\tau)\leq -\left(1-\frac{1}{2M}\right)\int_{\tau_0}^{\tau}W(\sigma)d\sigma
\end{equation*}
that follows from the inequalities :
\begin{equation*}
 \label{}
\begin{split}
  A_{n+1}(\tau)\leq&
 -\int_{\tau_0}^{\tau}W(\sigma)d\sigma+\int_{\tau_0}^{\tau}\left(\int_{\tau_0}^{\sigma}W(\rho)d\rho\right)^2d\sigma\\
\leq&
-\int_{\tau_0}^{\tau}W(\sigma)d\sigma+\left(\int_{\tau_0}^{\tau}W(\rho)d\rho\right)\left(\int_{\tau_0}^{\tau_+}\left(\int_{\tau_0}^{\sigma}W(\rho)d\rho\right)d\sigma\right)\\
\leq&-\left(1-\frac{1}{2M}\right)\int_{\tau_0}^{\tau}W(\sigma)d\sigma.
\end{split}
\end{equation*}
Now for $n\geq 1$ we evaluate
\begin{equation*}
 \label{}
 \begin{split}
\mid A_{n+1}(\tau)-A_n(\tau)\mid\leq &\sup_{\sigma\in[\tau_0,\tau]}\mid
A_n(\sigma)-A_{n-1}(\sigma)\mid\int_{\tau_0}^{\tau}\mid
A_n(\sigma)+A_{n-1}(\sigma)\mid d\sigma\\
\leq&2\sup_{\sigma\in[\tau_0,\tau]}\mid
A_n(\sigma)-A_{n-1}(\sigma)\mid
\int_{\tau_0}^{\tau}\left(\int_{\tau_0}^{\sigma}W(\rho)d\rho\right)d\sigma\\
\leq&\frac{1}{M}\sup_{\sigma\in[\tau_0,\tau]}\mid
A_n(\sigma)-A_{n-1}(\sigma)\mid.
\end{split}
\end{equation*}
We deduce that
$$
\sup_{\sigma\in[\tau_0,\tau]}\mid
A_n(\sigma)-A_{n-1}(\sigma)\mid\leq \frac{1}{M^n}\int_{\tau_0}^{\tau}W(\sigma)d\sigma.
$$
We conclude that we can define $A$ as
\begin{equation*}
 \label{}
 A(\tau):=\lim_{n\rightarrow\infty}A_n(\tau)=\sum_{n=0}^{\infty}\left(A_{n+1}(\tau)-A_n(\tau)\right),
\end{equation*}
and $A$ satisfies (\ref{ricattm}), (\ref{encadram}) and (\ref{glom}) follows from
(\ref{encadram}) and (\ref{prom}).

\fin

To end the proof of the theorem, we apply the previous Lemma with
$W(\tau)=-V(\tau)$, then we use
Lemma \ref{enerkoko} with $B^2=-\Delta_{\KK}+1$,
$G=1-\frac{d-1}{4d}R_{\gamma}$, and we achieve the proof as for the
Theorem \ref{bigrip}.

\fin


\section{At a Big Bang/Crunch/Rip (Non Conformal Coupling)}

We consider the case of a Big Crunch, $\eta_0^+> 0$, or a Big Rip $\eta_0^+<0$
and a general coupling $\xi\in\RR$,
$\xi\neq \frac{d-1}{4d}.$ Using the conformal time, we have to investigate the solutions of the linear
Klein Gordon equation (\ref{kgfifi}) with the potential $V(\tau)$ given
by (\ref{VQ}).
The estimates of the scale factor (\ref{applequn}) (\ref{appun})
and
(\ref{appgequn}) assure that: \\
- if $\eta_0^+>1$, we have
\begin{equation*}
 \label{}
 V(\tau)=\left(\xi-\frac{d-1}{4d}\right)\frac{d\eta_0^+}{(\eta_0^+-1)^2}\left(\eta_0^+(d+1)-2\right)\tau^{-2}+O(\tau^{-\frac{2\eta_0^+}{\eta_0^+-1}}),\;\;\tau\rightarrow+\infty,
\end{equation*}
- if $\eta_0^+=1$, we have
\begin{equation*}
 \label{}
 V(\tau)=(c_0^+)^2d(d-1)\left(\xi-\frac{d-1}{4d}\right)+O\left(e^{-c_0^+\tau}\right),\;\; \tau\rightarrow+\infty,
\end{equation*}
- if $\eta_0^+<1$, we have for some $\epsilon>0$
\begin{equation*}
 \label{}
 V(\tau)= \left(\xi-\frac{d-1}{4d}\right)\left[\eta_0^+(d+1)-2\right]d\frac{\eta_0^+}{1-\eta_0^+}(\tau_+-\tau)^{-2}+O\left((\tau_+-\tau)^{-2+\epsilon}\right),\;\;\tau\rightarrow\tau_+.
\end{equation*}

We begin with the case of a $C^1-$ Big Crunch ($\eta_0\geq 1$)
for which $\tau_+=+\infty$. If $\eta_0>1$, $V\in L^1([0,\infty))$
hence the situation
is analogous to those of the conformal coupling and
we compare the solutions of (\ref{kgfifi}) with those of (\ref{asssfr}).
If $\eta_0=1$, the asymptotic dynamics is defined by (\ref{asssmod}).
Like for the conformal coupling, we have to avoid the
increasing solutions by introducing a spectral cut-off. The Kato Rellich theorem  and (\ref{rb}) assure that  $-\Delta_{\mathbf
  K}+\xi R_{\gamma}$ endowed with the domain
$H^2(\mathbf{K})$ is selfadjoint on $L^2$. Given $\delta\in\RR$,
we introduce
\begin{equation}
 \label{Rpos}
 H_{\xi}^{2,\delta}(\mathbf K):=\left\{\varphi\in H^2(\mathbf K);\;\chi_{(-\infty,\delta)}\left(-\Delta_{\mathbf
  K}+\xi R_{\gamma}\right)\varphi=0\right\}
\end{equation}
where $\chi_{\Omega}$ is the characteristic function of
$\Omega\subset\RR$, and for $s\leq 2$ we define $H_{\xi}^{s,\delta}(\mathbf
K)$ as the closure of $H_{\xi}^{2,\delta}(\mathbf K)$ for the norm (\ref{normhs}).


\begin{Theorem}[$C^1-$Big Crunch]
We assume  that (\ref{condida}), (\ref{condidac}) and (\ref{rb})
hold with  $1\leq \eta_0^+$.
 Given $\xi\in\RR$, we take $\delta\in\RR$ such that
\begin{equation}
 \label{deltaxi}
 \delta>0\;\;if\;\;\eta_0^+>1,\;\;\;\delta>-(c_0^+)^2d(d-1)\left(\xi-\frac{d-1}{4d}\right)\;\;if\;\;\eta_0^+=1.
\end{equation}
For any $\theta\in[0,1]$,  $u_0\in H_{\xi}^{1-\theta,\delta}(\mathbf{K})$, $u_1\in
 H_{\xi}^{-\theta,\delta}(\mathbf{K})$, we consider the  unique $u\in C^0\left(I;
   H^{1-\theta}(\mathbf{K})\right)\cap C^1\left(I;
   H^{-\theta}(\mathbf{K})\right)$ solution of (\ref{kg}), (\ref{CI}). Then there exist unique $\varphi_0^+\in
 H_{\xi}^{1-\theta,\delta}(\KK)$,  $\varphi_1^+\in
 H_{\xi}^{-\theta,\delta}(\KK)$, such that
\begin{equation}
 \left\Vert
 \left[a(t)\right]^{\frac{d-1}{2}}u(t,.)-\varphi(\tau)\right\Vert_{H^{1-\theta}}+\left\Vert a(t)\partial_t\left[\left[a(t)\right]^{\frac{d-1}{2}}u(t,.)\right]-\partial_{\tau}\varphi(\tau)\right\Vert_{H^{-\theta}}\longrightarrow
   0,\;\;t\rightarrow t_+,
\end{equation}
where $\varphi$ is the solution of (\ref{asssfr}) if $\eta_0^+>1$ or
(\ref{asssmod}) if $\eta_0^+=1$, satisfying
$\varphi(0)=\varphi_0^+$, $\partial_{\tau}\varphi(0)=\varphi_1^+$. The map $W_+:(u_0,u_1)\mapsto (\varphi_0^+,\varphi_1^+)$ is an isomorphism
on $H^{1-\theta,\delta}_{\xi}(\mathbf{K})\times
H_{\xi}^{-\theta,\delta}(\mathbf{K})$.\\

If $\theta=\frac{1}{2}$ and $\sigma$ is given by
(\ref{charge}),  we have
\begin{equation}
 \label{siguemanc}
 \sigma(u,\tilde{u})=\left<\varphi_0^+,\tilde{\varphi}_1^+\right>-\left<\tilde{\varphi}_0^+,{\varphi}_1^+\right>
\end{equation}
where $(\tilde{\varphi}_0^+,\tilde{\varphi}_1^+)=W_+(\tilde{u}(t_0),\partial_t\tilde{u}(t_0))$.
 \label{theobbnc}
\end{Theorem}

{\it Proof.} The proof follows the same way used for the conformal
coupling. If $\eta_0^+>1$ we just replace $m^2\alpha^2(\tau)$ by
$V(\tau)$ taking advantage of its integrability. If $\eta_0^+=1$ we
replace $U_0(\tau)$ by the propagator of (\ref{asssmod}) that is
uniformly bounded on $H^{1-\theta,\delta}\times H^{-\theta,\delta}$,
and we get the existence and asymptotic completeness of $W_+$ by using
the integrability of
$V(\tau)-(c_0^+)^2d(d-1)\left(\xi-\frac{d-1}{4d}\right)$.

\fin

For a $C^0-$ Big Crunch or a Big Rip with a non conformal coupling, the
potential $V(\tau)$ has a singularity of $(\tau_+-\tau)^{-2}$ type
that forbids the existence of $W_+$.
\begin{Theorem}[$C^0-$Big Crunch, Big Rip]
 We assume  that (\ref{condida}), (\ref{condidac}) and (\ref{rb})
hold with  $ \eta_0^+\in (-\infty,0)\cup(0,1)$.
 Given $\xi\in\RR$ satisfying (\ref{condichi}), 
$\theta\in[0,1]$,  there exists $K>0$ such that for any $u_0\in H^{1-\theta}(\mathbf{K})$, $u_1\in
 H^{-\theta}(\mathbf{K})$,  the  unique solution $u\in C^0\left(I;
   H^{1-\theta}(\mathbf{K})\right)\cap C^1\left(I;
   H^{-\theta}(\mathbf{K})\right)$ of (\ref{kg}),
 (\ref{CI}) satisfies:
\begin{equation}
 \label{TOTU}
 (t_+-t)^{\frac{d-1}{2}\eta_0^+}\Vert u(t,.)\Vert_{H^{1-\theta}}\longrightarrow
   0,\;\;t\rightarrow t_+,
\end{equation}
\begin{equation}
 \label{TUTO}
 \sup_{t\in[t_0,t)}\left((t_+-t)^{\frac{d-1}{2}\eta_0^+}\Vert u(t,.)\Vert_{H^{1-\theta}} +(t_+-t)^{\eta_0^++1-\epsilon}\left\Vert \partial_t\left[\left[a(t)\right]^{\frac{d-1}{2}}u(t,.)\right]
\right\Vert_{H^{-\theta}}\right)\leq K\left(\Vert
u_0\Vert_{H^{1-\theta}}+\Vert u_1\Vert_{H^{-\theta}}\right).
\end{equation}

 \label{}
\end{Theorem}

\begin{Remark}
\label{ringostarnc}
Due to constraint (\ref{condichi}), this theorem cannot be applied to the
important case $d=3$, $\xi=0$, $\eta_0^+\in(0,1)$. This situation, with
supplementary hypotheses,  can
be treated in the general framework developped by H. Ringstr\"om for
the silent singularities of Bianchi spacetimes \cite{ringstrom2017},
\cite{ringstrom2018}. We assume that: (1) $\KK$ is a connected
3-dimensional Lie group (an elementary but interesting
example is $\KK=S^3=SU(2)$) and the
mean curvature of $\KK\times\{t\}$ is strictly positive; (2) $\eta_0^+\in(1/3,1)$ (therefore $t_+$ is a
``silent monotone volume singularity''). Then the main results of
\cite{ringstrom2017} imply that the solution $u$ of (\ref{KG})
with $m=0$, $\xi=0$ satisfies
$$
u(t)\sim
A(t_+-t)^{1-3\eta_0^+}+B,\;\;(t_+-t)^{3\eta_0^+}\partial_tu(t)\rightarrow
C,\;\;t\rightarrow t_+.
$$
These asymptotics are also obtained  by A. Alho, G. Fournodavlos and A. T. Franzen in \cite{alho} if $\KK=\RR^3$, $\eta_0^+\in(1/3,1)$.

\end {Remark}


{\it Proof.}
First we write the Klein-Gordon  equation (\ref{kgfifi}) as:
\begin{equation*}
 \label{}
 \left\{\partial_{\tau}^2+{\frak A}+\frac{q}{(\tau_+-\tau)^2}\right\}\varphi=p(\tau)\varphi,
\end{equation*}
where
\begin{equation*}
 \label{}
 q:=\left(\xi-\frac{d-1}{4d}\right)\left[\eta_0^+(d+1)-2\right]d\frac{\eta_0^+}{1-\eta_0^+}>\frac{1}{4},
\end{equation*}
\begin{equation*}
 \label{}
 p(\tau):=N+\frac{q}{(\tau_+-\tau)^2}-V(\tau),
\end{equation*}
and $N\geq 0$ is choosen such that
\begin{equation*}
 \label{}
 {\frak A}:=-\Delta_{\mathbf
     K}+\xi R_{\gamma}(\xx)+N\geq Id.
\end{equation*}
We remark that (\ref{applequn}) assures that $p\in C^0([0,\tau_+))$ satisfies for some
$\epsilon\in (0,\frac{1}{2})$ and $C_p>0$,
\begin{equation}
 \label{papy}
 \mid p(\tau)\mid\leq C_p(\tau_+-\tau)^{-2+\epsilon}.
\end{equation}
Using the spectral decomposition of $\frak{A}$, the issue of the
asymptotics of $\varphi$ will be reduced to the study of  the behaviour at $\tau_+$ of the solution of the equation
\begin{equation}
 \label{totomimi}
 \psi''(\tau)+\lambda^2\psi(\tau)+\frac{q}{(\tau_+-\tau)^2}\psi(\tau)=p(\tau)\psi(\tau),\;\;\tau\in[0,\tau_+).
\end{equation}
 
\begin{Lemma}
 \label{lemmapsipsi}
 Given $q>\frac{1}{4}$, $M>0$, there exists $K>0$ such as for any
 $\lambda\geq M$, $\psi_0,\,\psi_1\in\CC$, the solution
 $\psi\in C^0([0,\tau_+))$ of (\ref{totomimi}) with intial data
 $\psi(0)=\psi_0$, $\psi'(0)=\psi_1 $, satisfies
\begin{equation}
 \label{SUPPOZ}
 \sup_{0\leq\tau<\tau_+}\left(\lambda\mid\psi(\tau)\mid+(\tau_+-\tau)^{1-\epsilon}\mid\partial_{\tau}\psi(\tau)\mid\right)\leq K\left(\lambda\mid\psi_0\mid+\mid\psi_1\mid\right).
\end{equation}
Moreover we have
\begin{equation}
 \label{convo}
 \psi(\tau)\rightarrow 0,\;\;\tau\rightarrow\tau_+.
\end{equation}
\end{Lemma}

{\it Proof.}
Given $\lambda>0$, $q>\frac{1}{4}$, we consider the solution $\psi$ of
\begin{equation}
 \label{freeque}
 \psi''(\tau)+\lambda^2\psi(\tau)+\frac{q}{(\tau_+-\tau)^2}\psi(\tau)=0,\;\;\tau\in[0,\tau_+),\;\;\psi(0)=\psi_0,\;\;\psi'(0)=\psi_1.
\end{equation}
$\psi$ is expressed by using the Riemann functions
$R_k(\tau;\lambda)$, where $R_0$ is the solution for $\psi_0=1$,
$\psi_1=0$, and $R_1$ is the solution for $\psi_0=0$ and $\psi_1=1$,
\begin{equation*}
 \label{}
 \psi(\tau)=R_0(\tau;\lambda)\psi_0+R_1(\tau;\lambda)\psi_1.
\end{equation*}
$R_k$ can be written in terms of Bessel
functions. Writing $\psi(\tau)=(\tau_+-\tau)^{\frac{1}{2}}f(\lambda(\tau_+-\tau))$,
we can see that $\psi$ is a solution of (\ref{freeque}) iff $f(z)$
is  solution of the Bessel equation
$$
f''(z)+\frac{1}{z}f'(z)+\left(1+\frac{\mu^2}{z^2}\right)=0,\;\;\mu:=\sqrt{q-\frac{1}{4}}.
$$
Elementary and tedious computations give
\begin{equation}
 \label{Rzero}
\begin{split}
 R_0(\tau;\lambda)=
(\tau_+-\tau)^{\frac{1}{2}}\frac{\pi}{4}\tau_+^{-1}&\left[\left(2\lambda\tau_+Y'_{i\mu}(\lambda\tau_+)+Y_{i\mu}(\lambda\tau_+)\right)J_{i\mu}(\lambda(\tau_+-\tau))\right. \\
&\left.-
\left(2\lambda\tau_+J'_{i\mu}(\lambda\tau_+)+J_{i\mu}(\lambda\tau_+)\right)Y_{i\mu}(\lambda(\tau_+-\tau))\right],
\end{split}
\end{equation}
\begin{equation}
 \label{Run}
 R_1(\tau;\lambda)=
(\tau_+-\tau)^{\frac{1}{2}}\frac{\pi}{2}\tau_+^{\frac{1}{2}}\left[J_{i\mu}(\lambda\tau_+)Y_{i\mu}(\lambda(\tau_+-\tau)) -Y_{i\mu}(\lambda\tau_+)J_{i\mu}(\lambda(\tau_+-\tau))
\right].
\end{equation}
We estimate $R_k(\tau;\lambda)$ near
$\tau_+$. A characteristic time is given by
\begin{equation*}
 \label{}
 \tau_{\lambda}:=(1-M\lambda^{-1})\tau_+.
\end{equation*}
We know (see \cite{nist}) that $J_{i\mu}(x)$, $Y_{i\mu}(x)$, $xJ'_{i\mu}(x)$, $xY'_{i\mu}(x)$ are
bounded on $(0,1)$ and 
$J_{i\mu}(x)$, $J'_{i\mu}(x)$, $Y_{i\mu}(x)$, $Y'_{i\mu}(x)$ decay
as $x^{-\frac{1}{2}}$ as $x\rightarrow\infty$. Therefore,
given $M>0$, there exists $C>0$ such that for any $\lambda\geq M$ we have
 for $k=0,1,$
\begin{equation}
 \label{riper}
 \forall \tau \in[0,\tau_{\lambda}],\;\;\mid R_k(\tau;\lambda)\mid+\lambda^{-1}\mid\partial_{\tau}R_k(\tau)\mid\leq C\lambda^{-k},
\end{equation}
\begin{equation}
 \label{riperr}
 \forall \tau \in[\tau_{\lambda},\tau_+),\;\;\mid R_k(\tau;\lambda)\mid+(\tau_+-\tau)\mid\partial_{\tau}R_k(\tau)\mid\leq C\lambda^{\frac{1}{2}-k}(\tau_+-\tau)^{\frac{1}{2}},
\end{equation} 
(see \cite{delsanto} for another proof using the confluent hyperbolic functions).
The solution $\psi\in C^0([0,\tau_+))$ of (\ref{totomimi}) is given by
\begin{equation}
 \label{ecrifi}
 \psi(\tau)=R_0(\tau;\lambda)\psi_0+R_1(\tau;\lambda)\psi_1+\int_0^{\tau}K_{\lambda}(\tau,\sigma)p(\sigma)\psi(\sigma)d\sigma,
\end{equation}
where we have put
$$
K_{\lambda}(\tau,\sigma):=R_1(\tau;\lambda)R_0(\sigma;\lambda)-R_0(\tau;\lambda)R_1(\sigma;\lambda),\;\;0\leq\sigma,\tau<\tau_+.
$$
Thanks to (\ref{papy})
and  (\ref{riper}) we have for $\tau\in[0,\tau_{\lambda}]$ :
we have
\begin{equation}
 \label{crocpetit}
 0\leq\sigma\leq\tau,\;\;\left\vert K_{\lambda}(\tau,\sigma)p(\sigma)\right\vert\leq
2C^2C_p\lambda^{-1}(\tau_+-\sigma)^{-2+\epsilon}\leq 2C^2C_pM^{-1}(\tau_+-\sigma)^{-1+\epsilon},
\end{equation}
and with (\ref{riperr}) for $\tau\in[\tau_{\lambda},\tau_+)$ :
\begin{equation}
 \label{crococo}
 0\leq\sigma\leq\tau_{\lambda},\;\;\left\vert
   K_{\lambda}(\tau,\sigma)p(\sigma)\right\vert\leq
 2C^2C_p\lambda^{-\frac{1}{2}}(\tau_+-\tau)^{\frac{1}{2}}(\tau_+-\sigma)^{-2+\epsilon}\leq 2C^2C_pM^{-\frac{1}{2}}(\tau_+-\sigma)^{-1+\epsilon},
\end{equation}
\begin{equation}
 \label{crocgrand}
 \tau_{\lambda}\leq\sigma\leq\tau,\;\;\left\vert
   K_{\lambda}(\tau,\sigma)p(\sigma)\right\vert\leq2C^2C_p(\tau_+-\tau)^{\frac{1}{2}}(\tau_+-\sigma)^{-\frac{3}{2}+\epsilon}\leq 2C^2C_p(\tau_+-\sigma)^{-1+\epsilon}.
\end{equation}
We deduce that there exists $C'>0$ independent of $\lambda\geq M$ such
that for any $\tau\in[0,\tau_+)$ 
\begin{equation}
 \label{}
 \mid\psi(\tau)\mid\leq
 C'\left[\mid\psi_0\mid+\lambda^{-1}\mid\psi_1\mid+\int_0^{\tau}(\tau_+-\sigma)^{-1+\epsilon}\mid\psi(\sigma)\mid d\sigma\right].
\end{equation}
Then the Gronwall lemma implies that
\begin{equation}
 \label{estiper}
 \forall \tau\in[0,\tau_+),\;\;\mid \psi(\tau)\mid\leq C'(\mid\psi_0\mid+\lambda^{-1}\mid\psi_1\mid) \exp\left(C'\epsilon^{-1}\tau_+^{\epsilon}\right).
\end{equation}
Moreover since $R_k(\tau;\lambda)$ and
$K_{\lambda}(\tau,\sigma)$ tend to zero as $\tau\rightarrow\tau_+$,
(\ref{convo}) follows from the dominated convergence theorem applied
to (\ref{ecrifi}). To obtain the estimate for $\partial_{\tau}\psi$ we
note that
$$
0\leq\sigma\leq\tau\leq\tau_{\lambda},\;\;\mid \partial_{\tau}K(\tau,\sigma)\mid\leq 2C^2,
$$
$$
0\leq\sigma\leq\tau_{\lambda}\leq\tau<\tau_+, \;\;\mid \partial_{\tau}K(\tau,\sigma)\mid\leq 2C^2\lambda^{-\frac{1}{2}}(\tau_+-\tau)^{-\frac{1}{2}},
$$
$$
\tau_{\lambda}\leq\sigma\leq\tau<\tau_+,\;\;  \mid \partial_{\tau}K(\tau,\sigma)\mid\leq 2C^2 (\tau_+-\tau)^{-\frac{1}{2}}(\tau_+-\sigma)^{\frac{1}{2}},
$$
and we deduce from (\ref{papy}), (\ref{ecrifi})  and (\ref{estiper})
that there exists $C''>0$ independent of $\lambda\geq M$ such that
\begin{equation*}
 \label{}
 \forall \tau\in[0,\tau_+),\;\;\mid\partial_{\tau}\psi(\tau)\mid\leq C''(\lambda\mid\psi_0\mid+\mid\psi_1\mid)(\tau_+-\tau)^{-1+\epsilon}.
\end{equation*}
\fin

The spectral theorem assures the existence of a measure space $({\bf
  M},d\mu)$ with finite measure $d\mu$, an unitary operator $U$ from
$L^2(\KK)$ onto $L^2({\bf M},d\mu)$ and a real valued function
$f\in\bigcap_{p\geq 1}L^p({\bf M},d\mu)$ such that
$$
\psi\in D({\frak A})\Leftrightarrow f(.)(U\psi)(.)\in L^2({\bf
  M},d\mu),\;\;(U({\frak A}\psi))(\mu)=f(\mu)(U\psi)(\mu).
$$
Since ${\frak A}\geq Id$, we have $f\geq 1$ $a.e.$ and on $D({\frak A})$ the norms $H^s(\KK)$
are equivalent to
$$
\left(\int_{\bf M}\mid (U\psi)(\mu)\mid^2[f(\mu)]^sd\mu\right)^{\frac{1}{2}}.
$$
Therefore, for $s\geq 0$,  we may consider $U$ as an isometry from $H^s(\KK)$ onto  $L_s^2({\bf
      M})$ where $L^2_s({\bf M}):= L^2\left({\bf
      M},[f(\mu)]^sd\mu\right)$. Moreover, $U$ can be uniquely
extended into an isometry from  $H^s(\KK)$ onto  $L_s^2({\bf
      M})$ for any $s\leq 0$, denoted $U$ again.

For $\mu\in\bf M$ we introduce the solution $\psi(\tau;\mu)$ of the
equation (\ref{totomimi}) with the initial data
$\psi(0;\mu)=(U\varphi_0)(\mu)$,
$\partial_{\tau}\psi(0;\mu)=(U\varphi_1)(\mu)$ where  $(\varphi_0,\varphi_1)={\mathcal L}(u_0,u_1)$ given
by (\ref{ufifi}).
Lemma \ref{lemmapsipsi} with $\lambda=[f(\mu)]^{\frac{1}{2}}$ and $M=1$  implies that
$$
\psi\in C^0\left([0,\tau_+]; L^2_{1-\theta}({\bf
      M})\right)\cap C^1\left([0,\tau_+); L^2_{-\theta}({\bf M})\right),
$$
$$
\sup_{\tau\in[0,\tau_+)}\left(\Vert
\psi(\tau;.)\Vert_{L^2_{1-\theta}}+(\tau_+-\tau)^{1-\epsilon}\Vert\partial_{\tau}\psi(\tau;.)\Vert_{L^2_{-\theta}}\right)\leq K\left(\Vert\varphi_0\Vert_{H^{1-\theta}}+\Vert\varphi_1\Vert_{H^{-\theta}}\right),
$$
$$
\psi(\tau_+;.)=0.
$$
Then $\varphi(\tau,\xx):=(U^{-1}\psi(\tau;.))(\xx)$ belongs to
$C^0([0,\tau_+];H^{1-\theta}(\KK))\cap
  C^1([0,\tau_+);H^{-\theta}(\KK))$ and satisfies
\begin{equation*}
 \label{}
 \sup_{\tau\in[0,\tau_+)}\left(\Vert
\varphi(\tau;.)\Vert_{H^{1-\theta}}+(\tau_+-\tau)^{1-\epsilon}\Vert\partial_{\tau}\varphi(\tau;.)\Vert_{H^{-\theta}}\right)\leq K\left(\Vert\varphi_0\Vert_{H^{1-\theta}}+\Vert\varphi_1\Vert_{H^{-\theta}}\right),\;\;\varphi(\tau_+)=0.
\end{equation*}
Moreover $\varphi$ is solution of  (\ref{kgfifi}) and
  $\varphi(0,\xx)=\varphi_0(\xx)$,
  $\partial_{\tau}\varphi(0,\xx)=\varphi_1(\xx)$. Now the theorem
  follows from  the uniqueness of
  the solution of the Cauchy problem for (\ref{kg}).

\fin


\section{Quantum field}
In this section we investigate a peculiar aspect of the quantum fields  theory
in curved space-times: the particle creation  in a universe
beginning with a Big Bang ($\eta_0^->0$) and ending with a Big Crunch
($\eta_0^+>0$) or a Sudden Singularity ($\eta_0^+=0$). For the sake of
simplicity, we suppose that $\bf K$ is a compact manifold and the main
result of this part states that the number of cosmological particle
creation is finite under general assumptions. If $\bf K$ is a
homogeneous infinite space, this number obviously has to be infinite. We let
open the issue of the finitness of the local {\it density} of particle
creation if the volume of $\KK$ is infinite.\\

In the previous sections, we have proved the existence and asymptotic
completeness of the wave operators associated to the Klein-Gordon
equation (\ref{kgfifi}) if the potential $V(\tau)$ defined by
(\ref{VQ}) has finite limits
at $\tau_{\pm}$,
\begin{equation*}
 \label{}
 V_{\pm}:=\lim_{\tau\rightarrow\tau_{\pm}}V(\tau).
\end{equation*}

For us to be able to define the (anti-)particles, the asymptotic Hamiltonians
\begin{equation*}
 \label{}
 -\Delta_{\bf K}+\xi R_{\gamma}+V_{\pm}
\end{equation*}
have to be positive on a suitable subspace of $L^2(\bf{K})$. Hence we
introduce an infrared cut-off by choosing
$\delta\in\RR$ such that
\begin{equation*}
 \label{}
 \delta>-\min(V_-,V_+)
\end{equation*}
and using (\ref{Rpos}) we put
\begin{equation*}
 \label{}
 X:=H^{\frac{1}{2},\delta}_{\xi}({\bf K})\times H^{-\frac{1}{2},\delta}_{\xi}({\bf K})
\end{equation*}
that is a Hilbert space for the equivalent norms:
\begin{equation*}
 \label{}
 \Vert(\varphi_0,\varphi_1)\Vert_{\pm}:=\left(\left\Vert \left(-\Delta_{\bf K}+\xi
     R_{\gamma}+V_{\pm}\right)^{\frac{1}{4}}\varphi_0\right\Vert_{L^2(\bf K)}^2+\left\Vert \left(-\Delta_{\bf K}+\xi
     R_{\gamma}+V_{\pm}\right)^{-\frac{1}{4}}\varphi_1\right\Vert_{L^2(\bf K)}^2\right)^{\frac{1}{2}}.
\end{equation*}
Therefore the classical Scattering Operator
\begin{equation*}
 \label{}
 S:=W_+\circ W_-^{-1}
\end{equation*}
exists and it is an isomorphism on $X$
 in the following cases presented in Figures
\ref{figuescatconf} and \ref{figuescatnonconf}.


\begin{figure}[h!]
$$
\begin{array}{|c|c|c|c|}
\hline
\rm Singularity&\eta_0^{\pm}&\tau_{\pm}&V_{\pm}\\
\hline
\hline
C^0-Big\;Bang&0<\eta_0^-<1&-\infty<\tau_-&V_-=0\\
\hline
C^1-Big\;Bang&1\leq \eta_0^-&\tau_-=-\infty&V_-=0\\
\hline
\hline
Sudden\;Singularity&\eta_0^+=0&\tau_+<\infty&V_+=m^2(c_0^+)^2\\
\hline
C^0-Big\;Crunch&0<\eta_0^+<1&\tau_+<\infty&V_+=0\\
\hline
C^1-Big\;Crunch&1\leq\eta_0^+&\tau_+=\infty&V_+=0\\
\hline
\end{array}
$$
\caption{Existence of the Scattering Operator for the conformal
  coupling $\xi=\frac{d-1}{4d}$.}
\label{figuescatconf}
\end{figure}



\begin{figure}[h!]
$$
\begin{array}{|c|c|c|c|c|}
\hline
\rm Singularity&\eta_0^{\pm}&\eta_1^+&\tau_{\pm}&V_{\pm}\\
\hline
\hline
C^1-Big\;Bang&\eta_0^-=1&(\eta_0^-,\infty)&\tau_-=-\infty&V_-=(c_0^-)^2d(d-1)\left(\xi-\frac{d-1}{4d}\right)\\
\hline
C^1-Big\;Bang&\eta_0^->1&(\eta_0^-,\infty)&\tau_-=-\infty&V_-=0\\
\hline
\hline
C^1-Big\;Brake&\eta_0^+=0&\eta_1^+>2&\tau_+<\infty&V_+=m^2(c_0^+)^2\\
\hline
C^1-Big\;Crunch&\eta_0^+=1&(\eta_0^+,\infty)&\tau_+=\infty&V_+=(c_0^+)^2d(d-1)\left(\xi-\frac{d-1}{4d}\right)\\
\hline
C^1-Big\;Crunch&\eta_0^+>1&(\eta_0^+,\infty)&\tau_+=\infty&V_+=0\\
\hline
\end{array}
$$
\caption{Existence of the Scattering Operator for the non conformal
  coupling $\xi\neq\frac{d-1}{4d}$.}
\label{figuescatnonconf}
\end{figure}


To quantize the scattering operator we must introduce the one-particle
space $X^{\pm}_{pos}$ and the one-antiparticle $X^{\pm}_{neg}$ that  are
defined as
\begin{equation}
 \label{posneg}
 X^{\pm}_{pos(neg)}:=\left\{(\varphi_0,\varphi_1)\in
   X;\;\;\varphi_1=+(-)i\left(-\Delta_{\bf K}+\xi R_{\gamma}+V_{\pm}\right)^{\frac{1}{2}}\varphi_0\right\}.
\end{equation}
We denote $P_{pos(neg)}^{\pm}$ the projection on $X_{pos(neg)}^{\pm}$
along $X_{neg(pos)}^{\pm}$:
\begin{equation}
 \label{ppPROJJPO}
 P_{pos(neg)}^{\pm}=\frac{1}{2}
\left(
\begin{array}{cc}
Id&-(+)i\left(-\Delta_{\bf K}+\xi R_{\gamma}+V_{\pm}\right)^{-\frac{1}{2}}\\
+(-)i\left(-\Delta_{\bf K}+\xi R_{\gamma}+V_{\pm}\right)^{\frac{1}{2}}&Id
\end{array}
\right).
\end{equation}


A classic result of the
second quantization due to Shale \cite{shale62} (see so Theorem XI.108 in
\cite{reed-simon3}) assures that $S$ is a
Bogolioubov transform that is unitarily implementable on the Fock-Cook
spaces over $X$, iff its off diagonal parts defining the
particle-antiparticle mixing, 
\begin{equation}
 \label{offdiag}
P_{neg(pos)}^+SP_{pos(neg)}^-
\end{equation}
are Hilbert-Schmidt. The physical meaning of this result is that a {\it finite}
number $\mathcal{N}$ of creation of particle-antiparticle pairs occurs at
$\tau_+$ if the initial quantum state is the Fock vacuum at
$\tau_-$. This number is given by
\begin{equation}
 \label{N}
 \mathcal{N}=\left\Vert P_{neg(pos)}^+SP_{pos(neg)}^-\right\Vert^2_{HS}
\end{equation}
where $\Vert.\Vert_{HS}$ denotes the Hilbert-Schmidt norm (for
explanations of the meaning of this norm in terms of particles, see
{\it e.g.}
\cite{fulling}, \cite{fulling-79}, \cite{parker}, \cite{reed-simon3}).

Adopting the approach introduced by Fulling \cite{fulling-79}, a  key
ingredient to prove this property  is the Liouville-Green (WKB) approximation that needs a
sufficient regularity of the potential. In particular to apply the
famous theorems of Olver \cite{olver}, it is necessary to have
$V''\in L^1(\tau_-,\tau_+)$. Hence some supplementary constraints on
$\eta_0^{\pm}$ and $\eta_1^{\pm}$ will be necessary. Moreover the
Hilbert-Schmidt property of (\ref{offdiag}) is closely linked to the
convergence of the series of the Zeta function $\zeta(s)$ of the Laplacian
$\Delta_{\bf K}$, that is converging iff $\Re s>d/2$. Therefore the
constraint of regularity of the scale factor is increasing with the
dimension $d$ of $\bf{K}$. Hence we introduce:
\begin{equation}
 \label{M(D)}
 l(d):=\max(2,[d/2]),\;\;[d/2]\in\NN,\;\;[d/2]\leq d/2<[d/2]+1.
\end{equation}
The following theorem is established by proving that
\begin{equation}
 \label{NN}
 \mathcal{N}\lesssim \zeta(-l(d)-1):=\sum_n\lambda_n^{-l(d)-1}
\end{equation}
where $\lambda_n$ are the positive eigenvalues of 
$-\Delta_{\bf K}$. In some cases, the WKB approximation (more
precisely, Theorem 4 in \cite{olver}) allows to only obtain
\begin{equation}
 \label{NNweak}
 \mathcal{N}\lesssim \zeta(-3)=\sum_n\lambda_n^{-3}.
\end{equation}
Therefore the constraint of convergence, $3>d/2$, restricts these
results to $d\leq 5$.

\begin{Theorem}

The operators (\ref{offdiag})
are Hilbert-Schmidt on $X$ if we assume one of the following
hypotheses in which $a\in
C^l(t_-,t_+)$ satisfies (\ref{condida}) for any $k\leq l.$

\begin{itemize}
\item
A universe beginning with a Big Bang and ending with a Big Crunch:

\begin{equation}
 \label{}
 \xi=\frac{d-1}{4d},\;\; \frac{l-1}{l+1}\leq\eta_0^-,\eta_0^+,\;\;d\geq 3,\;\;l =l(d),
\end{equation}

\begin{equation}
 \label{nccbcc1c1}
 \xi\neq\frac{d-1}{4d},\;\; 1<
 \eta_0^-,\eta_0^+,\;\;d\geq 3, \;\; l=l(d)+2,
\end{equation}

\begin{equation}
 \label{nccbcc1c-111}
 \xi\neq\frac{d-1}{4d},\;\; 
 \eta_0^-=\eta_0^+=1,\;\;c_0^-=c_0^+,\;\;d\geq 3, \;\;l=l(d)+2,
\end{equation}

\begin{equation}
 \label{nccbcc1c1-1}
 \xi\neq\frac{d-1}{4d},\;\; 1=\min\left(
   \eta_0^-,\eta_0^+\right),\;\;d\leq 5,\;\;l=4,
\end{equation}

\item
A universe beginning with a Big Bang and ending with a Big Brake:
\begin{equation}
 \label{ccbb}
 \xi=\frac{d-1}{4d},\;\;\eta_0^-\in(1/3,\infty),\;\;\eta_0^+=0,\;\;\eta_1^+>1,\;\;d\leq 5,\;\;l=2,
\end{equation}

\begin{equation}
 \label{KOUCIKOUCA}
 \xi\neq\frac{d-1}{4d},\;\; 1\leq \eta_0^-,\;\;\eta_0^+=0,\;\;\eta_1^+>3,\;\;d\leq 5,\;\;l=4.
\end{equation}

\end{itemize}
 \label{}
\end{Theorem}


{\it Proof of the Theorem.}
We shall use the notation $A(\lambda)\sim B(\lambda)$ where $\lambda$
is an asymptotic parameter tending to $\infty$ to mean
$A(\lambda)=B(\lambda)+o(B(\lambda))$ as $\lambda\rightarrow\infty$.
Since $\mathbf K$ is a compact manifold, $-\Delta_{\bf K}+\xi
R_{\gamma}$ has compact resolvent. We denote $\left(\mu_n\right)_{n\in\NN}$ the eigenvalues
of this operator and we assume that $\mu_n\leq\mu_{n+1}$. We introduce
\begin{equation*}
 \label{}
 N:=\min\{n\in\NN;\;\delta<\mu_n\}.
\end{equation*}
We introduce
a Hilbert basis
$\left(\phi_n(\bf x)\right)_{n\in\NN}$ of $L^2(\bf K)$, composed of eigenfunctions, $-\Delta_{\bf K}\phi_n+\xi
R_{\gamma}\phi_n=\mu_n\phi_n$.
Then the vectors
$\Phi_{pos(neg),n}^{\pm}:=2^{-\frac{1}{2}}\left(\left(\mu_n+V_{\pm}\right)^{-\frac{1}{4}}\phi_n,+(-)i
  \left(\mu_n+V_{\pm}\right)^{\frac{1}{4}}\phi_n\right)$, $n\geq N$, form a Hilbert
  basis $\mathcal{B}_{pos(neg)}^{\pm}$ of
  $\left(X^{\pm}_{pos(neg)},\Vert.\Vert_{\pm}\right)$, and
  $\mathcal{B}_{pos}^{\pm}\cup \mathcal{B}_{neg}^{\pm}$ is a Hilbert
  basis of $\left(X,\Vert.\Vert_{\pm}\right)$. To establish the
  theorem, we have to prove that
\begin{equation}
 \label{zesom}
 \sum_{n=N}^{\infty}\left\Vert P^+_{neg(pos)}S\Phi_{pos(neg),n}^-\right\Vert_+^2<\infty.
\end{equation}
In the sequel, we consider only $P^+_{neg}SP^-_{pos}$, the case of
$P^+_{pos}SP^-_{neg}$ being analogous. We denote $\varphi_n(\tau,\bf{x})$ the solution of (\ref{kgfifi})
with initial data
\begin{equation*}
 \label{}
 (\varphi_n(0,.),\partial_{\tau}\varphi_n(0,.))=W_-^{-1}\Phi_{pos,n}^-,
\end{equation*}
that is to say, either
\begin{equation*}
 \label{}
 (\varphi(\tau_-,{\bf x}),\partial_{\tau}\varphi_n(\tau_-,{\bf
   x}))=\Phi_{pos,n}^-({\bf x})\;\;if\;\;\tau_->-\infty,
\end{equation*}
or
\begin{equation*}
 \label{}
 \lim_{\tau\rightarrow\tau_-}\left\Vert
 (\varphi_n(\tau),\partial_{\tau}\varphi_n(\tau))-e^{+i\tau\sqrt{\mu_n+V_-}}\Phi_{pos,n}^-\right\Vert_X=0\;\;if\;\;\tau_-=-\infty.
\end{equation*}
Since the perturbation $V$ does not depend on ${\bf x}$, the dynamics
of (\ref{kgfifi}) does not mix the modes $\phi_n$. Then
$\varphi_n(\tau,{\bf x})$ has the form
\begin{equation*}
 \label{}
 \varphi_n(\tau,{\bf x})=\psi_n(\tau)\phi_n({\bf x}),
\end{equation*}
where $\psi_n$ is solution of the differential equation
\begin{equation}
 \label{zedo}
 \psi''+\mu_n\psi+V(\tau)\psi=0,\;\;\tau\in(\tau_-,\tau_+),
\end{equation}
and satisfies the incoming data
\begin{equation}
 \label{damoins}
 if\;\tau_->-\infty,\;\;\psi_n(\tau_-)=2^{-\frac{1}{2}}\left(\mu_n+V_-\right)^{-\frac{1}{4}},\;\; \psi'_n(\tau_-)=2^{-\frac{1}{2}}i\left(\mu_n+V_-\right)^{\frac{1}{4}},
\end{equation}
\begin{equation}
 \label{DAMOINSINF}
 if\;\tau_-=-\infty,\;\;\psi_n(\tau)\sim 2^{-\frac{1}{2}}\left(\mu_n+V_-\right)^{-\frac{1}{4}}e^{i\tau\sqrt{\mu_n+V_-}},\;\tau\rightarrow-\infty.
\end{equation}
We put
\begin{equation*}
 \label{}
 S\Phi^-_{pos,n}=(\varphi_{0,n}^+,\varphi^+_{1,n})=(\lambda_{0,n}\phi_n,\lambda_{1,n}\phi_n)=\alpha_n\Phi^+_{pos,n}+\beta_n\Phi^+_{neg,n},\;\;\lambda_{j,n},\alpha_n,\beta_n\in\CC.
\end{equation*}
Then (\ref{zesom}) is equivalent to
\begin{equation}
 \label{sombeta}
 \sum_{n=N}^{\infty}\mid\beta_n\mid^2<\infty.
\end{equation}
On the one hand we have
\begin{equation*}
 \label{}
\alpha_n=\frac{1}{2}\left(\lambda_{0,n}+i(\mu_n+V_+)^{-\frac{1}{2}}\lambda_{1,n}\right),\;\;
 \beta_n=\frac{1}{2}\left(-i(\mu_n+V_+)^{\frac{1}{2}}\lambda_{0,n}+\lambda_{1,n}\right),
\end{equation*}
and on the other hand
\begin{equation*}
 \label{}
 if\;\tau_+<\infty,\;\;\psi_n(\tau_+)=\lambda_{0,n},\;\;\psi'_n(\tau_+)=\lambda_{1,n},
\end{equation*}
\begin{equation*}
 \label{}
 if\;\tau_+=\infty,\;\;\psi_n(\tau)\sim 2^{-\frac{1}{2}}(\mu_n+V_+)^{-\frac{1}{4}}\left(\alpha_ne^{i\tau\sqrt{\mu_n+V_+}}+\beta_ne^{-i\tau\sqrt{\mu_n+V_+}}\right),\;\tau\rightarrow\infty.
\end{equation*}
Now the strategy of the proof consists in proving  the following key estimate
\begin{equation}
 \label{keybeta}
 \beta_n=O\left(\mu_n^{-\frac{s}{2}}\right),\;\;s>d/2.
\end{equation}
Therefore to have (\ref{sombeta}) it is sufficient to establish that
for some $N_0\geq N$
\begin{equation}
 \label{somu}
 \sum_{n=N_0}^{\infty}\mu_n^{-s}<\infty.
\end{equation}
We take $M=\mid\xi\mid\Vert R_{\gamma}\Vert_{L^{\infty}}+1$. Then
(\ref{somu}) means that $\left(-\Delta_{\bf K}+\xi
  R_{\gamma}+M\right)^{-1}$ is a $s$th Shatten class operator. We write $\left(-\Delta_{\bf K}+\xi
  R_{\gamma}+M\right)^{-1}=\left(Id+(-\Delta_{\bf K}+M)^{-1}\xi
  R_{\gamma}\right)^{-1}(-\Delta_{\bf K}+M)^{-1}$. Since the space of
the $s$th Shatten class  operators is an ideal of the bounded operators, it is
sufficient to prove that $(-\Delta_{\bf K}+M)^{-1}$ is a $s$th
Shatten class  operator. This property is equivalent to
$$
\zeta(s)<\infty,\;\;\zeta(s):=\sum_{\lambda_n\in\sigma(-\Delta_{\bf K})\setminus\{0\}}\lambda_n^{-s}.
$$
We conclude by invoking a famous result of Minakshisundaram and
Pleijel on the Zeta Function of a Laplacian on a $d$-dimensional compact manifold (see
{\it e.g.} Theorem 5.2 in \cite{rosenberg}) that states that the series
  $\zeta(s)$ converges if $s>\frac{d}{2}$. In the sequel, we establish
  (\ref{keybeta}) with $s=l(d)+1$, by distinguishing the various cases according to
  $\tau_{\pm}$ is finite or not, and $V_-=V_+$ or not.\\

(1) We begin with the case $\tau_-=-\infty$, $\tau_+=\infty$, $V_-=V_+$.  First,
we consider the assumptions assuring that $V_-=V_+=0$, hence we assume
that $\xi=\frac{d-1}{4d}$ and $\eta_0^{\pm}\geq 1$, or
$\xi\neq\frac{d-1}{4d}$ and $\eta_0^{\pm}> 1$.
Then $V\in
C^{l(d)}(\RR)$, $V_-=V_+=0$ and $\frac{d^k}{d\tau^k}V\in L^1\cap L^{\infty}$ for
$k\leq l(d)$. Since $V$ is integrable and $\mu_n\neq 0$,  the integral equation
$$
\psi(\tau)=e^{i\sqrt{\mu_n}\tau}+\int_{-\infty}^{\tau}\frac{\sin(\sqrt{\mu_n}(\tau-\sigma))}{\sqrt{\mu_n}}V(\sigma)\psi(\sigma)d\sigma
$$
has a unique solution $\psi=\sqrt{2}\mu_n^{\frac{1}{4}}\psi_n$ that satisfies
\begin{equation*}
 \label{}
 \psi(\tau)\sim \alpha_ne^{i\sqrt{\mu_n}\tau}+\beta_ne^{-i\sqrt{\mu_n}\tau},\;\;\tau\rightarrow+\infty.
\end{equation*}
We approximate $\psi$ with a Liouville-Green function.
Following \cite{olver} we define recursively functions $A_k(\tau)$,
$0\leq k\leq l(d)+1,$ by the relations
$$
A_0=1,\;\;A_{k+1}(\tau)=-\frac{1}{2}A'_k(\tau)-\frac{1}{2}\int_0^{\tau}V(\sigma)A_k(\sigma)d\sigma.
$$
Then
Theorem 6 of \cite{olver} assures that for all $n\geq N_0$, $N_0$  large enough,
there exists a solution $w(\mu_n;.)$ of (\ref{zedo}) satisfying 
\begin{equation}
 \label{wkkb}
 w(\mu_n;\tau)=e^{i\sqrt{\mu_n}\tau}\sum_{k=0}^{l(d)}\frac{A_k(\tau)}{(i\sqrt{\mu_n})^k}+\epsilon(\mu_n;\tau),
\end{equation}
 with
\begin{equation}
 \label{epchout}
 \forall n\geq N_0,\;\;\forall \tau\in\RR,\;\;\mid\epsilon(\mu_n;\tau)\mid
 +\mu_n^{\frac{1}{2}}\mid\partial_{\tau}\epsilon(\mu_n;\tau)\mid\leq
 C\mu_n^{-\frac{l(d)+1}{2}}\int_0^{\infty}\mid A'_{l(d)+1}(\sigma)\mid d\sigma.
\end{equation}
We easily prove by
recurrence that for $1\leq p\leq l(d)+1$
\begin{equation}
 \label{}
 A'_p(\tau)=\sum_{k=0}^{p-1}\left(-\frac{1}{2}\right)^{k+1}\frac{d^k}{d\tau^k}\left(VA_{p-1-k}\right)(\tau),
\end{equation}
and it follows by recurrence that
\begin{equation}
 \label{AAAA}
 k\leq l(d)+1,\;\;A_k\in C^{l(d)+2-k}(\RR),\;\;1\leq p\leq
 l(d)+2-k,\;\;\frac{d^pA_k}{d\tau^p}\in L^1\cap L^{\infty}(\RR),
\end{equation}
In particular
$A'_{l(d)+1}\in L^1(\RR)$, hence (\ref{epchout}) implies
\begin{equation}
 \label{epchon}
\forall n>N_0,\;\forall \tau\in\RR,\;\mid\epsilon(\mu_n;\tau)\mid\leq C\mu_n^{-\frac{l(d)+1}{2}},
\end{equation}
We also deduce that for any
$k\leq l(d)$, $A_k(\tau)$ has finite limits as $\tau\rightarrow\pm\infty$.
Moreover, since $V\in L^1$ there exists
$\alpha^{\pm}(\mu_n),\beta^{\pm}(\mu_n)\in\CC$ such that
\begin{equation*}
 \label{}
 w(\mu_n;\tau)\sim \alpha^{\pm}(\mu_n)e^{i\sqrt{\mu_n}\tau}+\beta^{\pm}(\mu_n)e^{-i\sqrt{\mu_n}\tau},\;\;\tau\rightarrow\pm\infty.
\end{equation*}
We deduce from (\ref{epchon}) that for $n\geq N_0$,
\begin{equation}
 \label{betau}
 \mid \beta^{\pm}(\mu_n)\mid\leq C\mu_n^{-\frac{l(d)+1}{2}}.
\end{equation}
On the other hand we have
\begin{equation*}
 w(\mu_n;\tau)=\cos\left(\tau\sqrt{\mu_n}\right)w(\mu_n;0)+\frac{\sin\left(\tau\sqrt{\mu_n}\right)}{\sqrt{\mu_n}}\partial_{\tau}w(\mu_n;0)+\int_0^{\tau}\frac{\sin\left((\tau-\sigma)\sqrt{\mu_n}\right)}{\sqrt{\mu_n}}V(\sigma)w(\mu_n;\sigma)d\sigma,
 \end{equation*}
hence we obtain
\begin{equation*}
 \label{}
 \alpha^{\pm}(\mu_n)=\frac{1}{2}\left(w(\mu_n;0)+\frac{\partial_{\tau}w(\mu_n;0)}{i\sqrt{\mu_n}}-\frac{1}{2i\sqrt{\mu_n}}\int_0^{\pm\infty}e^{-i\sqrt{\mu_n}\sigma}V(\sigma)w(\mu_n;\sigma)d\sigma\right).
\end{equation*}
Since $A_0=1$ we deduce from (\ref{wkkb})  and (\ref{epchon}) that
\begin{equation}
 \label{}
 \alpha^{\pm}(\mu_n)=1+O\left(\mu_n^{-\frac{1}{2}}\right).
\end{equation}
Expressing $\psi$ on the basis $\{w(\mu_n;.),w^*(\mu_n;.)\}$ we obtain
\begin{equation*}
 \label{betaenne}
 \beta_n=\frac{\alpha^{-*}(\mu_n)\beta^+(\mu_n)-\alpha^{+*}(\mu_n)\beta^-(\mu_n)}{\mid\alpha^-(\mu_n)\mid^2-\mid\beta^-(\mu_n)\mid^2},
\end{equation*}
hence we deduce that
\begin{equation*}
 \label{}
 \beta_n=O\left(\mu_n^{-\frac{l(d)+1}{2}}\right),
\end{equation*}
and we conclude that (\ref{keybeta}) holds with $s=l(d)+1$. The
case (\ref{nccbcc1c-111}) is similar since $V_-=V_+$. We can employ
the same method replacing $\mu_n$ by $\mu_n+V_{\pm}$ and $V(\tau)$ by
$V(\tau)-V_{\pm}$.\\

(2) Now we consider the case $\tau_-=-\infty$, $\tau_+=\infty$ and possibly
$V_-\neq V_+$, {\it i.e.} we assume (\ref{nccbcc1c1-1}).
 If $\eta_0^-=\eta_0^+=1$
(and $c_0^-\neq c_0^+$), we have $\int_0^{\pm}\mid
V(\tau)-V_{\pm}\mid(1+\mid\tau\mid)d\tau<\infty$ and we can invoke the
Lemma 1.4 (iv) of \cite{cohen} that implies that (\ref{keybeta}) holds
with $s=3$. If one $\eta_0^{\pm}>1$, $V(\tau)-V_{\pm}$ decays
as $\tau^{-1}$ and we cannot applied
the known results. We consider an integer $N_0\geq N$ such that
\begin{equation*}
 \label{}
 \forall \tau\in(\tau_-,\tau_+),\;\;\mu_{N_0}>0,\;\;\mu_{N_0}+V(\tau)\geq 1.
\end{equation*}
For any $n\geq N_0$ we take the Liouville-Green solution of (\ref{zedo}) introduced
by Olver
(Theorem 4 in \cite{olver})  that satisfies
\begin{equation}
 \label{wolver}
 w(\mu_n;\tau)=2^{-\frac{1}{2}}\left(\mu_n+V(\tau)\right)^{-\frac{1}{4}}\exp\left(i\int_0^{\tau}\left(\mu_n+V(\sigma)\right)^{\frac{1}{2}}d\sigma\right)(1+\epsilon_n(\tau)),
\end{equation}
\begin{equation*}
 \label{}
 \mid\epsilon_n(\tau)\mid,\;\mu_n^{-\frac{1}{2}}\mid
 \epsilon_n'(\tau)\mid\leq \exp\left(\left\vert\int_0^{\tau}(\mu_n+V(\sigma))^{-\frac{1}{4}}\left\vert \frac{d^2}{d\sigma^2}(\mu_n+V(\sigma))^{-\frac{1}{4}} \right\vert d\sigma\right\vert\right)-1.
\end{equation*}
Since $V'\in L^2(\tau_-,\tau_+)$,
$V''\in L^1(\tau_-,\tau_+)$ we get
\begin{equation}
 \label{epstrois}
 \sup_{\tau\in(\tau_-,\tau_+)}\left(\mid\epsilon_n(\tau)\mid+\mu_n^{-\frac{1}{2}}\mid
 \epsilon_n'(\tau)\mid\right)\leq C\mu_n^{-\frac{3}{2}},
\end{equation}
As a consequence of $\int_0^{\tau_{\pm}}\mid
V(\tau)-V_{\pm}\mid d\tau<\infty$, there exists $\alpha^{\pm}(\mu_n),
\beta^{\pm}(\mu_n)\in\CC$ such that
\begin{equation*}
 \label{}
 w(\mu_n;\tau)\sim
 2^{-\frac{1}{2}}(\mu_n+V_{\pm})^{-\frac{1}{4}}\left(\alpha^{\pm}(\mu_n)e^{i\tau\sqrt{\mu_n+V_{\pm}}}+
\beta^{\pm}(\mu_n)e^{-i\tau\sqrt{\mu_n+V_{\pm}}}\right),\;\;\tau\rightarrow\pm\infty.
\end{equation*}
Using the fact that
$$
\alpha^{\pm}(\mu_n)=2^{\frac{1}{2}}(\mu_n+V_{\pm})^{\frac{1}{4}}\lim_{\tau\rightarrow\pm\infty}\frac{1}{\tau}\int_0^{\tau}e^{- i\sigma\sqrt{\mu_n+V_{\pm}}}w(\mu_n;\sigma)d\sigma,
$$
$$
\beta^{\pm}(\mu_n)=2^{\frac{1}{2}}(\mu_n+V_{\pm})^{\frac{1}{4}}\lim_{\tau\rightarrow\pm\infty}\frac{1}{\tau}\int_0^{\tau}e^{ i\sigma\sqrt{\mu_n+V_{\pm}}}w(\mu_n;\sigma)d\sigma,
$$
we get
\begin{equation}
 \label{alfar}
 \alpha^{\pm}(\mu_n)=\exp\left[i\int_0^{\pm\infty}\frac{V(\tau)-V_{\pm}}{(\mu_n+V(\tau))^{\frac{1}{2}}+(\mu_n+V_{\pm})^{\frac{1}{2}}}d\tau\right]+O\left(\mu_n^{-\frac{3}{2}}\right),
\end{equation}
\begin{equation}
 \label{betar}
 \beta^{\pm}(\mu_n)=O\left(\mu_n^{-\frac{3}{2}}\right).
\end{equation}
Now we express $\psi_n$ on the basis $\{w(\mu_n;.),w^*(\mu_n;.)\}$ by
putting $\psi_n(\tau)=Aw(\mu_n;\tau)+Bw^*(\mu_n;\tau)$,
$A,B\in\CC$. We have
$$
1=A\alpha^-(\mu_n)+B\beta^{-*}(\mu_n),\;\;0=A\beta^-(\mu_n)+B\alpha^{-*}(\mu_n),
$$
$$
\alpha_n=A\alpha^+(\mu_n)+B\beta^{+*}(\mu_n),\;\;\beta_n=A\beta^+(\mu_n)+B\alpha^{+*}(\mu_n),
$$
and $\beta_n$ is given by (\ref{betaenne}) again. Hence we obtain from
(\ref{alfar}) and (\ref{betar}) that (\ref{keybeta}) holds with
$s=3$.\\

(3) Now we consider the cases for which $\tau_+$ and
$\tau_-$ are finite and $V_-\neq V_+$. From the assumption
(\ref{ccbb}) with $\eta_0^-\in(1/3,1)$, we have $V\in
C^2(\tau_-,\tau_+)\cap C^0([\tau_-,\tau_+])$, $V'\in
L^2(\tau_-,\tau_+)$, $V''\in L^1(\tau_-,\tau_+)$.
We denote $\psi_+$ the solution of (\ref{zedo}) with data
\begin{equation}
 \label{daplouch}
 \psi_+(\tau_+)=2^{-\frac{1}{2}}\left(\mu_n+V_+\right)^{-\frac{1}{4}},\;\; \psi_+'(\tau_+)=2^{-\frac{1}{2}}i\left(\mu_n+V_+\right)^{\frac{1}{4}}.
\end{equation}
$\psi_+$ and $\psi_+^*$ are linearly independent and we have
\begin{equation*}
 \label{}
 \psi_n(\tau)=\alpha_n\psi_+(\tau)+\beta_n\psi_+^*(\tau).
\end{equation*}
Now we
choose the two complex-conjugate linearly independent  solutions
$w,w^*$ defined by (\ref{wolver}) and we
express $\psi_n$ and $\psi_+$ as:
\begin{equation}
 \label{psinw}
 \psi_n(\tau)=Aw(\tau)+Bw^*(\tau),\;\;
w(\tau)=\alpha\psi_+(\tau)+\beta\psi_+^*(\tau),\;\;A,B,\alpha,\beta\in\CC.
\end{equation}
We deduce that
\begin{equation}
 \label{alphanbetan}
 \alpha_n=A\alpha+B\beta^*,\;\;\beta_n=A\beta+B\alpha^*.
\end{equation}
Matching with the data (\ref{damoins}) and (\ref{daplouch}), we have
$$
A=2^{-\frac{1}{2}}(\mu_n+V_-)^{-\frac{1}{4}}\frac{\left(w'(\tau_-)+i(\mu_n+V_-)^{\frac{1}{2}}w(\tau_-)\right)^*}{[w,w^*]},
$$
$$
B=-2^{-\frac{1}{2}}(\mu_n+V_-)^{-\frac{1}{4}}\frac{w'(\tau_-)-i(\mu_n+V_-)^{\frac{1}{2}}w(\tau_-)}{[w,w^*]},
$$
$$
\alpha=-2^{-\frac{1}{2}}i(\mu_n+V_+)^{-\frac{1}{4}}\left(w'(\tau_+)+i(\mu_n+V_+)^{\frac{1}{2}}w(\tau_+)\right),
$$
$$
\beta=2^{-\frac{1}{2}}i(\mu_n+V_+)^{-\frac{1}{4}}\left(w'(\tau_+)-i(\mu_n+V_+)^{\frac{1}{2}}w(\tau_+\right),
$$
where $[f,g]$ is the Wronskian, $[f,g]:=fg'-f'g$.
Since $V'\in L^2(\tau_-,\tau_+)$,
$V''\in L^1(\tau_-,\tau_+)$ we have (\ref{epstrois}) again,
hence $[w,w^*]\sim 2i$, $n\rightarrow\infty$, and $A,\alpha=O(1)$ and
$B,\beta=O(\mu_n^{-\frac{3}{2}})$. Therefore  we obtain the key estimate (\ref{keybeta})  with $s=3$. We conclude that
(\ref{sombeta}) holds if $d\leq 5$.\\

(4)  Now we consider the cases for which $\tau_+$ and
$\tau_-$ are finite and $V_-= V_+=0$, {\it i.e.} $\xi=\frac{d-1}{4d}$,
$\eta_0^{\pm}\in\left(\frac{l-1}{l+1},1\right)$.
We apply the method of the previous case (3) but we choose another
Liouville-Green function $w$ by taking (\ref{wkkb}). We remark that
$\frac{d^kV}{d\tau^k}(\tau_{\pm})=0$ for $0\leq k\leq l(d)-1$ and 
$\frac{d^{l(d)}V}{d\tau^{l(d)}}\in L^1(\tau_-,\tau_+)$. We deduce that
$A'_k(\tau_{\pm})=0$ for $0\leq k\leq l(d)$ and $A'_{l(d)+1}\in
L^1(\tau_-,\tau_+)$. Then we have
\begin{equation*}
 \label{}
 w'(\tau_{\pm})-i\sqrt{\mu_n}w(\tau_{\pm})=O\left(\mu_n^{-\frac{l(d)}{2}}\right),
\end{equation*}
and we obtain $[w,w^*]\sim 2i\sqrt{\mu_n}$, $n\rightarrow\infty$,
$A=O\left(\mu_n^{-\frac{1}{4}}\right)$,
$\alpha=O\left(\mu_n^{\frac{1}{4}}\right)$,
$B=O\left(\mu_n^{-\frac{1}{4}-\frac{l(d)+1}{2}}\right)$,
$\beta=O\left(\mu_n^{-\frac{1}{4}-\frac{l(d)}{2}}\right)$, therefore
(\ref{keybeta}) holds with $s=\frac{l(d)+1}{2}$.\\

(5) We consider the case $\tau_-=-\infty$, $\tau_+<\infty$,
$V_-=V_+=0$, that occurs if $\xi=\frac{d-1}{4d}$, $ \eta_0^-\geq 1$,
$\eta_0^+\in\left(\frac{l-1}{l+1},1\right)$.
We have
$\frac{d^kV}{d\tau^k}(\tau_+)=0$ for $0\leq k\leq l(d)-1$ and 
$\frac{d^{l(d)}V}{d\tau^{l(d)}}\in L^1(-\infty,\tau_+)$. We deduce that
$A'_k(\tau_+)=0$ for $0\leq k\leq l(d)$ and $A'_{l(d)+1}\in
L^1(-\infty,\tau_+)$. We employ the Liouville-Green function
(\ref{wkkb}) that satisfies
\begin{equation*}
 \label{}
 w'(\tau_+)+i\sqrt{\mu_n}w(\tau_+)=O\left(\mu_n^{-\frac{l(d)}{2}}\right),
\end{equation*}
and $[w,w^*]\sim 2i\sqrt{\mu_n}$, $n\rightarrow\infty$. There exist
$\alpha^-(\mu_n),\beta^-(\mu_n)\in\CC$ such that
$$
w(\mu_n;\tau)\sim \alpha^-(\mu_n)e^{i\sqrt{\mu_n}\tau}+\beta^-(\mu_n) e^{-i\sqrt{\mu_n}\tau},\;\;\tau\rightarrow-\infty,
$$
and like in the case (1), they satisfy
$$
\alpha^-(\mu_n)=1+O\left(\mu_n^{-\frac{1}{2}}\right),\;\;\beta^-(\mu_n)=O\left(\mu_n^{-\frac{l(d)+1}{2}}\right).
$$
We use (\ref{psinw}) and
{\ref{alphanbetan}) in which 
$$
A=2^{-\frac{1}{2}}\mu_n^{-\frac{1}{4}}\frac{\left(\alpha^-(\mu_n)\right)^*}{\mid\alpha^-(\mu_n)\mid^2-\mid\beta^-(\mu_n)\mid^2}\sim
2^{-\frac{1}{2}}\mu_n^{-\frac{1}{4}},
\;\;n\rightarrow\infty,
$$
$$
B=2^{-\frac{1}{2}}\mu_n^{-\frac{1}{4}}\beta^-(\mu_n)=O\left(\mu_n^{-\frac{1}{4}-\frac{l(d)+1}{2}}\right),
$$
$$
\alpha\sim 2^{\frac{1}{2}}\mu_n^{\frac{1}{4}},\;n\rightarrow\infty,
\;\;
\beta=O\left(\mu_n^{-\frac{1}{4}-\frac{l(d)}{2}}\right).
$$
We conclude that we have
$\beta_n=O\left(\mu_n^{-\frac{l(d)+1}{2}}\right).$\\

(6) We consider the case $\tau_-=-\infty$, $\tau_+<\infty$,
$V_-\neq V_+$, that occurs if $d\leq 5$, $ \eta_0^-\geq 1$,
$\eta_0^+=0$, with $\eta_1^+>1$ and $l=2$ if  $\xi=\frac{d-1}{4d}$, or
$\eta_1^+>3$ and $l=4$ if  $\xi\neq\frac{d-1}{4d}$. With these
hypotheses, $V\in C^2(-\infty,\tau_+)\cap C^0((-\infty,\tau_+])$,
$V'\in L^2(-\infty,\tau_+)$, $V''\in L^1(-\infty,\tau_+)$,
$\int_{-\infty}^0\mid V(\tau)-V_-\mid d\tau<\infty$. We adapt the
method employed for the case (2). We take the Liouville-Green function
$w(\mu_n,\tau)$ defined by (\ref{wolver}) that satisfies
(\ref{epstrois}) again. We have
$$
 w(\mu_n;\tau)\sim
 2^{-\frac{1}{2}}(\mu_n+V_-)^{-\frac{1}{4}}\left(\alpha^{-}(\mu_n)e^{i\tau\sqrt{\mu_n+V_-}}+
\beta^{-}(\mu_n)e^{-i\tau\sqrt{\mu_n+V_-}}\right),\;\;\tau\rightarrow-\infty,
$$
$$
 \alpha^-(\mu_n)=\exp\left[i\int_0^{-\infty}\frac{V(\tau)-V_-}{(\mu_n+V(\tau))^{\frac{1}{2}}+(\mu_n+V_-)^{\frac{1}{2}}}d\tau\right]+O\left(\mu_n^{-\frac{3}{2}}\right),
\;\;
 \beta^-(\mu_n)=O\left(\mu_n^{-\frac{3}{2}}\right),
$$
$$
w'(\tau_+)+i\sqrt{\mu_n}w(\tau_+)=O\left(\mu_n^{\frac{1}{4}}\right),\;\;
w'(\tau_+)-i\sqrt{\mu_n}w(\tau_+)=O\left(\mu_n^{-\frac{5}{4}}\right).
$$
Using (\ref{psinw}) and
{\ref{alphanbetan}), we have $A,\alpha=O(1)$,
  $B,\beta=O\left(\mu_n^{-\frac{3}{2}}\right)$ and we conclude that
    $\beta_n=O\left(\mu_n^{-\frac{3}{2}}\right)$.\\

(7) The last case concerns $\tau_->-\infty$, $\tau_+=\infty$ that
occurs for $\eta_0^-\in\left(\frac{l-1}{l+1},1\right)$, $\eta_0^+\geq
1$. By a time reversing, it is equivalent to the case (5). Finally we
have investigated all the situations and the proof of the theorem is
complete.

\fin

\section{Conclusion}
In this work we have carried out a complete description of the
asymptotics of the solutions of the linear Klein-Gordon equation on a
FLRW universe beginning with a Big Bang and ending with a Big Crunch,
a Big Rip, or a Sudden Singularity. In these cases, the dynamics is
defined by a scalar equation with a time-dependent mass that can be  zero
or infinite at these singularities. We also have obtained similar
results for the semilinear Klein-Gordon equation in the simple case of subcritical exponents. The fundamental problem of the general
non-linearity $\mid u\mid^{p-1}u$ is open. This case is much more difficult and certainly
refined tools of harmonic analysis will be necessary, such as
Strichartz estimates or the $L^p-L^q$ continuity of the propagator. Due to
the time dependence of the coefficients and their singularity at the
Big Bang/Crunch/Rip, these properties are certainly hard to
get. Nevertheless, we note that in the case of the FLRW universes without
singularity or the De Sitter space-time,
similar results have be obtained \cite{galstian2009},
\cite{galstian2015}, \cite{galstian2017}, \cite{nakamura}. Finally we
have showed that the number of cosmological particle creation is
finite under rather general assumptions on the initial Big Bang and
the final Big Crunch or Big Brake. The concept of particles in the
dynamical universes is rather ambiguous. It would be interesting to
pursue the investigation of the quantum fields near a general
time singularity, by the study of fiducial quantum states in the
spirit of \cite{degner} or \cite{gerard}.

\section{Appendix}
This appendix is devoted to the asymptotics near a
time-future singularity,  of the scale factor 
expressed with the conformal time. The scale factor $a(t)$ is a positive  function in $C^l([t_0,t_+))$ and we
assume that near $t_+$ we have as $t\rightarrow t_+$
$$
\frac{d^ka}{dt^k}(t)=\frac{d^k}{dt^k}\left[c_0(t_+-t)^{\eta_0}+c_1(t_+-t)^{\eta_1}\right]+o\left((
  t_+-t)^{\eta_1-k}\right),\;\;k\leq l,
$$
where the coefficients $c_0,\;\eta_j\in\RR$ satisfy :
\begin{equation*}
 c_0>0,\;\eta_0<\eta_1.
\end{equation*}
The conformal time $\tau$ is defined by
$$
\tau:=\int_{t_0}^t\frac{1}{a(s)}ds,\;\;\tau_+:=\int_{t_0}^{t_+}\frac{1}{a(s)}ds\in(0,\infty],
$$
and the scale factor in this coordinate is expressed as
$$
\alpha(\tau):=a(t).
$$
We begin by the case $\eta_0<1$. Then $\tau_+<\infty$ and we write for
$t_+-t$ small enough
\begin{equation*}
\begin{split}
\tau_+-\tau&=c_0^{-1}\int_t^{t_+}(t_+-s)^{-\eta_0}\left[1-\frac{c_1}{c_0}(t_+-s)^{\eta_1-\eta_0}+o\left((t_+-s)^{\eta_1-\eta_0}\right)
\right]ds\\
&=c_0^{-1}(1-\eta_0)^{-1}\left(t_+-t\right)^{1-\eta_0}\left[1-\frac{c_1}{c_0}\frac{(1-\eta_0)}{(1-2\eta_0+\eta_1)}\left(t_+-t\right)^{\eta_1-\eta_0}+o\left(\left(t_+-t\right)^{\eta_1-\eta_0}\right)\right].
 \end{split}
\end{equation*}
We deduce that
\begin{equation*}
 \label{}
 t_+-t=\left[c_0(1-\eta_0)\right]^{\frac{1}{1-\eta_0}}\left(\tau_+-\tau\right)^{\frac{1}{1-\eta_0}}\left[1+O\left((\tau_+-\tau)^{\frac{\eta_1-\eta_0}{1-\eta_0}}\right)\right].
\end{equation*}
We conclude that the scale factor satisfies for $k\leq l$
\begin{equation}
 \label{applequn}
 \frac{d^k}{d\tau^k}\alpha(\tau)=\left[c_0(1-\eta_0)^{\eta_0}\right]^{\frac{1}{1-\eta_0}}\frac{d^k}{d\tau^k}\left((\tau_+-\tau)^{\frac{\eta_0}{1-\eta_0}}\right)+O\left((\tau_+-\tau)^{\frac{\eta_1}{1-\eta_0}-k}\right),\;\;\eta_0<1,
\end{equation}
and more specifically for $\eta_0=0$,
\begin{equation}
 \label{appzero}
  \frac{d^k}{d\tau^k}\alpha(\tau)= \frac{d^k}{d\tau^k}
  \left(c_0+c_1\left[c_0\right]^{\eta_1}(\tau_+-\tau)^{\eta_1}\right)+o\left((\tau_+-\tau)^{\eta_1-k}\right),\;\;k\leq
  l,\;\;\eta_0=0.
\end{equation}

If $\eta_0\geq 1$, then $\tau_+=+\infty$. First we consider the case
$\eta_0=1$ that is particular. We have for $t$ close to $t_+$
\begin{equation*}
\begin{split}
\tau&=c_0^{-1}\int_{t_0}^t\frac{1}{t_+-s}\left[1-\frac{c_1}{c_0}(t_+-s)^{\eta_1-1}+o\left((t_+-s)^{\eta_1-1}\right)\right]ds\\
&=-c_0^{-1}\ln(t_+-t)+k_0+O\left((t_+-t)^{\eta_1-1}\right),\;\;k_0\in\RR.
\end{split}
 \end{equation*}
We deduce that
$$
t_+-t=e^{-c_0(\tau-k_0)}+o\left(e^{-c_0\tau}\right),
$$
hence
\begin{equation}
 \label{appun}
 \frac{d^k}{d\tau^k}\alpha(\tau)=c_0
 e^{c_0k_0}\frac{d^k}{d\tau^k}\left(e^{-c_0\tau}\right)+o\left(e^{-c_0\tau}\right),\;\;k\leq
 l,\;\;\eta_0=1.
\end{equation}
Finally we consider the case $\eta_0>1$. We write
\begin{equation*}
\begin{split}
\tau&=c_0^{-1}\int_{t_0}^t(t_+-s)^{-\eta_0}\left[1-\frac{c_1}{c_0}(t_+-s)^{\eta_1-\eta_0}+o\left((t_+-s)^{\eta_1-\eta_0}\right)\right]ds\\
&=\frac{1}{c_0(\eta_0-1)}(t_+-t)^{1-\eta_0}+o\left((t_+-t)^{1-\eta_0}\right),
\end{split}
 \end{equation*}
hence
$$
(t_+-t)=\left[c_0(\eta_0-1)\tau\right]^{-\frac{1}{\eta_0-1}}+o\left(\tau^{-\frac{1}{\eta_0-1}}\right).
$$
We conclude that
\begin{equation}
 \label{appgequn}
  \frac{d^k}{d\tau^k}\alpha(\tau)=
  c_0^{\frac{1}{1-\eta_0}}(\eta_0-1)^{\frac{\eta_0}{1-\eta_0}}\frac{d^k}{d\tau^k}\left(\tau^{\frac{\eta_0}{1-\eta_0}}\right)+o\left(\tau^{\frac{\eta_0}{1-\eta_0}-k}\right),\;\;k\leq
  l,\;\;\eta_0>1.
\end{equation}




\begin{thebibliography}{10}

\bibitem{alho}
A. Alho, G. Fournodavlos, A. T. Franzen,
The wave equation near flat Friedmann-Lema\^{\i}tre-Robertson-Walker and
Kasner Big Bang singularities, 
{\it preprint} (2018), arXiv:1805.12558.


\bibitem{allen}
P. T. Allen, A. D.  Rendall,
Asymptotics of linearized cosmological perturbations,
{\it J. Hyperbolic Differ. Equ.} 7 (2010), no. 2, 255--277.

\bibitem{andersson}
L. Andersson, A. D. Rendall,
Quiescent cosmological singularities,
{\it Commun. Math. Phys.} 218 (2001), 479--511.


\bibitem{dodeca}
A. Bachelot, A. Bachelot-Motet,
Waves on accelerating dodecahedral universes,
{\it  Class. Quantum Grav. } 34   (2017),  no. 5, 055010, 39 pp. 


\bibitem{barrow2011}
J. D. Barrow, A. B. Batista, G. Dito, J. C. Fabris, M. J. S. Houndjo,
Sudden singularities survive massive quantum particle production,
{\it Phys. Rev. D} 84 (2011), 123518.

\bibitem{barrow2010}
J. D. Barrow, S. Cotsakis and A. Tsokaros,
A General Sudden Cosmological Singularity,
{\it  Class. Quantum Grav.  }27 (2010), 165017.

 \bibitem{barrow2015}
J. D. Barrow, A. A. H. Graham,
Singular Inflation,
{\it Phys. Rev. D} 91 (2015), 083513.

\bibitem{barrow2009}
J. D. Barrow, S. Z. W. Lip,
Classical Stability of Sudden and Big Rip Singularities,
{\it Phys. Rev. D} 80 (2009), 043518.


\bibitem{beyer2010}
F. Beyer, P. G. LeFloch,
Second-order hyperbolic Fuchsian systems and applications,
{\it  Class. Quantum Grav.  }27 (2010), 245012.

\bibitem{beyer2017}
F. Beyer, P. G. LeFloch,
Self-gravitating fluid flows with Gowdy symmetry near cosmological singularities,
{\it Comm. Partial Differential Equations} 42 (2017), no. 8, 1199--1248. 

\bibitem{calderon}
H. Calder\'on and W. A. Hiscock,
Quantum fields and ``big rip'' expansion singularities,
{\it Class. Quantum Grav.} 22 (2005), L23--L26.

\bibitem{caldwell}
R. R. Caldwell, M.  Kamionkowski, N. N. Weinberg,
Phantom Energy and Cosmic Doomsday,
{\it Phys.Rev.Lett. } 91 (2003), 071301.

\bibitem{visser}
C. Catto\"{e}n, M. Visser,
Necessary and sufficient conditions for big bangs, bounces, crunches,
rips, sudden singularities and extremality events,
{\it Class. Quantum Grav.} 22 (2005), 4913--4930.


\bibitem{chimento}
L. P. Chimento,  M. G. Richarte,
Big brake singularity is accommodated as an exotic quintessence field,
{\it Phys. Rev. D} 93 (2016), no. 4, 043524. 

\bibitem{choquet}
Y. Choquet-Bruhat,
{\it General Relativity and the Einstein Equations}
(Oxford University Press, 2009).


\bibitem{cohen}
A. Cohen, T. Kappeler,
Scattering and inverse scattering for steplike potentials in the Schr\"odinger equation,
{\it Indiana Univ. Math. J. } 34 (1) (1985), 127--180.

\bibitem{[1]}
T. Damour, M. Henneaux,  A. D. Rendall, M. Weaver,
Kasner-like behaviour for subcritical Einstein-matter systems,
{\it Ann. Henri Poincar\'e} 3 (2002), no. 6, 1049--1111.




\bibitem{degner}
A. Degner, R. Verch,
Cosmological particle creation in states of low energy,
{\it J. Math. Phys.} 51 (2010), 022302.

\bibitem{delsanto}
D. Del Santo, T. Kinoshita, M. Reissig,
Klein-Gordon Type Equations with a Singular Time-dependent Potential,
{\it Rend. Istit. Mat. Univ. Trieste} 39 (2007), 141--175.

\bibitem{ebert2017}
M. R. Ebert, W. N. Nascimento,
A classification for wave models with time-dependent mass and speed of
propagation, {\it preprint}
(2017), arXiv:1710.01212.

\bibitem{ebert2018}
M. R. Ebert, M. Reissig,
Regularity theory and global existence of small data solutions to
semi-linear de Sitter models with power non-linearity,
{\it Nonlinear Analysis: Real World Appl.} 40 (2018), 14--54.

\bibitem{eichhorn}
J. Eichhorn,
The Boundedness of Connection Coefficients and their Derivatives,
{\it Math. Nachr.} 152 (1991), 145--158.

\bibitem{eichhorn2}
J. Eichhorn,
The Banach manifold structure of the space of metrics on noncompact manifolds,
{\it Differential Geom. Appl.} 1 (1991), no. 2, 89--108. 

\bibitem{elizalde}
E. Elizalde, S. Nojiri, S.D. Odintsov,
Late-time cosmology in (phantom) scalar-tensor theory: dark energy and
the cosmic speed-up,
{\it Phys. Rev. D}, 70 (2004), 043539.

\bibitem{fernandez}
L. Fern{\'a}ndez-Jambrina, R. Lazkoz,
Classification of cosmological milestones,
{\it Phys. Rev. D} (3) 74 (2006), no. 6, 064030.

\bibitem{[2]}
H. Friedrich,
On the existence of n-geodesically complete or future complete
solutions of Einstein's field equations with smooth asymptotic
structure,
{\it Comm. Math. Phys.} 107 (1986), no. 4, 587--609.


\bibitem{fulling}
S. A. Fulling,
{\it Aspects of Quantum Field Theory in Curved Space-Time},
London Mathematical Society Student Texts, 17 (Cambridge University
Press, 1989).

\bibitem{fulling-79}
S. A. Fulling,
Remarks on positive frequency and Hamiltonians in expanding universes. 
{\it Gen. Relativity Gravitation} 10 (1979), no. 10, 807--824. 

\bibitem{galstian2009}
A. Galstian, T. Kinoshita, K. Yagdjian,
A note on wave equation in Einstein and de Sitter space-time,
{\it J. Math. Phys. } 51 (2010), no. 5, 052501, 18 pp. 

\bibitem{galstian2015}
A. Galstian, K. Yagdjian,
Global solutions for semilinear Klein-Gordon equations in FLRW
spacetimes,
{\it Nonlinear Analysis} 113 (2015), 339--356.

\bibitem{galstian2017}
A. Galstian, K. Yagdjian,
Global in time existence of self-interacting scalar  field in de Sitter
spacetimes,
{\it Nonlinear Analysis: Real World Appl.} 34 (2017), 110--139.

\bibitem{gerard}
C. G\'erard, O. Oulghazi, M. Wrochna,
Hadamard States for the Klein-Gordon Equation on Lorentzian Manifolds
of Bounded Geometry,
{\it Commun. Math. Phys.} 352 (2017), 519--583.

\bibitem{gorini}
V. Gorini, A. Y. Kamenshchik, U. Moschella, V. Pasquier,
 Tachyons, scalar fields and cosmology,
 {\it Phys. Rev. D} 69 (2004), 123512.

\bibitem{grib}
A. A. Grib, Yu. V. Pavlov,
Particle creation in the early Universe: achievements and problems
{\it Grav. Cosmol.} 22 (2016), 107--115.

\bibitem{grosse}
N. Gro{\ss}e, C. Schneider,
Sobolev spaces on Riemannian manifolds with bounded geometry: General coordinates and traces,
{\it Math. Nach.} 286, 16 (2013), 1586--1613.


\bibitem{[3]}
M. Hadzic, J. Speck,
The global future stability of the FLRW so- lutions to the
dust-Einstein system with a positive cosmological constant,
{\it J. Hyperbolic Differ. Equ.} 12 (2015), no. 1, 87--188.


\bibitem{[4]}
S. Kichenassamy, A. D. Rendall,
Analytic description of singularities in Gowdy spacetimes,
{\it Classical Quantum Gravity} 15 (1998), no. 5, 1339--1355.



\bibitem{leray}
J.  Leray,
\newblock {\it Hyperbolic differential equations}
(Princeton University Press, 1953).

\bibitem{lions}
J-L.  Lions, E. Magenes,
\newblock {\it Non-Homogeneous Boundary Value Problems and
  Applications, volume I},
Die Grundlehren der mathematischen Wissenschaften, 181
(Springer-Verlag, 1972).

\bibitem{nakamura}
M. Nakamura,
The Cauchy problem for semi-linear Klein-Gordon equations in de Sitter spacetime,
{\it J. Math. Anal. Appl.} 410 (2014), 445--454.

\bibitem{nist}
F. W. J. Olver,
D. W. Lozier,
R. F. Boisvert,
C. W. Clark,
{\it NIST Handbook of Mathematical Functions}
(Cambridge University Press, 2010).


\bibitem{nojiri}
S. Nojiri, S. D. Odinstov, S. Tsujikawa,
Properties of singularities in (phantom) dark energy universe,
{\it Phys. Rev. D.} 71 (2005), 063004.

\bibitem{olver}
F. W. J. Olver,
Error bounds for the Liouville-Green (or WKB) approximation,
{\it Proc. Cambridge Philos. Soc.} 57 (1961) 790--810. 

\bibitem{parker}
L. E. Parker, D. J. Toms,
{\it Quantum Field Theory in Curved Spacetime, Quantized Fields and
  Gravity,}
Cambridge Monographs on Mathematical Physics (Cambridge University
 Press 2009).

\bibitem{reed-simon3}
M. Reed, B. Simon,
{\it Methods of modern mathematical physics III, Scattering Theory}
(Academic Press, 1975).


\bibitem{[5]}
A.D. Rendall,
Fuchsian analysis of singularities in Gowdy
spacetimes beyond analyticity,
{\it Classical Quantum Gravity} 17 (2000),
no. 16, 3305--3316.

\bibitem{[6]}
H. Ringstr\"om,
Future stability of the Einstein non-linear scalar field system,
{\it Invent. math.} 173 (2008), 123--208.


\bibitem{[7]}
H. Ringstr\"om,
Strong cosmic censorship in $T^3$-Gowdy
spacetimes,
{\it Ann. of Math.} (2) 170 (2009), no. 3, 1181--1240.


\bibitem{[8]}
H. Ringstr\"om,
{\it On the topology and future stability of the universe},
 Oxford Mathematical Monographs (Oxford University Press,  2013).


\bibitem{ringstrom2017}
H. Ringstr\"om,
Linear systems of wave equations on cosmological backgrounds with
convergent asymptotics,
{\it preprint} (2017), arXiv:1707.02803.

\bibitem{ringstrom2018}
H. Ringstr\"om,
A unified approach to the Klein-Gordon equation on Bianchi
backgrounds,
{\it preprint} (2018), arXiv:1808.00786.



\bibitem{[10]}
I. Rodnianski,  J. Speck,
The nonlinear future stability of the FLRW family of solutions to the
irrotational Euler-Einstein system with a positive cosmological
constant,
{\it J. Eur. Math. Soc.} 15 (2013), no. 6, 2369--2462.

\bibitem{[11]}
I. Rodnianski,  J. Speck,
A regime of linear stability for the Einstein-scalar field system with
applications to nonlinear big bang formation,
{\it Ann. of Math.} (2) 187 (2018), no. 1, 65--156.


\bibitem{[12]}
I. Rodnianski, J. Speck,
Stable Big Bang formation in near-FLRW solutions to the
Einstein-scalar field and Einstein-stiff fluid systems,
{\it Selecta Math. (N.S.)} 24 (2018), no. 5, 4293--4459.


\bibitem{[13]}
I. Rodnianski, J. Speck,
On the nature of Hawking's incompleteness for the Einstein-vacuum
equations: The regime of moderately spatially anisotropic initial
data,
{\it preprint} (2018) arXiv:1804.06825.


\bibitem{rosenberg}
S. Rosenberg,
{\it The Laplacian on a Riemannian manifold. An introduction to
  analysis on manifolds,}
 London Mathematical Society Student Texts, 31 (Cambridge University
 Press 1997).

\bibitem{shale62}
D. Shale,
Linear symmetries of free boson fields,
{\it Trans. Amer. Math. Soc.} 103 (1962), 149--167.

\bibitem{[14]}
J.  Speck,
The nonlinear future stability of the FLRW family of solutions to
the Euler-Einstein system with a positive cosmological constant,
{\it Selecta Math.}18 (2012), no. 3, 633--715.

\bibitem{[15]}
J. Speck,
The Maximal Development of Near-FLRW Data for the Einstein-Scalar
Field System with Spatial Topology $S^3$,
{\it Comm. Math. Phys.} 364 (2018), no. 3, 879--979.


\bibitem{strauss66}
W. A. Strauss,
On continuity of functions with values in various Banach spaces,
{\it Pacific J. Math.}, 19 (3) (1966), 543--551.



\bibitem{strichartz}
R. S. Strichartz,
Analysis of the Laplacian on the complete Riemannian manifold,
{\it J. Funct. Anal. } 52 (1983), 48--79.

\bibitem{[16]}
C.  Svedberg,
 Future Stability of the Einstein-Maxwell-Scalar Field System,
{\it Ann. Henri Poincar\'e} 12, No. 5, (2011) 849--917.


\bibitem{tanabe}
H. Tanabe,
{\it Functional analytic methods for partial differential equations},
Pure and Applied Mathematics, 204
(Marcel Dekker, 1997).


\end{thebibliography}
\end{document}